\newcommand{\dd}{{\rm d}}
\newcommand{\by}{{\bf y}}
\newcommand{\bsigma}{\boldsymbol\sigma}

\documentclass[11pt,preprint]{aastex}

\usepackage{enumerate,natbib,hyperref,graphicx,amsmath,amssymb}

\shorttitle {Using X-ray burst oscillations to determine neutron star masses and radii}
\shortauthors {Lo et al.}

\begin{document}

\title {Determining neutron star masses and radii using energy-resolved waveforms of X-ray burst oscillations}

\author {Ka Ho Lo\altaffilmark{1}, M. Coleman
Miller\altaffilmark{2}, Sudip Bhattacharyya\altaffilmark{3}, Frederick K. Lamb\altaffilmark{1,4}}

\affil{
{$^1$}{Center for Theoretical Astrophysics and
Department of Physics, University of Illinois at
Urbana-Champaign, 1110 West Green Street, Urbana, IL 61801-3080, USA; fkl@illinois.edu}\\
{$^2$}{Department of Astronomy and Joint Space-Science Institute, University of Maryland, 
College Park, MD 20742-2421, USA}\\
{$^3$}{Department of Astronomy and Astrophysics, Tata Institute of Fundamental Research, Mumbai 400005, India}\\
{$^4$}{Department of Astronomy, University of Illinois at
Urbana-Champaign, 1110 West Green Street, Urbana, IL 61801-3074, USA}
}

\begin{abstract}
\noindent  
Simultaneous, precise measurements of the mass $M$ and radius $R$ of neutron stars can yield uniquely valuable information about the still uncertain properties of cold matter at several times the density of nuclear matter. One method that could be used to measure $M$ and $R$ is to analyze the energy-dependent waveforms of the X-ray flux oscillations seen during some thermonuclear bursts from some neutron stars. These oscillations are thought to be produced by X-ray emission from hot regions on the surface of the star that are rotating at or near the spin frequency of the star. Here we explore how well $M$ and $R$ could be determined by analyzing energy-resolved X-ray data obtained using a future space mission having 2--30 keV energy coverage and an effective area of 10~m$^2$, such as the proposed \textit{LOFT} or \textit{AXTAR} missions. We do this by generating energy-dependent synthetic observed waveforms for a variety of neutron star and hot spot properties and then using a Bayesian approach and Markov chain Monte Carlo sampling methods to determine the joint posterior probability distributions of the parameters in our waveform model, given each synthetic waveform. We use the resulting posterior distributions to determine Bayesian confidence regions in the $M$-$R$ plane by marginalizing the other parameters in our model. We explore how the sizes and positions of these confidence regions depend on the inclinations of the hot spot and the observer, the background count rate, and deviations in the actual shape of the hot spot, radiation beaming pattern, and spectrum from those assumed in the waveform model. We also explore the effect on the confidence regions if the distance to the star or the inclination of the observer are known from other measurements, if a resonance scattering line is observed in the burst oscillation spectrum, or if the properties of the background are independently known. We assume that about 10$^6$ counts are collected from the hot spot and that all sources of background contribute about $0.3\times10^6$, 10$^6$, or $9\times10^6$ counts.

We find that the uncertainties in the measured values of $M$ and $R$ depend strongly on the inclination of the hot spot relative to the spin axis. If the hot spot is within 10$^\circ$ of the rotation equator, both $M$ and $R$ can usually be determined with an uncertainty of about 10\%. If instead the spot is within 20$^\circ$ of the rotation pole, the uncertainties are so large that waveform measurements alone provide no useful constraints on $M$ and $R$. Observation of an identifiable atomic line in the hot-spot emission always tightly constrains $M/R$; it can also tightly constrain $M$ and $R$ individually if the spot is within about 30$^\circ$ of the rotation equator. These constraints can usually be achieved even if the burst oscillations vary with time and data from multiple bursts must be used to obtain 10$^6$ counts from the hot spot. Independent knowledge of the observer's inclination can greatly reduce the uncertainties, as can independent information about the background. Knowledge of the star's distance can also help, but not as much. Modest deviations of the actual spectrum from that assumed in the waveform model have little effect on the accuracies or uncertainties of $M$ and $R$ estimates. Large deviations of the actual shape of the hot spot or the radiation beaming pattern from those assumed can sometimes increase the uncertainties significantly, and can in some cases bias $M$ and $R$ estimates by moderate amounts. Our results show that if a sufficient number of burst oscillations produced by hot spots at high inclinations are observed using the next generation of X-ray timing satellites, one can constrain $M$ and $R$ tightly enough to discriminate strongly between competing models of cold, high-density matter.
\end{abstract}

\keywords {stars: neutron --- stars: rotation --- X-rays: bursts --- X-rays: stars}


\section{INTRODUCTION}
\label{sec:introduction}

The properties of matter at extremely high densities are among the most important currently unresolved questions in physics and astronomy. Neutron stars contain large quantities of cold matter at densities that are otherwise inaccessible. Studies of these stars can therefore help determine the properties of such matter. In particular, simultaneous measurements of the mass $M$ and radius $R$ of neutron stars could provide tight constraints on the equation of state of ultradense matter \citep[see, e.g.][]{latt07a, latt07b, read09, ozel09b, hebe10}.

Following the discovery of thermonuclear X-ray bursts from neutron stars in the mid-1970s (\citealt{grin76, lewi76}; for a review, see \citealt{lewi93}) and the discovery two decades later that some of these bursts produce X-ray flux oscillations at or near the star's spin frequency (\citealt{stro96}; for a review, see \citealt{watt12}), several approaches have been proposed for using observations of X-ray bursts to determine $M$ and $R$.

The first attempts to constrain neutron star properties using X-ray bursts used their apparent peak luminosity and spectral temperature
\citep{para79, gold79, hosh81, mars82}. With improved understanding of the complexities of burst emission, this approach was developed into a method for measuring $M$ and $R$ \cite[see][]{lewi93}. 
This method relies on five assumptions:
(1)~that the radius of the photosphere at ``touchdown'' (defined as the moment after the photosphere has reached its maximum inferred radius when the temperature $T_{\rm bb}$ derived from fitting a Planck spectrum to the observed spectrum is highest) is the stellar radius; (2)~that the emitting area at ``touchdown'' is the entire surface of the star;
(3)~that the flux seen at Earth at touchdown is the Eddington flux
diluted by the distance to the burst source; 
(4)~that the distance to the source is known; and
(5)~that the ratio of $T_{\rm bb}$ to the effective temperature
$T_{\rm eff}$ is known.

This first method has been widely used to constrain $M$ and $R$ \citep[see, e.g.,][]{para82, pacz86, para90}. Tight constraints on $M$ and $R$ have recently been derived by using this method to analyze X-ray burst data obtained using the \textit{Rossi X-ray Timing Explorer} (\textit{RXTE}) \citep{ozel06, ozel09a, ozel10, guve10a, guve10b}. The derived constraints on $M$ and $R$ are substantially smaller than the uncertainties in the observed values of the input parameters because most of the observed values are discarded, since they would produce values of $M$ and $R$ that are complex numbers (\citealt{ozel10}; see also \citealt{stei10}). When this inconsistency is addressed and systematic errors are fully included, the uncertainties in $M$ and $R$ are likely to be substantially larger \citep[see, e.g.,][]{stei10}.

A second approach to measuring $M$ and $R$ using X-ray burst observations is to fit detailed spectral models to high-precision measurements of X-ray burst spectra \citep{majc05a, mill11, mill13}. This approach assumes that the spectrum predicted by the atmospheric model is an accurate physical description of the observed spectrum. If it is, fitting a grid of spectral models to the observed X-ray spectra yields the radiation temperature $T_{\rm co}$ in the locally comoving frame at the stellar surface as well as $M$, $R$, and the composition \citep{mill11, sule12}. This approach can only be used if high-precision spectral data and accurate, high-precision spectral models are both available. At present, adequate data are available only for the \mbox{4U~1820$-$30} superburst \citep{mill11, mill13}.

A third approach for determining $M$ and $R$ using X-ray burst observations is to fit models of burst oscillation waveforms to observations of these waveforms (\citealt{stroh97, mill98a}; see also \citealt{wein01}). As \citet{mill98a} demonstrated, fitting energy-resolved waveform data can improve the constraints. Burst oscillations are thought to be produced by X-ray emission from a region on the surface of the star that is hotter than the rest of the surface and is rotating at or near the spin frequency of the star (\citealt{stro96}; for a review, see \citealt{watt12}). Such a  hotter region could be present either because a part of the stellar surface has retained more heat than other parts after thermonuclear burning has occurred or because a disturbance in the outer layers of the star, such as a surface normal mode, has made a localized region hotter. The observed amplitude of the waveform constrains the inclination of the hot spot relative to the spin axis, the observer's inclination, and the compactness ($M/R$) of the star, the last primarily via general relativistic light-bending effects. The observed asymmetry and harmonic content of the waveform constrain the component of the velocity of the emitting region in the observer's direction, primarily via special relativistic Doppler boosts and aberration. Because the rotation frequency of the emitting region is accurately known from the oscillation frequency, knowing the surface velocity constrains the stellar radius.

\cite{nath02} explored the constraints that could be derived on the compactness of 4U~1636$-$536 and 4U~1728$-$34 by analyzing \textit{RXTE} observations of the bolometric flux oscillations that occur during the rise of X-ray bursts from these two neutron stars. They modeled the bolometric waveforms during the burst rise using small hot spots that expand linearly with time, neglected the relativistic Doppler shifts and aberration produced by the rotational motion of the hot spots, and assumed that the background was known. They found that they could not determine whether the oscillations are produced by a single spot or two antipodal spots, or constrain the hot spot and viewing geometry assuming either of these alternatives, using bolometric \textit{RXTE} waveform data. The primary reason is that the changes in the waveform produced by changes in different parameters of the model are very similar. As a result of this degeneracy, they were unable to constrain the stellar compactness for models with a single hot spot, even if they assumed that the spot and the observer were known from other information to be in the rotation equator. They were able to obtain interesting upper bounds on the compactness for models with two antipodal hot spots and argued that it would be possible to simultaneously constrain the stellar compactness and the hot spot and viewing geometries with a count rate $\sim\,$10--20 times higher than the \textit{RXTE} rate. However, the 290~Hz subharmonic in the 4U~1636$-$536 waveform that appeared significant in one burst \citep{mill99, stro01}, and suggested consideration of two antipodal spots, was subsequently found not to be significant in a more detailed analysis of additional data \citep{stro01}.

\cite{bhat05} investigated the constraints on the stellar compactness, hot spot properties, and system geometry that could be obtained by fitting model pulse profiles to an average energy-resolved oscillation profile produced by folding and stacking \textit{RXTE} observations of 22 burst oscillations in three groups of bursts from XTE~J1814$-$338. Stable oscillations occur throughout the bursts produced by this neutron star \citep{stro03}. In generating their energy-resolved oscillation profile models, \citeauthor{bhat05} assumed a single hot spot and an oscillation-phase-independent background chosen so that the model produces the observed total number of counts in each energy channel, when summed over oscillation phase. They included the relativistic Doppler shifts, aberration, and frame dragging produced by the rotation of the star, as well as gravitational redshift and light bending effects. In order to include frame dragging, \citeauthor{bhat05} had to construct general relativistic stellar models numerically and therefore considered just two illustrative equations of state. They were able to achieve acceptable fits to the average oscillation profile using these equations of state, but obtained only weak constraints on the stellar compactness, hot spot properties, and system geometry.

\cite{stro04} explored the constraints on $M$ and $R$ that might be achieved by analyzing burst oscillation observations with much higher count rates than were achieved using \textit{RXTE}. He did this by analyzing a synthetic bolometric waveform similar to a waveform seen in 4U~1636$-$536. 
\citeauthor{stro04} generated his synthetic and model waveforms using a code that is similar to the one used by \cite{nath02} but included the relativistic Doppler shifts and aberration produced by the rotational motion of the hot spots. The synthetic observed waveform did not include a background of any kind. 
\citeauthor{stro04} then assumed that the absence of any background is known to the observer by means other than fitting the 
waveform.\footnote{We note that attempting to remove the background by subtracting the pre-burst count rate assumes that the pre-burst background, which is presumably produced at least in part by accretion onto the star, persists unchanged throughout the burst. This introduces possible systematic errors, because luminous bursts are expected to alter the accretion flow significantly \citep[see, e.g.,][]{mill96b, worp13}. Even if the pre-burst background persists unchanged throughout the burst, subtracting it from the count rate during the burst, rather than including the background as a component of the model, incorrectly neglects the fluctuation in the number of counts produced by the background and the corresponding uncertainties in the model parameters. This conceptual error is built into the current version of the XSPEC analysis package, so all data analysis done using XSPEC makes this error.}
He also assumed that the spot and the observer are both in the rotation equator, which is the most favorable possible geometry, and made the strong assumptions that the hot spot size and location, the radiation spectrum and beaming pattern, and the observer's inclination are known independently of the waveform, so that only $M$ and $R$ have to be estimated using the waveform data. These assumptions eliminate the strong degeneracies between $M$, $R$, and the other model parameters that greatly increase the uncertainties in more realistic situations. Based on these assumptions, \citeauthor{stro04} concluded that an X-ray timing mission with a collecting area $\sim\!10$ times larger than \textit{RXTE} would be able to determine $M$ and $R$ to within a few percent, thereby placing interesting constraints on the equation of state of neutron star matter.

\cite{muno03} examined averages of the energy-resolved flux oscillations observed from several X-ray burst sources using \textit{RXTE}. They found that folded oscillations observed in their higher energy bands arrived later than those in their lower energy bands, although the energy dependence varied significantly with epoch and source. Their analysis of folded and averaged \mbox{4U~1636$-$536}  oscillation profiles showed the clearest evidence for such a trend, which would be inconsistent with a simple rotating spot model of burst oscillations. Recently, \citet{arti13} have analyzed the same \mbox{4U~1636$-$536} data in much more detail and find that the data are entirely consistent with a simple rotating spot model, although the parameters of such models are poorly constrained by these data.

New mission concepts are now being proposed that would provide much larger count rates than previous missions. Plans for the proposed Large Observatory for X-ray Timing (LOFT) include a detector with an effective area $\sim\!10$~m$^2$ at 8~keV and 2--30~keV energy coverage \citep{fero10, mign12, delm12}, more than an order of magnitude larger than the area of the \textit{RXTE} Proportional Counter Array (PCA). The Advanced X-ray Timing Array (\textit{AXTAR}) concept includes a detector with an effective area greater than 3~m$^2$ and 2--50~keV energy coverage \citep{chak08, ray11}. One of the main motivations for these missions is to determine the properties of neutron stars with much higher precision than has been possible previously, by analyzing high-precision observations of the waveforms of burst oscillations. Although we focus here on exploring the constraints on $M$ and $R$ that can be obtained by analyzing the waveforms of X-ray burst oscillations, our results are equally relevant for measuring $M$ and $R$ using the waveforms produced by X-ray emission from the heated polar caps of isolated rotation-powered millisecond pulsars \citep[see, e.g.][]{bogd07, bogd08}, which is the goal of the proposed NICER mission \citep{gend12}. Using methods similar to ours, \cite{bogd13} has recently derived a lower limit of 11.1~km on the radius of the isolated pulsar PSR~J0437$-$4715, assuming its mass is 1.76~$M_\odot$ and that systematic errors can be neglected.

In this paper we explore the constraints on $M$ and $R$ that could be derived by analyzing energy-resolved burst oscillation waveforms obtained using a future, satellite-borne detector with 2--30~keV energy coverage and an effective area 10 to 20 times larger than the \textit{RXTE PCA}. We do this by first generating energy-dependent synthetic observed waveforms for a variety of neutron star and hot spot properties. We then use a Bayesian approach and Markov chain Monte Carlo (MCMC) sampling methods to determine the constraints on $M$ and $R$ that can be obtained by analyzing these synthetic observed waveforms. Specifically, we determine the joint posterior probability distribution of the parameters in our waveform model, given the synthetic waveform of interest, use this distribution to determine the joint posterior distribution of $M$ and $R$ by marginalizing the other parameters in the waveform model, and then use the joint distribution of $M$ and $R$ to determine the most probable values of $M$ and $R$ and Bayesian confidence regions in the $M$-$R$ plane. 

Our analysis applies to any waveforms produced by emission from a single region of hotter gas that is rotating about the star. Such a region could be present either because a part of the stellar surface has retained more heat than other parts after thermonuclear burning has occurred or because a disturbance in the outer layers of the star, such as a surface normal mode, has made a localized region hotter \citep[see][]{watt12}. Oscillations with the amplitudes $\sim\,$10\% that are required to obtain significant constraints on $M$ and $R$ are very probably produced by a single hotter region \citep[see][]{lamb09a}.

We explore how the sizes and positions of these confidence regions depend on the inclinations of the hot spot and the observer and the background count rate. We also explore the effect on the confidence regions if the distance to the star or the inclination of the observer are known from other measurements, if a resonance scattering line is observed in the burst oscillation spectrum, or if the properties of the background are independently known. Finally, we explore the effects of deviations in the actual shape of the hot spot, radiation beaming pattern, and spectrum from those assumed in the fitted model. We assume that about 10$^6$ counts are collected from the hot spot and that all sources of background contribute about $0.3\times10^6$, 10$^6$, or $9\times10^6$ counts. Our treatment of the background is very conservative, in the sense that we usually make no assumptions about the magnitude or spectrum of the background. We do not even assume that the background is constant, only that it does not vary at frequencies commensurate with the burst oscillation frequency.

We find that the uncertainties in the measured values of $M$ and $R$ depend strongly on the inclination of the hot spot relative to the stellar rotation axis. If the hot spot is within 10$^\circ$ of the rotation equator, both $M$ and $R$ can usually be determined with an uncertainty of about 10\%. If instead the spot is within 20$^\circ$ of the rotation pole, the uncertainties are so large that waveform measurements alone provide no useful constraints on $M$ and $R$. The uncertainties in $M$ and $R$ are affected little by background count rates less than or comparable to the count rate from the hot spot, but become significantly larger for higher background count rates. The precisions we report here can usually be achieved even if the burst oscillations vary with time and data from multiple bursts must be combined to obtain 10$^6$ counts from the hot spot.

Observation of an identifiable atomic line in the hot-spot emission always tightly constrains $M/R$; it can also tightly constrain $M$ and $R$ individually, if the spot is within about 30$^\circ$ of the rotation equator. Independent knowledge of the observer's inclination can greatly reduce the uncertainties, as can independent information about the background. Knowledge of the star's distance can also help, but not as much.

Modest deviations of the actual spectrum from that assumed in the fitted model have little effect on the accuracy or uncertainty of $M$ and $R$ estimates. 
Large deviations of the actual radiation beaming pattern from the pattern assumed in the waveform model can increase the uncertainties of $M$ and $R$ measurements substantially.
In some cases, but not always, large deviations of the actual shape of the hot spot from the circular shape assumed in the waveform model can increase the uncertainties of $M$ and $R$ estimates and bias them by moderate amounts. The physical conditions that produce tight constraints on $M$ and $R$ (relatively small spots far from the rotation pole) are the conditions in which the shape of the spot is unimportant.

Our results show that if a sufficient number of burst oscillations produced by hot spots at high inclinations are observed using the next generation of X-ray timing satellites, one can constrain $M$ and $R$ tightly enough to discriminate strongly between competing models of cold, high-density matter.

The remainder of this paper is organized as follows. In Section~\ref {sec:approach}, we outline our approach to generating synthetic observed waveforms and producing the model waveforms that we fit to the synthetic waveform data. In Section~\ref{sec:methods}, we describe the computational methods we used to produce synthetic waveform data and construct model waveforms, and the MCMC computational methods we used to determine the posterior probability distributions of the parameters in the model, given a synthetic waveform. In Section~\ref{sec:results}, we describe our results and in Section~\ref{sec:conclusions}, we summarize our conclusions. In Appendix~\ref{app:code-validation}, we summarize the suite of test problems and  solutions that we use to validate our waveform and MCMC codes. In Appendix~\ref{app:jointfits}, we discuss the constraints on system parameters that can be obtained by jointly fitting many segments of waveform data, from a single burst or from multiple bursts.

\clearpage
\newpage


\newpage
\section{APPROACH}
\label{sec:approach}

\subsection{Generation of ``observed'' and model waveforms}
\label{sec:approach:waveform-assumptions}

In the present work we fit energy-resolved model waveforms to  energy-resolved synthetic ``observed'' waveforms, using a Bayesian approach. For both waveforms, we assume that the oscillation is produced by emission from a single, uniformly emitting, hotter region on the stellar surface. Such a region could be present either because a part of the stellar surface has retained more heat than other parts after thermonuclear burning occurred or because a disturbance in the outer layers of the star has made a localized region hotter \citep[see][]{watt12}. 

We consider synthetic observed waveforms generated by circular and elongated hot spots (see 
Section~\ref{sec:results:waveform-analysis:synthetic-waveforms}), but to reduce the number of fitted parameters our model waveforms assume the hot spot is circular
(see Section~\ref{sec:results:waveform-analysis:model-waveforms}). For simplicity, we usually assume the hotter region emits radiation with a Planck spectrum and 100\% efficiency but with the beaming pattern appropriate for an electron scattering atmosphere, assumptions that are mutually inconsistent. This inconsistency can be avoided when fitting real data by including an appropriate color factor in the emission model used to construct fitted waveforms.
We also assume that the radiation propagating from the emitting area on the stellar surface reaches the observer without interacting with any ambient or intervening matter, but we do include background counts to illustrate the effects of possible background emission from the stellar surface, an accretion disk, and other sources in the field, as well as the instrumental background. 

In generating synthetic and model waveforms, we assume that the hotter region has a constant size and shape, is located at a fixed stellar rotational latitude, and rotates at a constant frequency. These assumptions are not as restrictive as they might at first appear. The reason is that constraints on $M$ and $R$ similar to those obtained for waveforms that satisfy these assumptions can usually be obtained by appropriately analyzing data from a single burst or from multiple bursts, provided the data contains the same number of counts as assumed in our analysis. In particular, constraints on $M$ and $R$ similar to those found here can usually be achieved even if the oscillation frequency and other physical parameters vary during the burst or from burst to burst (see Section~\ref{sec:results:constraints:combining-data} and~Appendix~\ref{app:jointfits}).

In constructing the synthetic waveforms, we include a constant background component. This component is a catch-all for all counts not produced by radiation from the hot spot. These counts could be produced by emission from unassociated sources in the field, the accretion disk, the non-spot portion of the star, instrumental backgrounds, or any combination of these. For simplicity, we model this background by adding emission from the entire stellar surface with the beaming pattern expected for an electron scattering atmosphere and a spectrum having the shape of a Planck spectrum with a temperature lower than the temperature of the hot spot. We normalize the background spectrum to achieve the desired number of background counts. The number of counts contributed by this emission is important, but not their detailed properties.

In fitting model waveforms to the synthetic waveform, we treat the background component in the model waveform very conservatively, in the sense that we usually make no assumptions about its magnitude or spectrum. We do not even assume that the background is constant, only that it does not vary at frequencies commensurate with the hot spot rotation frequency. Any prior knowledge of the properties of the background can be used to restrict the background model and usually tighten the constraints on the values of the system parameters that can be derived from waveform observations.

We compute the time- and energy-resolved waveforms that would be seen by a distant observer using the Schwarzschild plus Doppler (S+D) approximation introduced by \citet[][]{mill98a}. The S+D approximation treats exactly all special relativistic effects (such as relativistic Doppler boosts and aberration) produced by the rotational motion of the hot spot, but treats the star as spherical and uses the Schwarzschild spacetime to compute the general relativistic redshift, trace the propagation of light from the stellar surface to the observer, and calculate light travel-time effects. The S+D approximation does not include the effects of stellar oblateness or frame dragging. However, for the stars considered here, and indeed for any stars that do not both rotate rapidly and have very low compactness, the effects of stellar oblateness and frame dragging are minimal and are negligible compared to the uncertainties in the X-ray emission \citep[see][]{cade07}. 

The S+D approximation allows us to exploit the spherical symmetry of the spacetime to reduce substantially the number of integrations needed to trace rays from the emitting area to the observer. As a result, this code is much faster than codes that take into account the oblateness of the star, approximate the star's exterior spacetime using the Kerr spacetime, or use numerically computed stars and exterior spacetimes. This speed is essential, because finding the high-probability region of the parameter space of the waveform model and then determining the posterior probability density in this region with some precision requires accurate computation of the likelihood at a large number of points in a high-dimensional parameter space, and therefore requires many ray tracings.
 
The waveform code we use is based on the code used and validated in the analysis of accretion-powered millisecond X-ray pulsar waveforms by \citet{lamb09a, lamb09b}. We report the results of further code validation tests in 
Appendix~\ref{app:code-validation}.

To determine the constraints on $M$ and $R$ that can be derived from burst oscillation waveforms, we compute the posterior probability distribution of all the parameters in the waveform model, for each of a variety of synthetic observed waveform data sets, using a standard Bayesian approach. We use an MCMC algorithm to sample the parameters of the waveform model and compute the likelihood of each set of parameters, given the synthetic waveform being considered. 

We construct each synthetic observed waveform data set by Poisson sampling the counts in each phase and energy bin of each synthetic observed waveform that we computed using the S+D approximation, to mimic the statistical fluctuations that would be present in actual data.

We generate the model waveforms that we compare with the synthetic waveform data using the same code that we use to generate the synthetic waveforms. The model waveforms include a possible background component with an arbitrary energy spectrum.

\subsection{Burst rise or burst tail?}
\label{sec:approach:rise-or-tail}

Previous interest in using X-ray burst oscillations to determine the masses and radii of neutron stars has focused largely on using oscillations observed during the rise of bursts 
\citep[see, e.g.,][]{nath02, stro04}, primarily because of the large fractional modulation observed during the rise of some bursts and the evidence that emission early in the burst rise comes from a small, hotter region on the stellar surface. Although the oscillations observed during burst tails usually have smaller fractional amplitudes, tail oscillations usually last much longer than the oscillations during burst rises. We therefore consider the merits of using oscillations observed during burst tails as well as burst rises.

Key factors to consider in evaluating the relative merits of using burst rise oscillations and burst tail oscillations include (a)~the total number of oscillating counts, (b)~the total number of counts collected during the observation, (c)~the information contained in these counts, including any information encoded in the constant X-ray flux from the star, (d)~the variation of the oscillation waveform with time during the measurement interval, and (e)~the complexity of the emission pattern.

We show here that for a next-generation timing instrument with a collecting area 20 times that of the \textit{RXTE} PCA, the uncertainties in $M$ and $R$ estimates obtained by analyzing oscillations during the tails of bursts are likely to be as small or smaller than the uncertainties derived by analyzing the oscillations observed during the rise of bursts. Hence burst tail oscillations are likely to provide results that are equally good or even superior to the results obtained from burst rise oscillations.

\subsubsection{Statistical uncertainties in parameter estimates}
\label{sec:statistical-uncertainties}

Consider first the uncertainties in $M$ and $R$ estimates produced by the fluctuations in the observed waveform caused by photon counting noise. We expect the fractional uncertainties in $M$ and $R$ estimates to decrease as the ratio ${\cal R}$ of the number of modulated counts to the fluctuation in the number of counts increases. For the burst oscillation waveforms, we use as a figure of merit the quantity
\begin{equation}
{\cal R} \equiv N_{\rm osc}/\sqrt{N_{\rm tot}}
= 1.4\, f_{\rm rms} \, N_{\rm tot} \;,
\label{eqn:R-value}
\end{equation}
where $N_{\rm osc}$ is the number of counts in the oscillating component of the waveform, defined as the integral of the semi-amplitude of the oscillating count rate over the duration of the data segment; $N_{\rm tot}$ is the integral of the total count rate, including the background count rate, over the duration of the data segment; and $f_{\rm rms}$ is the fractional rms amplitude of the oscillation during the data segment. Our Bayesian analysis 
(see Section~\ref{sec:results:constraints:statistical-uncertainties})
shows that the fractional sizes of the confidence regions in the $M$--$R$ plane decrease with increasing ${\cal R}$, approximately as ${\cal R}^{-1}$. The value of ${\cal R}$ is therefore a useful figure of merit when comparing different data sets.

We illustrate the implications of this result for the relative merits of analyzing oscillations during burst rises and tails by scaling from the count rates observed during the rise and tail of a well-observed X-ray burst from 4U~1636$-$536 (oscillations were detected using \textit{RXTE} in the rise of this burst but not in the tail). We first summarize the observed \textit{RXTE} PCA count rates (\citealt{stro98}; see also \citealt{nath02}) and then use these count rates to estimate the count rates that might be observed using a next-generation large-area X-ray timing instrument like those mentioned in Section~\ref{sec:introduction}. 

\textit{Oscillations during the burst rise}.
During the first 1/16~s of the burst, the average total count rate (including a background assumed equal to the \mbox{$\sim\!2,000$~counts~s$^{-1}$} pre-burst background) was \mbox{$\sim\!4,500$}~counts~s$^{-1}$, while the average oscillation semi-amplitude was $\sim\!2,000$~counts~s$^{-1}$. Thus, during the first 1/16~s of the burst a total of $\sim\!280$ counts was collected, including $\sim\!125$ modulated counts. Our postulated next-generation X-ray timing instrument would therefore collect $\sim\!5,600$ total counts, including $\sim\!2,500$ modulated counts, yielding an ${\cal R}$-value $\sim\!33$ when data from this part of the burst rise is used. 
 
During the first 1/4~s of the burst, which includes most of the burst rise, the average total count rate (including the assumed background) was \mbox{$\sim\!8,000$~counts~s$^{-1}$}, while the average oscillation semi-amplitude was \mbox{$\sim\!2,000$~counts~s$^{-1}$}. Thus, during the first 1/4~s of the burst a total of $\sim\!2,000$ counts was collected, including $\sim\!500$ modulated counts. Our postulated next-generation X-ray timing instrument would therefore collect $\sim\!40,000$ total counts, including $\sim\!10,000$ modulated counts, yielding an ${\cal R}$-value $\sim\!50$ when all the data from the burst rise is used.

These results indicate that the 1/4~s data set would provide constraints similar to or even better than the 1/16~s data set, even though the average fractional amplitude of the oscillation was much smaller during the longer interval, because the total number of counts during the longer interval was much greater.

\textit{Oscillations during the burst tail}.
During the last 5~s of the burst, the \textit{RXTE} PCA collected a total of $\sim\!80,000$ counts. According to \citet{gall08b}, the mean fractional rms amplitude of a typical burst oscillation in the \textit{RXTE} sample is $\sim\!10$\%, implying a semi-amplitude $\sim \!14$\% for a sinusoidal waveform. Our postulated next-generation X-ray timing instrument would therefore collect \mbox{$\sim\!1.6\times10^6$} total counts during the last 5~s of a similar burst, including \mbox{$\sim\!200,000$} modulated counts if an oscillation with an rms amplitude of $\sim\!10$\% is present, yielding an $\cal R$-value $\sim\!160$.

$\cal R$ scales as the square root of the number of counts, so combining data from the early part of 25 bursts like this example could yield an $\cal R$-value $\sim\!250$, while combining data from the tails could yield an $\cal R$-value $\sim\!800$.

These results suggest that analyses of data from X-ray burst tails may provide more precise constraints than analyses of data from burst rises. Even if the tail oscillations last somewhat less than 5~s or the rms amplitude is somewhat less than $\sim\!10$\%, this example suggests that analyses of burst tail data may provide results comparable to those obtained by analyzing data from burst rises.

In these estimates we have assumed that the unmodulated (constant in time) component of the X-ray flux time series contains no information about $M$ and $R$, i.e., that only the modulated component of the waveform contains such information. If instead the unmodulated component contains some information about the system (such as the compactness of the star, which tends to increase the unmodulated count rate due to increased gravitational lensing) and one can extract this information, then using the data collected during the burst tail could be more useful than suggested by the preceding estimates. 

To see this, suppose first that the X-ray waveform is constant in time (i.e., unmodulated) and that the magnitude of this constant count rate is important. Then the fractional uncertainties in estimates of $M$ and $R$ will depend only on the total number of counts. Now suppose that in addition to this same unmodulated component there is an infinitesimal modulated component. Clearly, this modulated component will not add much extra information about the system. This case may, in fact, apply to observations of the tails of some bursts. In this case, considering only the oscillations would underestimate our ability to constrain the model parameters using data from the burst tail compared to using data from the burst rise.

These considerations suggest that using oscillations observed during the burst tail to estimate $M$ and $R$ is likely to be at least as effective as using oscillations observed during the burst rise, other things being equal.

\subsubsection{Other considerations}

In addition to the figure of merit ${\cal R}$ of the data set, other important considerations for assessing the relative merits of burst rise and burst tail oscillations include the time-dependence of the waveform and the size and possibly complicated shape of the emitting region that produces the oscillations. Here we consider these two factors.

\textit{Time-dependence of the waveform}. The frequencies and amplitudes of tail oscillations usually change relatively slowly \citep[see][]{watt12}. In contrast, the oscillations seen during the rise of bursts often show large and rapid changes in amplitude and often in frequency, including substantial deviations from the stellar spin frequency\citep[again see][]{watt12}. Hence, waveform models and procedures for fitting oscillations during the onset of bursts must be able to handle much more rapid and greater changes in the size, inclination, and longitude of the emitting region than are encountered in fitting tail oscillations.
 
We have investigated the problem of analyzing burst oscillation waveforms that change with time (see Section~\ref{sec:results:constraints:combining-data} and~Appendix~\ref{app:jointfits}). If the oscillation frequency is changing but the change can be modeled accurately enough to maintain the correct oscillation phase when folding successive periods of the oscillation, then the constraints on $M$ and $R$ that can be obtained by analyzing the resulting folded waveform will be nearly the same as those that could be obtained by analyzing a similar waveform with a fixed oscillation frequency and the same number of counts. If the oscillation frequency varies too rapidly or irregularly during the burst rise or tail to be described accurately by a simple frequency model, the full burst oscillation data set can be divided into smaller time segments and analyzed using standard Bayesian techniques. This approach can also be used if other physical properties of the system, such as the size and inclination of the emitting region, vary significantly. The computational burden of this kind of analysis increases only linearly with the number of segments. Consequently, variations of the burst oscillation waveform on timescales shorter than the burst rise or burst tail (but substantially longer than the burst oscillation period) do not appear to pose an insurmountable analysis problem.

\textit{Complexity of the emission pattern}. Previous modeling of the waveforms of accretion-powered millisecond X-ray pulsars \citep{lamb09a, lamb09b} showed that the size and shape of the emitting region have only a weak effect on the amplitude and shape of the waveform, unless the emitting region is very large (angular radius $\gtrsim 50^\circ$ for inclinations $\gtrsim 60^\circ$). The hotter emitting regions that produce oscillations early in the rise of a burst are thought to cover a small fraction of the stellar surface \citep{watt12}. In this case the detailed shape of the emitting region does not significantly affect the properties of the waveform. Oscillations observed late in the rise of a burst or in burst tails are thought to be produced by larger emitting regions \citep{watt12}, but these regions may still be small enough that their detailed shape does not significantly affect the properties of the waveform. Our fits of waveform models that assume a circular emitting region to synthetic observed waveforms produced by large regions elongated in the east-west and north-south directions suggest that if the emitting region is large and distorted, this will increase the uncertainties in estimates of $M$ and $R$, although not necessarily by a very large amount, but will not bias these estimates significantly (see 
Section~\ref{sec:results:constraints:model-errors}). Our experience in fitting waveforms suggests that results similar to those we report here are likely to be achievable, even if there are moderate temperature variations across the hot spot.

\clearpage
\newpage


\section{COMPUTATIONAL METHODS}
\label{sec:methods}

In this section we describe the computational methods we use to determine the accuracy and precision with which $M$ and $R$ could be determined using the energy- and time-resolved burst oscillation waveform data that could be obtained by future large-area X-ray timing missions. We apply standard Bayesian inference methods to compute the best-fit values and confidence intervals for the parameters $M$ and $R$ in our waveform model, for a variety of synthetic observed waveforms of interest. We first explain how we compute the oscillation waveforms that we use. We then discuss our Bayesian analysis and sampling methods. Finally, we describe how we integrate over uninteresting parameters to determine the joint posterior distribution of $M$ and $R$ and determine their best-fit values and confidence regions.

\subsection{Waveform Computation}
\label{sec:methods:waveformcomputation}

The burst oscillation waveform model that we use here has eight parameters. The neutron star is described by its gravitational mass $M$ and circumferential radius $R$. We represent the emitting area as a circular, uniformly radiating hot spot on the stellar surface, with an angular radius $\Delta\theta_{\rm spot}$. We assume that the center of the spot is inclined at an angle $\theta_{\rm spot}$ relative to the star's spin axis and that the spot rotates uniformly around the spin axis of the star with a frequency $\nu_{\rm rot}$. We also assume that the spot emits radiation that has a blackbody spectrum with a temperature $T_{\rm co}$ when measured in an inertial frame at the stellar surface that is momentarily comoving with the surface. We assume that the values of these parameters are constant in time; the waveform is then perfectly periodic. Finally, we assume that the observer views the system from a distance $d$, at an inclination $\theta_{\rm obs}$ relative to the star's spin axis. 

We use as our global coordinate system Schwarzschild coordinates $(r,\theta,\varphi,t)$ centered on the star, with $\theta=0$ aligned with the spin axis and $\varphi=0$ in the plane containing the spin axis and the observer. We choose the zero of the Schwarzschild time coordinate $t$ so that a light pulse that propagates radially from a point on the stellar surface immediately below the observer (i.e., at $\theta=\theta_{\rm obs}$ and $\varphi=0$) arrives at the observer at $t=0$.

We use our waveform code to compute the phase- and energy-resolved photon number flux that would be seen by a distant observer during a single rotation of the hot spot. This code is based on the code we used previously to compute the waveforms of accreting millisecond X-ray pulsars \citep{lamb09a,lamb09b} and uses the Schwarzschild plus Doppler (S+D) approximation \citep{mill98a} (see 
Section~\ref{sec:approach:waveform-assumptions}). In this approximation, the exterior spacetime is specified completely by the stellar compactness, $GM/Rc^2$ (hereafter, for conciseness we write $M/R$ for $GM/Rc^2$). The computational speed made possible by this approximation makes it practical to compute the required likelihood distributions of the model parameters on a large, modern computer cluster.

It is convenient to specify the radiation from a point on the stellar surface in a local inertial frame located at the surface and momentarily comoving with it. We assume that the beaming pattern of the radiation emitted from the stellar surface is axisymmetric about the normal to the surface, when measured in the comoving frame. The specific intensity $I_0'$ in the comoving frame at the stellar surface can then be expressed as $I_0'(E_0',\alpha')$, where $E_0'$ and $\alpha'$ are the photon energy and the angle between the photon direction and the normal to the surface measured in the comoving frame at the stellar surface.
We assume that $I_0'(E_0',\alpha')$ can be written as the product of a beaming function $g(\alpha')$ and a spectral function $f(E_0')$, i.e.,
\begin{equation}
I_0'(E_0',\alpha') = g(\alpha')f(E_0')\;.
\end{equation}

In the present work, we usually consider the waveforms produced by the beaming pattern expected for emission from an electron scattering atmosphere, but we sometimes consider isotropic beaming. For the former beaming function, we use the quadratic expression
\begin{equation}
g'(\alpha') = a + b\cos\alpha' + c\cos^2\alpha' \;,
\end{equation}
with
\begin{equation}
a=0.42822,~b=0.92236,~{\rm and}~c=-0.085751  \;.
\end{equation}
This expression is a least-squares fit to the beaming function for emission from a uniform, semi-infinite, Thomson scattering atmosphere, taking polarization into account, and agrees with the actual beaming function \citep[Table XXIV]{chan60} to better than 1\% for $0.02 \le \cos\alpha^\prime \le 1$.

In the present work, we usually set $f'(E_0')$ equal to the Planck function $B(E_0',T')$, where $T'$ is the radiation color temperature at the stellar surface, as measured in the comoving frame. We expect the actual spectral function to have a shape similar to but not exactly the same as a Planck spectrum and an efficiency $\sim 20$\%, rather than 100\% (see, e.g., \citealt{sule12}). Using a color factor to produce the appropriate lower efficiency would be important in analyzing real data.

In the few cases where we consider the presence in the spectrum of an atomic scattering line, we model the line by multiplying the continuum spectrum by the transmission factor $\exp[-\tau(E-E_c)]$, where $E_c$ is the centroid energy of the line and $\tau(x)$ is a Gaussian profile having a maximum value $\tau(0)$ and a specified FWHM.

The specific intensity $I_0'(E_0',\alpha')$ measured in the comoving frame at the stellar surface can be converted to the specific intensity $I_0(E_0,\alpha)$ measured in the static frame at the stellar surface using the invariance of $I(E)/E^3$ under a Lorentz boost. The result is
\begin{equation}
I_0(E_0,\alpha) = I_0'(E_0',\alpha')\gamma^{-3}[1-({\bf v}/c)\cdot{\bf\hat k}]^{-3} \;,
\label{eqn:intensitytransformation}
\end{equation}
where $E_0$ is the photon energy, $\alpha$ is the angle between the unit vector ${\bf\hat k}$ in the direction of the light ray and the normal to the surface, and ${\bf v}$ is the linear velocity of the stellar surface at the point of emission, all measured in the static frame, and $\gamma = [1-(v/c)^2]^{-1/2}$. The quantities $E_0$ and $\alpha$ are related to the corresponding quantities $E_0'$ and $\alpha'$ measured in the comoving frame by
\begin{equation}
E_0 = \delta E_0'
\end{equation}
and
\begin{equation}
\cos\alpha' = \delta\cos\alpha \;,
\end{equation}
where
\begin{equation}
\delta = \frac{1}{\gamma[1-({\bf v}/c)\cdot{\bf\hat k}]}
\end{equation}
is the Doppler factor. Here ${\bf v}\cdot{\bf\hat k} = v\cos\zeta$, where $\zeta$ is given by
\begin{equation}
\cos\zeta=\sin\alpha\sin\beta \;,
\end{equation}
in terms of $\alpha$ and the angle $\beta$ between ${\bf\hat k}$ and the direction to the star's spin pole projected onto the plane tangent to the stellar surface at the point of the emission.   

In order to compute the waveform seen by a distant observer located at an inclination $\theta_{\rm obs}$ relative to the star's spin axis, we divide the hot spot on the stellar surface into a fine grid in colatitude and longitude. We then consider the flux from an infinitesimal emitting area $\dd A_i'$ in the comoving frame around each grid point. We transform this flux into the flux in the static frame using equation~(\ref{eqn:intensitytransformation}) and the relation
\begin{equation}
\dd A_i = \delta \dd A_i'
\end{equation}
between the infinitesimal area $\dd A_i'$ measured in the comoving frame at the stellar surface and the infinitesimal area $\dd A_i$ measured in the static frame at the stellar surface. Next we use spherical trigonometry and ray-tracing in the Schwarzschild spacetime to determine the direction of emission ($\alpha_i,\beta_i$), measured in the static frame at the stellar surface, that is required for a light ray originating at a given grid point to reach the observer. We determine the angular separation $\psi_i$ between the location $(\theta_i,\varphi_i)$ of the grid point and the direction to the observer using the spherical trigonometric relation
\begin{equation}
\cos\psi_i = \cos\theta_i\cos\theta_{\rm obs} + \sin\theta_i\sin\theta_{\rm obs}\cos\varphi_i
\end{equation}
and then use the implicit relation
\begin{equation}
\psi_i\left(\alpha_i,\frac{M}{R}\right) = \int^1_0 \frac{\sin\alpha_i~\dd x}{\sqrt{\left(1-2M/R\right)-\left(1-2Mx/R\right)x^2\sin^2\alpha_i}}
\label{eqn:psi}
\end{equation}
between $\psi_i$ and $\alpha_i$ to determine $\alpha_i$. We accurately evaluate the integral in equation~(\ref{eqn:psi}) using a combination of analytical and numerical methods (see Section~\ref{sec:tests:WF:deflection}). Finally, we determine $\beta_i$ using the spherical trigonometric relation
\begin{equation}
\cos\beta_i = (\cos\theta_{\rm obs} - \cos\theta_i \cos\psi_i)/(\sin\theta_i \sin\psi_i) \;.
\end{equation}
Given the position of the observer and the emitting point on the star, this algorithm allows us to solve in a single step for the angles $\alpha_i$ and $\beta_i$ at which the emitted ray leaves the stellar surface. 

Knowing $\alpha_i$ and $\beta_i$, we can evaluate the energy-resolved photon number flux arriving at the distant observer from the infinitesimal element $\dd A_i$, using the expression \cite[][eq.~13]{pout03}
\begin{equation}
\dd F_i(E) = (1+z)^{-1}\frac{I_0(E_0,\alpha_i)}{E} \frac{\dd A_i \cos\alpha_i}{D^2}\left|\frac{\sin\psi}{\sin\alpha}\frac{\partial\psi}{\partial\alpha}\right|_i^{-1} \;,
\label{eq:flux}
\end{equation}
where $E=(1+z)^{-1}E_0$ is the photon energy measured in the static frame at infinity and $1+z = (1-2M/R)^{-1/2}$ is the gravitational redshift from the stellar surface to infinity.

In order to compute efficiently the waveform produced by the emission from the entire hot spot, we proceed in two steps. First, we determine each of the waveforms produced by emission from a set of emitting areas $\{\dd A_i'\}$ with $\varphi=0$ that span the colatitudes within the hot spot. In doing this, we take into account the fact that light emitted from different locations $(R,\theta_i,\varphi_i)$ takes different lengths of time to reach the observer. However, the waveform seen by the observer depends only on the \textit{difference} in the light travel time from different emitting points. (A different choice for the zero of time or a different total propagation time would shift the arrival time of waveform but would not affect the shape of the waveform, which is what concerns us here.) Hence, we  need to compute only the differences in the arrival times of the different rays that reach the observer. We choose to compute the arrival time of each ray relative to the arrival time of a radial ray emitted from a point on the stellar surface immediately below the observer, which in our time coordinate is $t=0$ (see above). The arrival time of a photon that leaves the surface at an angle $\alpha$ to the normal as measured in the static frame is then
\begin{equation}
\Delta t = \int^1_0 \dd x~\frac{R/c}{x^2\left(1-2Mx/R\right)}\left[\frac{1}{\sqrt{1-\sin^2\alpha\left(1-2M/R\right)^{-1}\left(1-2Mx/R\right)x^2}}-1\right] \;.
\end{equation}
We accurately evaluate this integral using a combination of analytical and numerical methods, as described in Appendix~\ref{app:code-validation}. Once we have determined the waveform produced by emitting areas at $\varphi=0$ that span the colatitudes within the hot spot, we compute the full waveform by adding the waveforms produced by emission from the grid points that have different values of $\varphi$. These waveforms can be generated quickly by appropriately shifting the phase of the waveform produced by the corresponding emitting element at $\varphi=0$.

We have validated our waveform code using a suite of code tests. These tests, and the results, are discussed in 
Appendix~\ref{app:code-validation}. We have carefully chosen values for the integration, hot spot, and angular resolution parameters that we use in our waveform code to provide resolutions fine enough to meet our accuracy requirements, but no finer, so that our code runs as fast as possible.

\subsection{Bayesian Analysis and Sampling Methods}
\label{sec:methods:bayesian&sampling}

We wish to determine both the most probable (``best-fit'') values of the parameters in our waveform model, given an observed burst oscillation waveform, and the confidence regions for the values of these parameters. Both goals can be accomplished simultaneously and efficiently using Bayesian inference and an MCMC algorithm to sample the parameter space. Once we have computed the best-fit values of the model parameters, we can determine the accuracy of the fit by comparing them with the values that were used to generate the synthetic observed waveform.

The most probable values of the parameters $\by$ in our waveform model and their confidence regions can be determined using the posterior probability distribution $p(\by|D,I)$, where $D$ is the synthetic energy- and oscillation phase-resolved waveform data of interest and $I$ is any information obtained prior to the measurements under consideration. The desired posterior probability distribution can be obtained from the likelihood of the data, given the parameter values, using Bayes' theorem
\begin{equation}
p(\by|D,I) \propto {p(D|{\bf y},I) p(\by|I)} \;,
\end{equation}
where $p({\bf y}|I)$ is the prior probability distribution of the parameter values and the constant of proportionality is the inverse of the normalization factor. This constant of proportionality is irrelevant when estimating the values of the parameters in a given model. In the present analysis, we use the most uninformative prior, namely, we assume that $p({\bf y}|I)$ is uniform for parameter values within the physical ranges we consider. Then, for parameter values within these ranges,
\begin{equation}
p(\by|D,I) \propto {p(D|{\bf y},I)} \;.
\label{eqn:pyDproptopDy}
\end{equation}

In performing the MCMC sampling of the parameter space, we use first-order Markov chains and start from a random point in the parameter space. At each step, we generate a proposed new set of parameter values $\by'$ based on the current set of parameter values $\by^{(n)}$ by drawing from a proposal distribution $p_p(\by'|\by^{(n)})$. The proposed new set of parameter values is then accepted or rejected with specified probabilities. We use the basic Metropolis algorithm \citep[see, e.g.,][]{tous11}, drawing the new set of parameter values from a joint-normal distribution
\begin{equation}
p_p(\by'|\by^{(n)}) = p_p(\by^{(n)}|\by') \sim N(\by'-\by^{(n)}, \bsigma) \;,
\end{equation}
where the elements of $\bsigma$ are the standard deviations for each parameter, to be specified. We adopt the acceptance probability (which is also the transition probability)
\begin{equation}
T(\by', \by^{(n)}) = \min\left\{1,\frac{p(\by'|D,I)}{p\left(\by^{(n)}|D,I\right)}\right\} \;.
\end{equation}
With this choice, we always accept the proposed new set of parameter values if its probability is higher than that of the current set; otherwise, we accept the proposed new set with probability $p(\by'|D,I)/p(\by^{(n)}|D,I)$. As required, this transition probability satisfies the detailed balance condition, as can be readily verified. Using relation~(\ref{eqn:pyDproptopDy}), this ratio can be written
\begin{equation}
\frac{p(\by'|D,I)}{p\left(\by^{(n)}|D,I\right)} = \frac{p(D|\by',I)}{p\left(D|\by^{(n)},I\right)}\;.
\end{equation}
Hence, at each step the acceptance probability for the proposed set of parameter values is determined by calculating the ratio of the likelihoods of the data given the current parameters and given the proposed parameter values.

If Poisson noise is the only source of fluctuations in the data, the likelihood of the ``observed'' data, given a particular set $\by$ of values for the model parameters, is
\begin{equation}
\mathcal{L} \equiv p(D|\by,I) = \prod_i\frac{m_i(\by)^{d_i}}{d_i!}e^{-m_i(\by)} \;,
\end{equation}
where the product is over all the oscillation phase and energy bins, $d_i$ is the measured number of counts in the $i^{\rm th}$ bin, and $m_i($\by$)$ is the number of counts in the $i^{\rm th}$ bin predicted by the model for the trial set $\by$ of parameter values. 

In our MCMC algorithm, we use the log likelihood only to determine the transition probability, which depends only on the \textit{difference} of log likelihoods, so the $d_i!$ terms in the log likelihood cancel out, producing the expression
\begin{equation}
\log\mathcal{L} = \sum_i d_i\log m_i(\by) - N_{\rm model}(\by) \;,
\end{equation}
where 
\begin{equation}
N_{\rm model}(\by) = \sum_i m_i(\by)
\end{equation}
is the total number of photon counts in the model spectrum. Unlike many situations, where the normalization of the model is a free parameter and hence the total number of counts can be adjusted to be the same for every set of parameter values, here the total number of counts, and hence the normalization, depends explicitly on the distance to the star, the angular diameter of the spot, and other model parameters. The normalization of the model is therefore a key quantity for discriminating between different sets of parameter values.

\subsection{Estimating \texorpdfstring{$M$}{M} and \texorpdfstring{$R$}{R}}
\label{sec:methods:estimatingMR}

We wish to determine the best-fit values and confidence regions of the parameters $M$ and $R$ in our waveform model, given a synthetic observed waveform of interest. With a uniform distribution over the allowed values of the model parameters as our prior, the posterior probability of a particular set of parameter values $\by$, given the data $D$, is proportional to $p(D|\by)$, the likelihood of the data given those values of the parameters. There are two main computational tasks: (1)~computing the relative likelihoods of the data over a set of trial model waveforms chosen to span and adequately sample the parameter space of the model, and (2)~marginalizing the resulting posterior distribution by  integrating over all the parameters in the model except $M$ and $R$.

\subsubsection{Construction of synthetic observed waveforms} 
\label{sec:methods:estimatingMR:synthetic-waveforms}

The data $D$ representing a \textit{synthetic observed waveform} is a list $d_i$ of the (integer) number of counts observed in each oscillation phase-energy bin. We produce each synthetic observed waveform in three steps. First, we use the waveform generating code described in Section~\ref{sec:methods:waveformcomputation} to compute the oscillation-phase- and energy-resolved waveform for a set of model parameter values of interest. Second, we add phase-independent (but energy-dependent) counts from our background model, which as we explained in Section~\ref{sec:approach:waveform-assumptions} is a catch-all intended to mimic possible emission from the entire stellar surface, an accretion disk, and other sources in the field of view, as well as the instrumental background. Finally, we Poisson-sample the total number of photons in each of the phase-energy bins.

\subsubsection{The computational problem}
\label{sec:methods:estimatingMR:computational-problem}

The straightforward way to determine the joint posterior probability distribution of $M$ and $R$ would be to integrate the joint distribution of all the parameters in our waveform model over every parameter except $M$ and $R$. This integral,
\begin{equation}
p(M,R|D,I) = \int\dd V~p(\by|D,I) \;,
\label{eqn:marginalizationintegral}
\end{equation}
could in principle be computed using a Monte Carlo algorithm. Here $V$ is the volume of the model parameter space when the $M$-$R$ subspace is excluded. If we were to sample the integrand at $N$ points $\{x_i\}$ picked randomly but uniformly within $V$, the uncertainty in $p(M,R|D,I)$ would be (Press et al.\ 1999, Section~7.6)
\begin{equation}
\Delta p(M,R|D,I) \approx V\sqrt{\frac{\langle p^2 \rangle - \langle p \rangle^2}{N}} \;,
\label{eqn:uncertaintyintheposterior}
\end{equation}
where
\begin{equation}
\langle p \rangle \equiv \frac{1}{N}\sum_{i=0}^N p(x_i | D,I)
\end{equation}
and
\begin{equation}
\langle p^2 \rangle \equiv \frac{1}{N}\sum_{i=0}^N p(x_i | D,I)^2 \;.
\end{equation}
Expression~(\ref{eqn:uncertaintyintheposterior}) shows that the uncertainty in the marginalized posterior probability distribution obtained by Monte Carlo integration over the full posterior probability distribution decreases as the number $N$ of sample points increases, but only as $1/\sqrt{N}$.

\subsubsection{Computational procedure}
\label{sec:methods:estimatingMR:computational-procedure}

In this work we seek to determine the most probable values of $M$ and $R$ and their confidence intervals by comparing our waveform model with a synthetic observed waveform, using a Bayesian approach. Each model waveform is a list of the expected number of photons $m_i$ in each oscillation phase-energy bin. We construct a \textit{complete} model waveform by computing a \textit{hot spot} waveform (i.e., a model waveform without any background) and then adding a model of the background. For the waveform model we use here, specifying a complete waveform requires specifying 38 model parameters: $M$, $R$, the triplet of angles $\theta_{\rm spot}$, $\Delta\theta_{\rm spot}$, and $\theta_{\rm obs}$ (which define the parameter subspace $\by'$), plus the color temperature $T_{\rm co}$ of the emission at the stellar surface measured in the comoving frame, the distance $d$ to the star, the absolute phase of the oscillation $\phi_0$, and the background counts in 30 energy channels. Determining the joint posterior probability distribution of $M$ and $R$ using equation~(\ref{eqn:marginalizationintegral}) therefore requires accurate computation of the posterior probability distribution over a high-dimensional parameter space and subsequent computation of the marginalization integral over this space. We found that the computational effort needed to achieve sufficient accuracy using this approach was excessive. We therefore sought a more efficient approach, which we now describe. We presume that when the data from a large-area timing mission become available, the computational resources needed to do full Bayesian analyses of these data will also be available.

The purpose of our analysis here is not to reproduce all the steps that would be needed for a full Bayesian analysis of a real observed waveform, but rather to determine the precision and accuracy with which such a full analysis could determine $M$ and $R$. Our initial exploration of the computational problem revealed several shortcuts that we could use for the current study that substantially reduce the computational burden but do not alter the results significantly. Using these shortcuts, the most probable values of $M$ and $R$ and their confidence intervals can usually be computed for a single synthetic observed waveform in 50--100 clock hours, running a parallel code on 150 nodes of a fast CPU cluster.

The procedure we use to determine the most probable values of $M$ and $R$ and their confidence intervals has seven steps: (1)~construct an initial grid of points in the $M$-$R$ plane; (2)~for each $M$-$R$ pair in this grid, choose values for the spot inclination $\theta_{\rm spot}$, the spot angular radius $\Delta\theta_{\rm spot}$, and the observer inclination $\theta_{\rm obs}$; (3)~determine the color temperature $T_{\rm co}$ of the emission from the hot spot measured in the comoving frame at the stellar surface, the absolute phase $\phi_0$ of the waveform, the distance $d$ to the star, and the background model that maximize the likelihood of the observed waveform; (4)~sample the likelihood distribution $p(D|\by')$ over the three-dimensional parameter space $\by'$ consisting of the three angles $\theta_{\rm spot}$, $\Delta\theta_{\rm spot}$, and $\theta_{\rm obs}$, using the most probable values of $T_{\rm co}$, $\phi_0$, $d$, and the background for each angle triplet; (5)~integrate the posterior probability distribution $p(\by|D)$ over $\by'$, using the most probable values of the other model parameters except $M$ and $R$, and thus associate an integrated probability density $p(M,R|D)$ with each point in the $M$-$R$ grid; (6)~refine and extend the $M$-$R$ grid as needed to determine the most probable values of $M$ and $R$ and their confidence intervals with the desired accuracy; (7)~use this final integrated probability density $p(M,R|D)$ to determine the most probable values of $M$ and $R$ and their $1\sigma$ and $3\sigma$ Bayesian confidence intervals. We now explain several aspects of this procedure in more detail.

\textit{Sampling $M$ and $R$}. To compute the posterior probability distribution $p(\by|D)$ efficiently, we first construct an initial grid of points in the $M$-$R$ subspace surrounding the values of $M$ and $R$ that were used to generate the synthetic observed waveform being analyzed. This approach will of course not be possible when analyzing real data, because the actual values of $M$ and $R$ will not be known in advance. Instead, one will have to search the entire $M$-$R$ parameter space, greatly increasing the computational effort required to adequately sample $p(\by|D)$. Once we have computed $p(\by|D)$ on the initial $M$-$R$ grid, we use the results to extend and refine the grid, repeatedly if necessary, until the probability density has been determined over the relevant portion of the $M$-$R$ plane with sufficient accuracy that we can determine the best-fit values of $M$ and $R$ and their uncertainties with the desired accuracy.

\textit{Determining the emission color temperature $T_{\rm co}$}. Our preliminary analysis showed that $T_{\rm co}$ is strongly constrained by the spectrum of the phase-dependent part of the observed waveform, and is much more strongly constrained than the angles $\theta_{\rm spot}$, $\Delta\theta_{\rm spot}$, and $\theta_{\rm obs}$. Hence the range of $T_{\rm co}$ values where the probability density is appreciable is very small and using the MCMC algorithm to sample this parameter is therefore very inefficient. Consequently we used a different approach to determine the value of $T_{\rm co}$ in our trial waveforms.

All the \textit{trial} waveforms used in this paper were generated assuming that the spectrum of the emission from the hot spot has the same shape as the Planck spectrum. Most of the synthetic \textit{observed} waveforms that we consider here were generated assuming that the spectrum of the emission from the hot spot has the same shape as a Planck spectrum. However, a few of the synthetic observed waveforms considered here were generated assuming a hot spot emission spectrum having the same shape as a Bose-Einstein spectrum, to explore the effects on the estimates of $M$ and $R$ of fitting a model that assumes a spectrum somewhat different from actual spectrum. We use different approaches to generate trial values of $T_{\rm co}$ in these two cases.

When the comoving emission spectrum used to generate the synthetic waveform has a Planck shape, we found that using the redshift relation for a non-rotating star, namely,
\begin{equation}
T_{\rm co} = T_{\infty}(1-2M/R)^{-1/2} \;,
\label{eqn:Tco-Tinf-relation}
\end{equation}
to estimate $T_{\rm co}$ using the observed radiation temperature $T_{\infty}$ gives $T_{\rm co}$ to within 5\%, for spot rotation frequencies $\le 600$~Hz. Consequently, to reduce the computational burden during our large-scale runs, for this synthetic spectrum we assume relation~(\ref{eqn:Tco-Tinf-relation}) and use as our trial value of $T_{\rm co}$ the value given by
\begin{equation}
T_{\rm co} = T_{{\rm co},0}
\frac{(1-2M_0/R_0)^{1/2}}{(1-2M/R)^{1/2}} \;,
\label{eqn:Tco-Tco-relation}
\end{equation}
where the quantities with `0' subscripts have the values that were used in generating the synthetic waveform. Using this approach decreases the time required to compute $p(M,R|D)$ by a factor of five compared to the time required to solve for the most probable value of $T_{\rm co}$.

When the emission spectrum used to generate the synthetic waveform has a Bose-Einstein shape, we included $T_{\rm co}$ as one of the parameters determined by maximizing the likelihood of the observed waveform, as we discuss below. 

Changing the blackbody spectrum in the model waveform takes very little computational effort. The reason is that changing from one Planck spectrum with a given temperature and normalization to another with a different temperature and normalization can be done simply by mapping the spectrum to the different energy and multiplying it by the factor needed to renormalize it. Because photons of all energies are Doppler shifted, gravitationally redshifted, and deflected in the same way, once the energy-dependent waveform for one comoving emission temperature has been computed, the waveform for any other comoving emission temperature can be computed using the appropriate mapping and renormalization factor.  Thus, the very time-consuming step of tracing rays needs to be done only once. To preserve accuracy when performing the mapping and renormalization, it is important to use tables that span the required energy range, so that interpolation can be used and extrapolation is not required.

If there is an atomic scattering line in the synthetic hot spot spectrum, we assume that we know the rest energy, width, and optical depth of the line and use this information to include a line with these properties in the spectrum of the waveform model. 

\textit{Determining the most probable values of $T_{\rm co}$, $d$, $\phi_0$, and the background}. Given an $M$-$R$ pair, there are 35 additional waveform parameters, if the synthetic emission has the shape of a Planck spectrum and $T_{\rm co}$ is determined as described above, or 36 additional parameters, if the synthetic emission has the shape of a Bose-Einstein spectrum and $T_{\rm co}$ has to be determined by a likelihood analysis.

As noted above, our preliminary analysis showed that $T_{\rm co}$ is strongly constrained by the spectrum of the phase-dependent part of the observed waveform. Hence, if $T_{\rm co}$ must be determined, we do so by maximizing the likelihood of the observed waveform, given the model waveform, using bisection.

In our preliminary investigation, we found that when the other waveform parameters are held fixed, the distance $d$ to the star (which is equivalent to the normalization factor of the Planck spectral shape) is also very tightly constrained by the total number of observed counts. The total number of observed counts is in turn very tightly constrained by the data. Consequently, determining $d$ by MCMC sampling and marginalization is very inefficient. We therefore determine $d$ by maximizing the likelihood of the data using bisection. This method determines the most probable value of the distance to within 1\%.

Our preliminary analysis showed that the phase $\phi_0$ of the oscillation is very tightly constrained by the observed waveform. We therefore determine $\phi_0$ by maximizing the likelihood using bisection. The first step is to apply the same, tentative phase shift to the model waveform in each photon-energy bin. We do this by Fourier transforming the trial waveform at every photon energy, applying the tentative phase shift to all the harmonic components, and then inverting the Fourier transform to obtain the shifted model waveform. We perform the phase shift in the frequency domain rather than in the time domain to increase the accuracy: a phase shift applied in the time domain would have to be at least as large as the frequency times the width of a time bin, whereas even a very small phase shift can be applied to every harmonic in the frequency domain. Then, when the Fourier-transform is inverted the full waveform will be shifted by this small amount. 

We compute the background model using bisection to determine the number of background counts at each energy that, when added to the phase-shifted hot spot waveform, produces a complete model waveform that maximizes the likelihood of the observed waveform.

\textit{Sampling the three angles}. The values of the three angles $\theta_{\rm spot}$, $\Delta\theta_{\rm spot}$, and $\theta_{\rm obs}$ are much less tightly constrained than are $T_{\rm co}$, $d$, $\phi_0$, and the background. Therefore, in our initial sampling of these angles we use the MCMC algorithm outlined above. In this step, we typically generate 30 chains, each with a length of 40, for a total of 1200 triplets. For each triplet of angles, we simultaneously determine the values of the remaining 32 (or 33) parameters that maximize the likelihood, using bisection. We then select the triplets that have a log likelihood within 20 of the maximum log likelihood. Finally, we use this collection of triplets to define the ranges of the three angles that we sample in the next step of our procedure.

For our second, final sample of the three angles, we typically choose 10,000 angle triplets randomly and uniformly within the ranges determined in the previous step. The number of triplets was chosen to provide enough resolution to obtain an acceptably accurate value of the integrated probability density. For each triplet of angles, we again determine the values of the remaining 32 (or 33) parameters that maximize the likelihood using bisection. We then sum the probabilities at all these triplets, to associate an integrated probability density $p(M,R|D)$ with this particular $M$-$R$ pair.

\textit{Completing the sampling of $M$ and $R$}. Having computed the integrated probability density for a particular $M$-$R$ pair, we then choose a different $M$-$R$ pair in our grid and repeat the process. If necessary, we extend and refine the $M$-$R$ grid, in order to determine the most probable values of $M$ and $R$ and their $1\sigma$ and $3\sigma$ confidence intervals with the desired accuracy. We use a common normalization for the integrated probability density of different $M$-$R$ pairs, so it can be compared for different pairs. Finally, we use the resulting probability distribution in the $M$-$R$ subspace to determine the most probable values of $M$ and $R$ and their $1\sigma$ and $3\sigma$ confidence intervals. We tested this algorithm in numerous cases and found that it does a good job of determining the most probable $M$-$R$ pair and the corresponding confidence regions.

\clearpage
\newpage


\section{WAVEFORM ANALYSIS AND RESULTS}
\label{sec:results}

Our burst oscillation waveform analysis proceeds in three steps:

\begin{enumerate}\itemsep0pt

\item Generate a synthetic observed waveform, using the method described in Section~\ref{sec:methods:waveformcomputation}.

\item Compute the joint posterior distribution of the mass and radius parameters in our waveform model, given this synthetic waveform, using the computational methods described in Sections~\ref{sec:methods:bayesian&sampling} 
and~\ref{sec:methods:estimatingMR}.

\item Use this posterior distribution to determine the most probable values of $M$ and $R$ and the $1\sigma$ and $3\sigma$ confidence regions in the $M$-$R$ plane, given this synthetic waveform.

\end{enumerate}
We use this procedure to explore how the accuracy and precision of $M$ and $R$ estimates depend on the physical characteristics of the observed system by generating a number of different synthetic waveforms corresponding to systems with different physical characteristics, such as spot rotation rate, spot inclination, and viewer inclination. We then explore the effects on the uncertainties in $M$ and $R$ of having additional information about the observed system that is independent of that provided by the burst oscillation waveform. We consider knowledge of the distance, the observer's inclination, and the size and spectrum of the background, measurement of the properties of an atomic scattering line in the emission from the hot spot, and high-precision measurements and modeling of the spectrum of the X-ray bursts from the star in question. We also explore the effects on $M$ and $R$ estimates if the actual spot shapes, X-ray spectra, and radiation beaming patterns differ from those assumed in our waveform model. Finally, we investigate the constraints that can be obtained by analyzing data on burst oscillations with changing waveforms, collected during a single burst or during multiple bursts.

\subsection{Waveform Analysis}
\label{sec:results:waveform-analysis}

All the synthetic waveforms used in this work assume $M=1.6\,M_\odot$ and $R=5.0\,M$. We compute the joint posterior distribution of the parameters $M$ and $R$ in our waveform model, given the synthetic waveform, for $3.5 \le R/M \le 6.8$. Given that all the synthetic waveforms assumed $R/M=5.0$, this range in $R/M$ is adequate to evaluate all cases that provide relatively tight constraints on $M$ and $R$.

\subsubsection{Model burst oscillation waveforms}
\label{sec:results:waveform-analysis:model-waveforms}

The hot spot emission model that we usually use in fitting the synthetic waveform data assumes that the spot is a uniformly emitting circular area on the stellar surface that is rotating around the star and that each point on the surface of the hot spot radiates with a spectrum that has the shape of a Planck spectrum and the angular pattern expected for emission from a semi-infinite, nonrelativistic, electron scattering atmosphere (see Section~\ref{sec:methods:waveformcomputation} for further details). 

The full waveform model has 9 parameters that describe the waveform produced by the emission from the hot spot (the star's gravitational mass $M$, circumferential radius $R$, and the spot rotation frequency $\nu_{\rm rot}$; the spot inclination or colatitude $\theta_{\rm spot}$ and angular radius $\Delta\theta_{\rm spot}$; the color temperature $T_{\rm co}$ of the spot emission as measured in the comoving frame at the stellar surface; the observer's inclination $\theta_{\rm obs}$ relative to the rotation axis and the distance $d$ to the star; and the phase of the waveform). The waveform model has an additional 30 parameters that describe the background model (the number of background counts in each of 30 energy channels). Thus there are in principle a total of 39 parameters in the model.

In practice, the spot rotation frequency can be determined separately and much more accurately than any of the other parameters in the waveform model. Therefore, to speed up the computations for the present study we assume that $\nu_{\rm rot}$ is already known. 
As discussed in
Section~\ref{sec:methods:estimatingMR:computational-procedure}, $T_{\rm co}$ can be determined very accurately using the color temperature observed at infinity when the spectrum of the emission from the hot spot that produces the observed waveform has the shape of a Planck spectrum. As discussed there, in this case we use this fact to determine the value of $T_{\rm co}$ separately from the fitting procedure, to speed up the computations by a factor of five. When the spectrum from the hot spot does not have the shape of a Planck spectrum, $T_{\rm co}$ must be included as one of the fitted parameters. Thus, the hot spot component of the waveform model has 7 adjustable parameters, if the spectrum has the shape of a Planck spectrum, or 8 free parameters, if it does not. The full waveform model, which includes the background, therefore has either 37 or 38 adjustable parameters. Each synthetic oscillation waveform has 480 data points (see below), so the number of degrees of freedom in the fits described here is either $480-37=443$ or $480-38=442$, depending on the hot spot spectrum that was assumed in generating the synthetic waveform data.

If the properties of the observed neutron star are consistent with the properties assumed here, we expect the constraints on $M$ and $R$ to improve as the ratio ${\cal R}$ of the number of modulated counts to the fluctuations in the total number of counts (see equation~(\ref{eqn:R-value})) increases. For a fixed value of ${\cal R}$, we expect the constraints to be tighter when the spot and observer are closer to the star's rotational equator, because the relativistic Doppler shift and aberration are then greater, producing a waveform that depends more sensitively on the radius of the star.

\subsubsection{Synthetic burst oscillation waveforms}
\label{sec:results:waveform-analysis:synthetic-waveforms}

Each synthetic observed waveform is represented by the number $d_i$ of counts in each of 30 energy bins of width 0.3~keV covering 3.5~keV to 12.5~keV at each of 16 rotational phases. There are thus a total of 480 data points in each synthetic waveform. 

The physical models we use to generate the component of these waveforms that is produced by the hot spot are characterized by the gravitational mass $M$ and circumferential radius $R$ of the star; the spot rotation frequency $\nu_{\rm rot}$; the location, size, and shape of the hot spot; the spectrum and beaming pattern of the emission from the hot spot; and the distance $d$ to the observer and the observer's inclination $\theta_{\rm obs}$ relative to the star's rotation axis. The waveforms used in this work assume that the emission from the hot spot has the shape of a Planck spectrum or a Bose-Einstein spectrum with a temperature $kT_{\rm co}=2.0$~keV, as measured in a frame comoving with the surface of the rotating hot spot. To facilitate comparison of different cases, we adjust the distance to the star so that the expected number of counts from the hot spot is $10^6$ in all cases.

To explore the effects of background counts, we add a background component to the waveform produced by the hot spot. As discussed in 
Section~\ref{sec:approach:waveform-assumptions}, this component is a catch-all for all counts not produced by radiation from the hot spot. These counts could be produced by emission from unassociated sources in the field, the accretion disk, the non-spot portion of the star, instrumental backgrounds, or any combination of these. We assume that the background does not vary at frequencies commensurate with the spot rotation frequency. We model the background by assuming uniform emission from the entire surface of the star with a spectrum having the shape of a Planck spectrum with a temperature of 1.5~keV as seen in a frame comoving with the surface of the star, which for this purpose is assumed to be rotating with the same frequency as the hot spot. To study the effects of different background levels, we adjust the normalization of the background component so that the number of counts expected from the background is either 0.3, 1.0, or 9.0 times the $10^6$ counts expected from the hot spot. We refer to these as our low, medium, and high backgrounds.

Once the preliminary synthetic waveform is generated, it is Poisson sampled to produce the complete synthetic waveform. Because the waveforms are Poisson-sampled, the actual counts contributed to the complete waveform by the hot spot and the background are slightly different from their expected values.

\textit{Standard synthetic waveforms}. We first analyze synthetic waveforms generated using burst oscillation models with the same qualitative properties as the model that we fit to the data. These ``standard'' synthetic waveforms assume that the hot spot is circular and uniformly emitting, that the spectrum of the emission from the hot spot has the shape of a Planck spectrum, as measured in the comoving frame at the stellar surface, and that the beaming pattern of the emission from the hot spot is that expected for a nonrelativistic election scattering atmosphere, which is described by the Hopf beaming function 
(see Section~\ref{sec:methods:waveformcomputation}). The standard waveforms we analyze here assume that the angular radius $\Delta\theta_{\rm spot}$ of the hot spot is $25^\circ$.

\textit{Reference cases}. The hot spot and observer inclinations have the biggest effects on the uncertainties in estimates of $M$ and $R$. To facilitate comparisons, we constructed several standard waveforms for a ``low-inclination'' reference case and for a ``high-inclination'' reference case.
The model parameters that define these two cases are $\nu_{\rm rot}$, $\theta_{\rm obs}$, and $\theta_{\rm spot}$. The values of these parameters for the two reference cases are listed in 
Table~\ref{table:results:reference-cases}.

\begin{deluxetable}{lccc}
\tablewidth{0pt}
\tablecaption{
System parameters for the two reference cases\tablenotemark{a}
}
\tablehead{
\colhead{Case label} & \colhead{$\nu_{\rm rot}$ (Hz)} 
& \colhead{$\theta_{\rm obs}$ (deg)} &
\colhead{$\theta_{\rm spot}$ (deg)}
}
\startdata
low inclination & 400 & 60 & 20   \\
high inclination & 600 & 90 & 90   \\
\enddata
\vskip-10pt
\tablenotetext{a}{
Here $\nu_{\rm rot}$ is the rotation frequency of the hot spot, $\theta_{\rm obs}$ is the inclination of the observer relative to the rotation axis, and $\theta_{\rm spot}$ is the inclination (colatitude) of the center of the hot spot. 
}
\label{table:results:reference-cases}
\end{deluxetable}

The low-inclination reference case is motivated by the evidence that the magnetic poles of many accreting neutron stars with millisecond spin periods are close their spin poles \citep[see][]{lamb09a, lamb09b}. This evidence includes the low fractional amplitudes of the X-ray oscillations of accretion-powered millisecond X-ray pulsars, their highly sinusoidal waveforms, and their intermittency in some stars. In the low-inclination reference case, the hot spot and the observer are at inclinations of $20^\circ$ and $60^\circ$, respectively, and the rotation frequency is 400~Hz.

Even if the star's magnetic poles are close to its spin pole, the region heated by nuclear burning or surface modes may not be at or even near a magnetic pole, and hence may not be close to the spin pole. As an example, the very high-amplitude X-ray oscillations found near the beginning of a burst from 4U~1636$-$536 by \citet{stro98} are difficult to understand if the heated region is near the spin pole. We therefore also consider a high-inclination reference case in which the hot spot and observer inclinations are both $90^\circ$ (i.e., both are in the rotational equator) and the spot rotation frequency is 600~Hz. These choices produce the largest expected Doppler shifts and relativistic aberration, and should therefore illustrate the tightest possible constraints on $M$ and $R$.

\textit{Nonstandard synthetic waveforms}. In addition to our ``standard'' synthetic waveforms, which assume a circular hot spot with a 25$^\circ$ angular radius, a Planck emission spectrum, and the radiation beaming pattern appropriate for emission from a Thomson scattering atmosphere (described by the Hopf function), we also generated several other waveforms, to explore the consequences of fitting an incorrect waveform model to the data. These ``nonstandard'' waveforms were generated for spots elongated in the north-south and east-west directions, for a Bose-Einstein emission spectrum, or for isotropic beaming from the surface of the hot spot.

\subsubsection{Independent additional information}
\label{sec:results:waveform-analysis:additional-info}

We explore the improvements in constraints on $M$ and $R$ that become possible with independent additional information about important system properties.
There are good prospects for determining the angle of our
line of sight relative to the axis of the binary orbit using optical burst reflection mapping \citep{casa10} and Fe-line modeling 
\citep{cack10,egro11}. The neutron stars in bursting systems are thought to have acquired nearly all their angular momentum via accretion \citep{lamb08a}, so the spin axes of these stars are likely to be closely aligned with the orbital axes of their systems, allowing us to infer the inclination of the observer relative to the stellar spin axis.
The distance of a bursting neutron star is often tricky to measure,
but it can be constrained if the system 
is in a globular cluster. Parallax distances for such stars are likely to become available in the future from \textit{GAIA}
\citep{eyer13}.
The properties of the non-spot background are likely to be difficult to constrain independently of the waveform fitting process, but if the spectrum of the background is sufficiently
distinct from the spectrum of the spot radiation (e.g., if both have Planck spectral shapes but have easily
distinguishable temperatures), it may be possible to restrict
the properties of the background using spectral measurements. It is also possible that 
an atomic resonance scattering
line could be identified in the emission from the hot spot. We expect that having information like this would improve the constraints on $M$ and $R$.

It may be possible to obtain other independent information by 
fitting high-precision spectral models to the observed spectra of 
long bursts or jointly fitting such models to the spectra of many
bursts from the same source. At high burst temperatures, the burst atmosphere is expected to be
scattering-dominated and to produce spectra with shapes that are very close to the shape of a Bose-Einstein spectrum (Lo et al., in preparation). This expectation is consistent with the observed spectra of such bursts \citep{bout10b}. If the observed spectra had exactly a Bose-Einstein spectral shape, they would provide no information about the properties of the star, because a thermodynamic equilibrium spectrum provides no such information. However, we have found that detailed, high-precision model atmosphere spectra provide a much better description of the best available spectral data (the \textit{RXTE} PCA data on the superburst from 4U~1820$-$30) than do Bose-Einstein spectra, so some information about the properties of the star can be obtained. Information about the mass and radius of the star comes mainly from the small deviation of the spectrum from a Bose-Einstein shape at low photon energies that is caused by the free-free opacity. This deviation allows one to determine $(1+z)/g^{2/9}$, where $z$ is the surface redshift and $g$ is the surface gravity of the star. This relation between $M$ and $R$ complements the constraints on $M$ and $R$ obtained from fits to the burst oscillation waveform.

\subsubsection{Errors in the waveform model}
\label{sec:results:waveform-analysis:model-errors}

Most of the results we present here assume that we know the shape of the hot spot and the spectrum and beaming pattern of the radiation from the spot. However, in practice these properties may not be accurately known. If any important properties of the actual star differ significantly from the properties assumed in the model being fit, we are attempting to fit the data using an incorrect model. In this situation we expect the fit to be poor and the best-fit values of $M$ and $R$ to differ from their actual values. Of particular concern is the possibility of situations in which the fit is formally good but the best-fit values of $M$ and $R$ are significantly different from their actual values. A full analysis of possible systematic errors is beyond the scope of a this work, but we briefly explore the effects of errors in the model of the spot emission.

\subsection{Constraints on \texorpdfstring{$M$}{M} and \texorpdfstring{$R$}{R}}
\label{sec:results:constraints}

\begin{deluxetable}{ccclrr}
\tablewidth{0pt}
\tablecaption{
Synthetic waveforms, fitted models, and corresponding figures\tablenotemark{a}
}
\tablehead{
\colhead{Figure} &
\colhead{Inclinations\,\tablenotemark{b}} &
\colhead{Background\,\tablenotemark{c}} &
\multicolumn{1}{l}{Additional description} &
\colhead{$\delta M_1$(\%)\,\tablenotemark{d}} &
\colhead{$\delta R_1$(\%)\,\tablenotemark{d}}
} 
\startdata
1 & high & medium & all parameters known except $M$ and $R$ & 1.9 \tablenotemark{e} & 1.3\,\tablenotemark{e} \\
2a & high & low & all parameters unknown & 2\, & 5\, \\
2b & low & low & all parameters unknown & 30\, & 30\, \\
2c & high & medium & all parameters unknown &\, 3 & 4\, \\
2d & low & medium & all parameters unknown & 50\, & 40\, \\
2e & high & high & all parameters unknown & 9\, & 8\, \\
2f & low & high & all parameters unknown & ---\,\tablenotemark{f} & ---\,\tablenotemark{f} \\
3a & high & medium & $\theta_{\rm spot} = 80^{\circ}$, all parameters unknown & 6\, & 6\, \\
3b & high & medium & $\theta_{\rm spot} = 60^{\circ}$, all parameters unknown & 5\, & 5\, \\
4a & high & high & observer inclination known & 5\, & 5\, \\
4b & low & high & observer inclination known & ---\,\tablenotemark{f} & ---\,\tablenotemark{f} \\
4c & high & high & distance known & 7\, & 6\, \\
4d & low & high & distance known & ---\,\tablenotemark{f} & ---\,\tablenotemark{f} \\
4e & high & high & atomic line present in the data & 4\, & 4\, \\
4f & low & high & atomic line present in the data & ---\,\tablenotemark{f} & ---\,\tablenotemark{f} \\
5a & high & medium & background known & 3\, & 5\, \\
5b & low & medium & background known & 20\, & 20\, \\
6a & high & medium & hot spot elongated longitudinally & 40\, & 40\, \\
6b & medium & medium & hot spot elongated latitudinally & 8\, & 17\, \\
7a & high & medium & Bose-Einstein spectrum in the data & 6\, & 6\, \\
7b & low & medium & Bose-Einstein spectrum in the data & ---\,\tablenotemark{f} & ---\,\tablenotemark{f} \\
7c & high & medium & isotropic beaming in the data & ---\,\tablenotemark{g} & ---\,\tablenotemark{g} \\
7d & low & medium & isotropic beaming in the data & ---\,\tablenotemark{f} & ---\,\tablenotemark{f} \\
\enddata
\vskip-10pt
\tablenotetext{a}{All synthetic waveforms were normalized to produce $\sim\,$$10^6$ counts from the hot spot.
$^b$See Table~\ref{table:results:reference-cases} for the parameters used in generating the high- and low-inclination synthetic waveforms. The elongated spots are defined in the text.
$^c$The backgrounds in all synthetic waveforms are phase-independent, with the spectral shape of a Planck spectrum having a temperature at infinity of $kT=1.5$~keV. Our low, medium, and high backgrounds produce $\sim 3 \times 10^5$, $\sim 1 \times 10^6$, and $\sim 9 \times 10^6$ detected counts. 
$^d$Approximate $1\sigma$ fractional uncertainties in $M$ and $R$ estimated by projecting the $1\sigma$ contours onto the $M$ and $R$ axes (see text).
$^e$These purely statistical uncertainties assume that all relevant properties of the actual system are accurately described by the waveform model, that the values of these properties lie within the range of values covered by models considered in the waveform fitting process, and that the values of all the model parameters except $M$ and $R$ are known exactly by some means that is independent of the waveform fitting process, thereby eliminating the degeneracies of the model parameters.
$^f$Reliable estimates of the uncertainties are not possible because the confidence regions are too large.
$^g$The reduced $\chi^2$ of this fit indicates that the model is incorrect.
}
\label{table:results:summary}
\end{deluxetable}

Consider now the constraints that can be obtained by fitting a waveform model to an observed burst oscillation waveform. Our results show that the uncertainties in $M$ and $R$ estimates produced by statistical fluctuations in the waveform are typically only a percent or two, if $\sim\,$$10^6$ counts are collected from the hot spot of a given bursting neutron star and the values of all other parameters are known independently of the waveform analysis. However, for realistic situations in which some of the other parameters in the model must be determined from the waveform, the statistical uncertainties are usually much larger. The reason for this is that the effects on the waveform of changing different parameters in the waveform model are often very similar. These degeneracies of the waveform with respect to changes in the values of the model parameters are an inherent property of any physical model based on a rotating hot spot and cannot be removed by ``improving'' the model. These degeneracies inflate the uncertainties in $M$ and $R$ estimates produced by statistical fluctuations.

The synthetic waveforms we consider here and the resulting $1\sigma$ uncertainties in $M$ and $R$ are summarized in 
Table~\ref{table:results:summary}. Plots of the $1\sigma$ and $3\sigma$ uncertainty contours for each of these waveforms are presented and discussed below. The $\chi^2$/dof for the best-fit model is shown on each plot.

The uncertainties in $M$ and $R$ quoted in this section were estimated by projecting the extent of the two-dimensional uncertainty regions onto the $M$ and $R$ axes. This approach neglects the fact that the probability distributions of $M$ and $R$ are often correlated. It also somewhat overestimates the individual uncertainties in $M$ and $R$, because the central peak in the distribution of the unwanted quantity would centrally weight the probability distribution of the quantity being considered, if the unwanted quantity were actually marginalized. This effect is neglected when the uncertainty region is simply projected. We note that the quoted uncertainties are themselves slightly uncertain, due to incomplete sampling of the high-dimensional parameter space.

\begin{figure*}[!t]
\begin{center}
\includegraphics[height=.26\textheight]{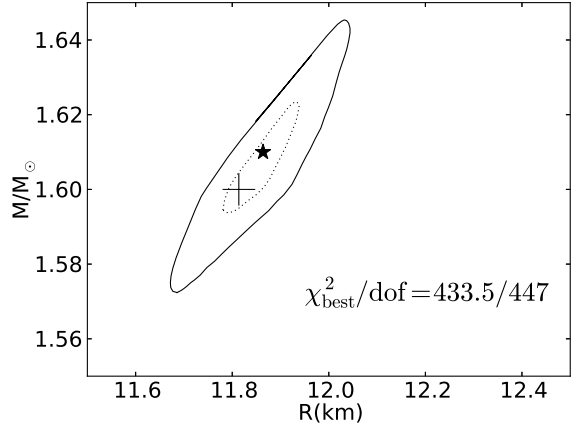}
\end{center}
\vspace{-0.7cm}
\caption{
These contours are based on the highly unrealistic assumptions that 
all the relevant properties of the actual system are correctly included in the waveform model, that the actual values of these properties lie within the range of values possible in the waveform model, and that the values of all the parameters in the waveform model except $M$ and $R$ are known exactly, independent of the waveform fitting process. This procedure eliminates completely the effects of the degeneracy of the waveform with respect to variations in the values of the system parameters.
The + symbol indicates the radius and mass that were used in generating the synthetic waveform and the star symbol shows where the marginalized posterior probability density is highest. The dotted and solid curves show, respectively, the $1\sigma$ and $3\sigma$ confidence contours. The reduced $\chi^2$ indicates that the fit is acceptable. See the text for further details.
}
\label{fig:results:nuisance_params_known}
\end{figure*}

\subsubsection{Effects of statistical fluctuations}
\label{sec:results:constraints:statistical-uncertainties}

We can estimate the uncertainties in estimates of $M$ and $R$ produced by the statistical fluctuations in the observed waveform by considering the (highly unrealistic) situation in which all the relevant properties of the actual system are correctly included in the waveform model, the actual values of these properties lie within the range of values possible in the model, and the values of all the parameters in the model except $M$ and $R$ are known exactly by some means that is completely independent of the waveform fitting process. 

Figure~\ref{fig:results:nuisance_params_known} shows an example in which the assumptions in the model used to fit the synthetic waveform data are identical to the assumptions in the model that was used to produce the waveform data. The waveform data were generated using the parameter values of the high-inclination reference case and our medium background. In fitting the model to the waveform data, the values of all the parameters in the model except $M$ and $R$ were set to the values used to produce the synthetic waveform. This procedure completely eliminates the effects of the degeneracy of the waveform under variations of the model parameters. The most probable values values of $M$ and $R$ were then determined using our standard fitting and marginalization procedures. The $1\sigma$ uncertainties in $M$ and $R$ are about 1.9\% and 1.3\%, respectively, whereas the $3\sigma$ uncertainties in $M$ and $R$ are about 5\% and 3\%. The value of $\chi^2$/dof indicates that the fit is acceptable.

These results can be compared with those of \cite{stro04}, who investigated the constraints on $M$ and $R$ that could be obtained by fitting a bolometric waveform model to synthetic bolometric waveform data. He assumed that the hot spot and the observer are both in the rotation equator (their most favorable locations) and that the size, shape, and location of the hot spot, the spectrum and beaming pattern of the radiation from the hot spot, and the inclination of the observer are all known independently of the waveform fitting analysis, so that only $M$ and $R$ have to be estimated from the waveform data. \citeauthor{stro04} also assumed there is no background. With these assumptions and approximations, \cite{stro04} found that fitting a bolometric waveform model to bolometric waveform data from an X-ray instrument with an effective area $\sim\,$10~m$^2$ could provide estimates of $M$ and $R$ with uncertainties of a few percent.

\begin{deluxetable}{lrrrr}
\tablewidth{0pt}
\tablecaption{
Scaling of uncertainties in $M$ and $R$ with total counts\tablenotemark{a}
}
\tablehead{
\colhead{$N_{\rm tot}$}
&\colhead{\qquad $\delta M_1$(\%)}    &\colhead{$\delta M_3$(\%)}
&\colhead{~~$\delta R_1$(\%)}         &\colhead{$\delta R_3$(\%)}
}
\startdata
$10^4$ &   15~~~~~   &  42~~~~~   &  12~~~~~    &   27~~~~~   \\
$10^5$ &   6.8~~~    &  15~~~~~   &   4.7~~~    &   11~~~~~ \\
$10^6$ &   1.8~~~    &   5.0~~~   &   1.3~~~    &    3.1~~~   \\
\enddata
\vskip-10pt
\tablenotetext{a}{
$N_{\rm tot}$ is the total number of counts in the high-inclination, medium-background synthetic waveforms that were analyzed, $\delta M_1$ and $\delta M_3$ are the $1\sigma$ and $3\sigma$ uncertainties in $M$, and $\delta R_1$ and $\delta R_3$ are the $1\sigma$ and $3\sigma$ uncertainties in $R$. 
These uncertainties assume all system parameters are known except $M$ and $R$ (see text).
The fractional rms amplitudes of all three waveforms are $\sim\,$54\% (see Table~\ref{table:results:R&rms}).
}
\label{table:results:uncertainty-scaling}
\end{deluxetable}

Our simulations show that the statistical uncertainties in $M$ and $R$ scale roughly as $1/{\cal R}$, where ${\cal R} \equiv N_{\rm osc}/\sqrt{N_{\rm tot}}$ in terms of the number $N_{\rm osc}$ of counts in the oscillating component of the synthetic waveform and the total number $N_{\rm tot}$ of counts in the waveform (see equation~(\ref{eqn:R-value}) and the discussion following it). This is illustrated in Table~\ref{table:results:uncertainty-scaling}. In these particular examples, the expected number $N_{\rm back}$ of counts in the background is equal to the number $N_{\rm spot}$ of counts from the hot spot, so $N_{\rm osc} \propto N_{\rm spot} \propto N_{\rm spot} + N_{\rm back} =  N_{\rm tot}$ and hence ${\cal R}$ is proportional to $\sqrt{N_{\rm tot}}$.

\subsubsection{Effects of parameter degeneracies}
\label{sec:results:constraints:parameter-degeneracies}

Comparison of the uncertainties listed in 
Table~\ref{table:results:summary} and comparison of the confidence regions shown in Figure~\ref{fig:results:nuisance_params_known}
with the confidence regions shown in the other figures in this section shows that when parameters in the waveform model other than $M$ and $R$ must be determined as part of the waveform fitting process, the uncertainties in the resulting estimates of $M$ and $R$ are generally much larger than when only $M$ and $R$ are unknown. The reason the uncertainties are so large is that the waveform is degenerate with respect to changes in many of the parameters in the waveform model, i.e., the effects on the waveform of changing the value of one of the system parameters can be partially or wholly compensated by changing the value of one or more other parameters:
\begin{itemize}

\item The effect on the stellar compactness and surface redshift of an increase in $R$ can be compensated almost exactly by an increase in $M$, because the special relativistic effects on the waveform of the rotational surface velocity are $\sim\,$5\% or less for the stellar models considered here (see Section~\ref{sec:methods}). This is a very strong degeneracy.

\item Changes in the spot inclination, viewer inclination, and stellar radius affect the waveform in similar ways and can therefore compensate for one another. The increase in the relativistic Doppler boost and aberration produced by an increase in the spot inclination can be compensated in part by a decrease in the observer's inclination and/or a decrease in the stellar radius. The change in the projected emitting area seen by the observer produced by an increase in the spot inclination can be partially compensated by an increase in the observer's inclination.

\item An increase in the stellar compactness can be partially compensated for by an increase in the spot radius, but as shown by \citet{lamb09a}, the dependence of the waveform on the spot radius is weak unless the spot is very large, while its dependence on the compactness is weak unless the star is very compact ($M/R > 0.25$). Hence this is a weak degeneracy.

\item Changing the number and spectrum of the background counts affects the inferred spectrum and modulation fraction of the waveform and therefore can partially compensate for changes in any of the other parameters that affect the spectrum and modulation fraction of the waveform, including the spot inclination and size, the stellar compactness, and the observer inclination.

\end{itemize}
As noted previously, these degeneracies are an inherent property of any physical model based on a rotating hot spot and cannot be removed by ``improving'' the model.

As a result of these degeneracies, the uncertainties in estimates of $M$ and $R$ depend sensitively on the physical properties of the system. If the hot spot and the observer are at high inclinations, the effects on the waveform caused by the rotation of the star will be maximal, reducing the effects of these degeneracies. If instead the hot spot and the observer are at low inclinations, the effects caused by the rotation of the star will be much less and these degeneracies will much more important.

A large background exacerbates the degeneracy problem, because it increases the statistical fluctuations in the observed waveform. As a result, a wider range of models will adequately fit the waveform data. On the other hand, independent additional knowledge about the system can reduce or eliminate degeneracies. For example, accurate a priori knowledge of the observer's inclination can significantly improve the constraints, if the spot and observer inclinations are high; a priori knowledge of the distance to the system can also help. Independent knowledge of the background can be used to tighten the constraints on $M$ and $R$ substantially. If the spot and observer inclinations are high, measurement of an identifiable atomic spectral line in the burst oscillation spectrum can be used to tighten the constraints.

\subsubsection{Effects of spot and observer inclinations}
\label{sec:results:constraints:inclination-effects}

Figure~\ref{fig:results:diff_bkg} (see also 
Table~\ref{table:results:summary}) shows that the constraints on $M$ and $R$ are fairly tight when the hot spot and the observer are both at high inclinations but very loose when they are both at low inclinations, regardless of the background level. Comparison of Figures~\ref{fig:results:diff_bkg}c,
\ref{fig:results:high_incl-ispot80_60}a,
and~\ref{fig:results:high_incl-ispot80_60}b (again see Table~\ref{table:results:summary}) shows that for high observer inclinations, the constraints on $M$ and $R$ weaken slowly as the spot inclination decreases from $90^{\circ}$ to $60^{\circ}$.

If the hot spot and observer inclinations are both high, the $1\sigma$ uncertainties in $M$ and $R$ range from a few percent for our low and medium backgrounds to about 10\% for our high background (see Table~\ref{table:results:summary}). The $3\sigma$ uncertainties are much larger, ranging from about 15\% for our low or medium backgrounds to about 50\% for our high background. 
If instead the spot and observer inclinations are both low, the $1\sigma$ uncertainties in $M$ and $R$ are large, ranging from 30\% to 50\% or more (again see Table~\ref{table:results:summary}), and the $3\sigma$ contours span much or all of the $M$-$R$ domain that was searched. The importance of high spot and observer inclinations becomes especially clear if one compares the confidence regions for the high-inclination case shown in
Figure~\ref{fig:results:diff_bkg}d with the confidence regions for the low-inclination case shown in Figure~\ref{fig:results:diff_bkg}e. The fractional rms amplitudes of the oscillations are the same in these two cases; only the inclinations  are different.

For high hot spot and observer inclinations, the oscillation  amplitude and the effects on the energy dependence and harmonic content of the waveform of the relativistic Doppler boost and aberration are responsible for the lower bound on $R$. The upper bound on $R$ is provided by the requirement that the amplitude be sufficiently large but the effects of rotational motion on the waveform must not be too large. This requirement also imposes a lower bound on the observer's inclination. 

For low spot and observer inclinations, the waveform is nearly sinusoidal and the modulation fraction is low. Therefore much wider ranges of the parameters other than $M$ and $R$ (such as the spot and observer inclinations) give a modulation fraction low enough to be consistent with the observed waveform, if the compactness is high. Consequently, when we marginalize over these other parameters, $M$-$R$ pairs near the high compactness boundary are given more weight. As a result, the highest posterior probability density is typically near the highest compactness that was searched, which means it is typically close to the high-compactness boundary of the search domain, even though comparable fits would be possible for many values of $M/R$ if we were to optimize the other parameters.

\subsubsection{Effects of the background}
\label{sec:results:constraints:background-effects}

\begin{figure*}[!t]
\begin{center}
\includegraphics[height=.26\textheight]{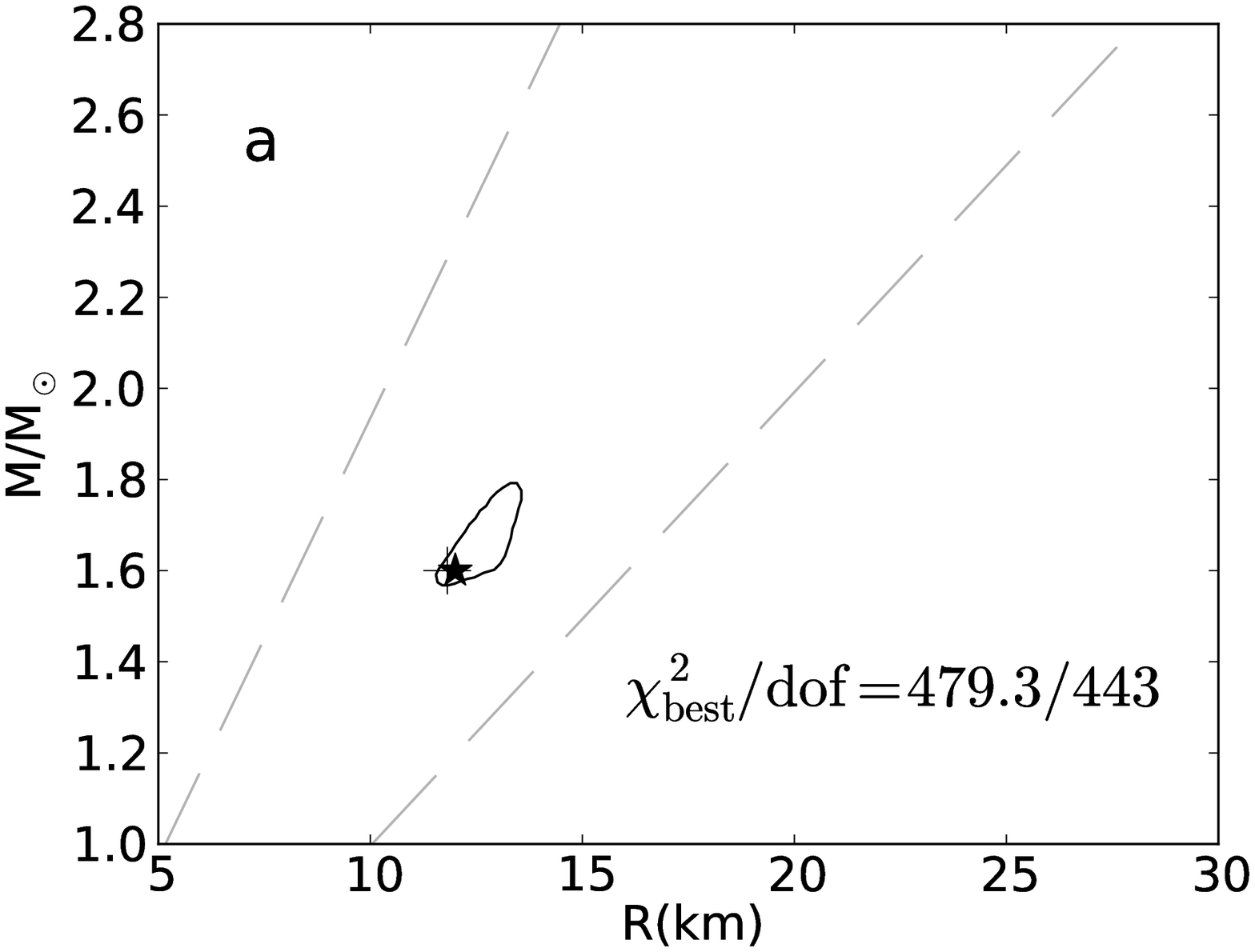}
\includegraphics[height=.26\textheight]{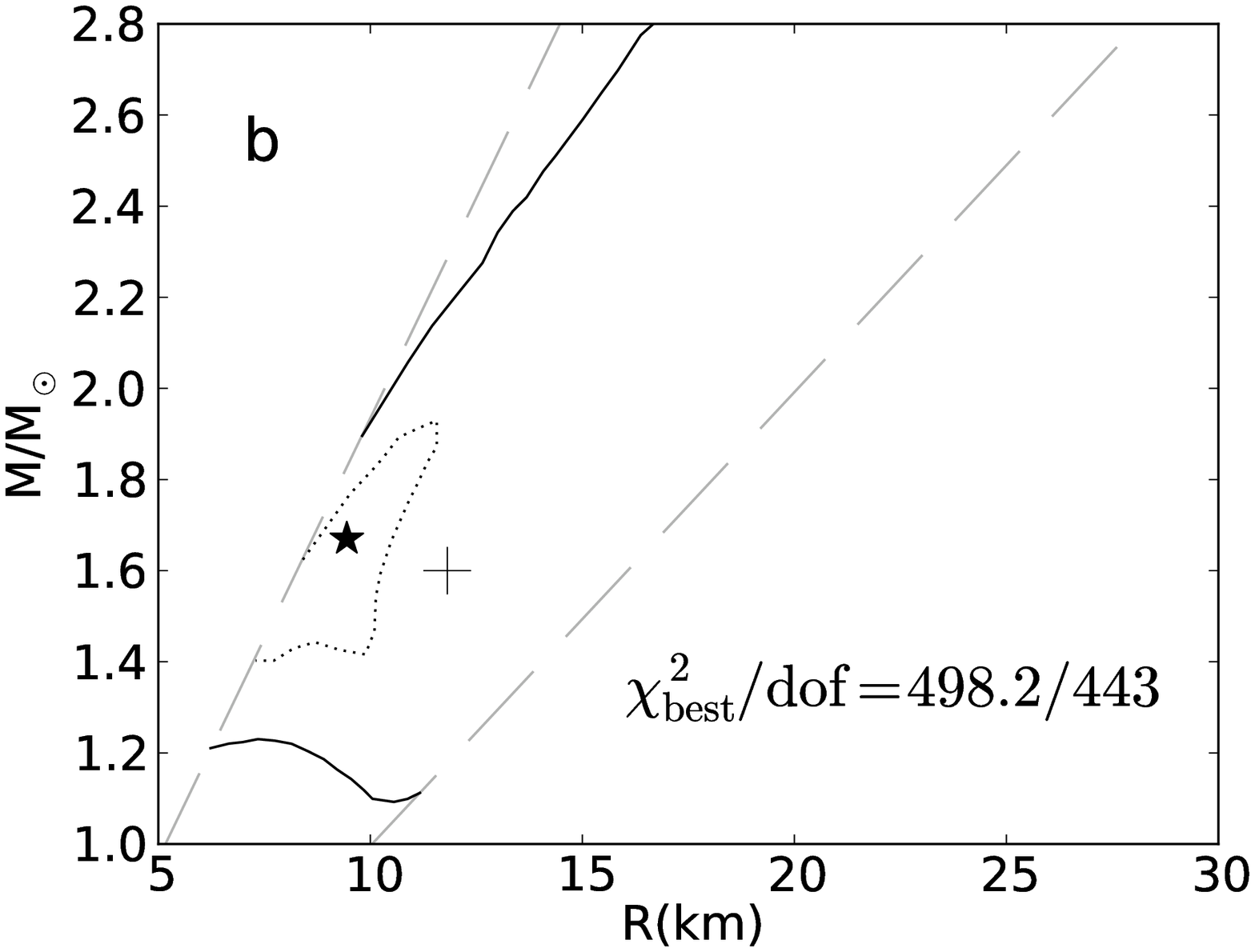} 
\vskip-2pt
\includegraphics[height=.26\textheight]{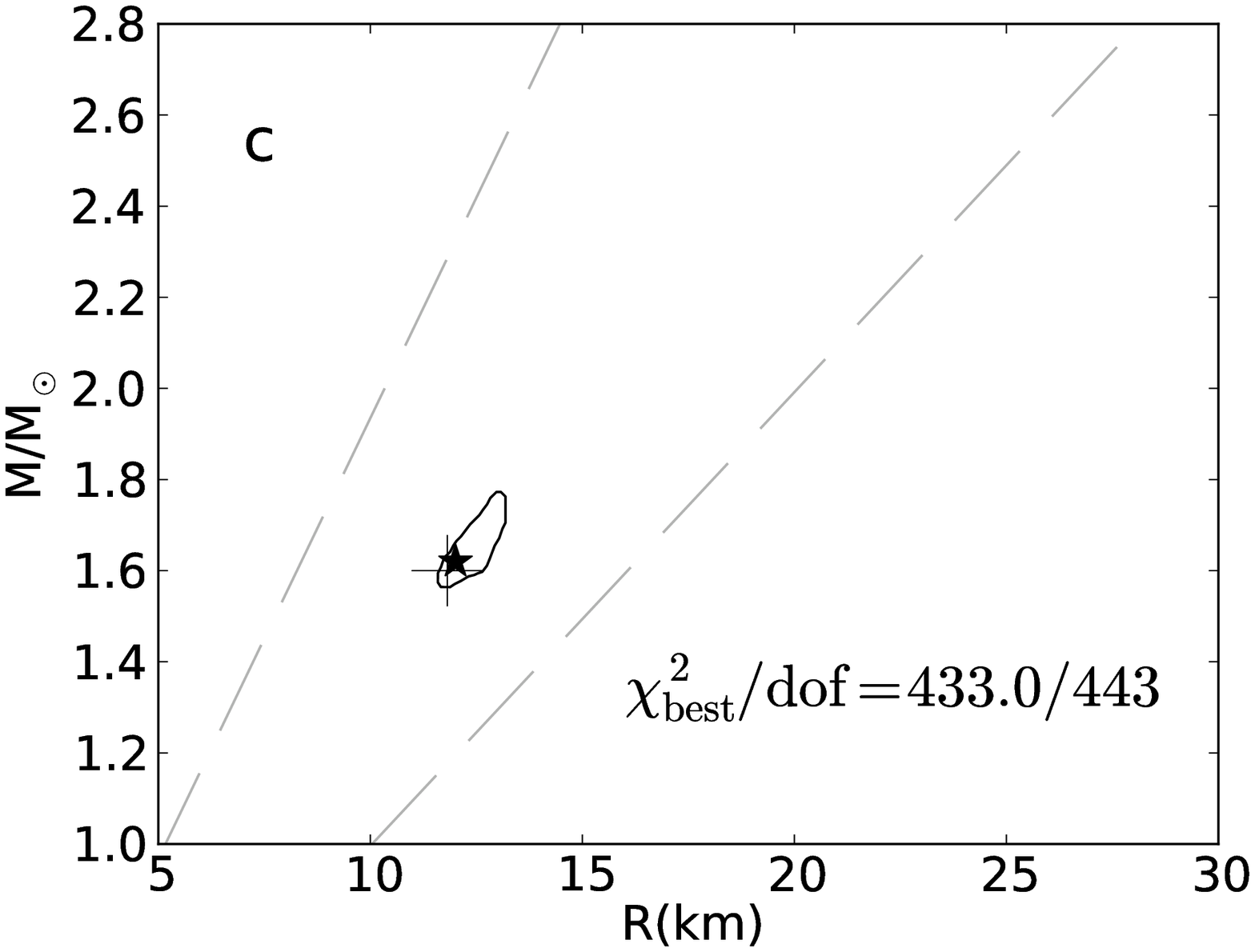}
\includegraphics[height=.26\textheight]{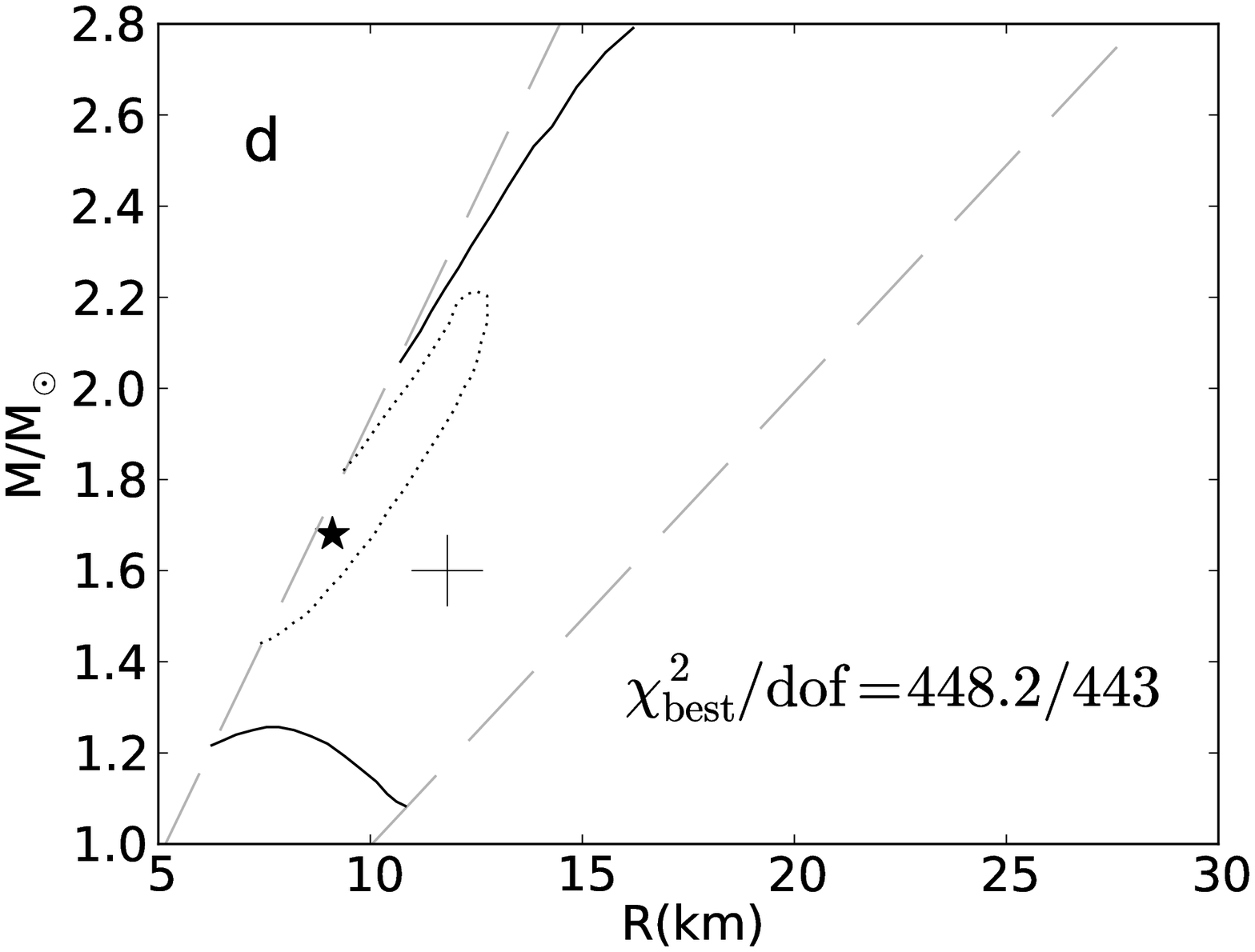}
\vskip-2pt
\includegraphics[height=.26\textheight]{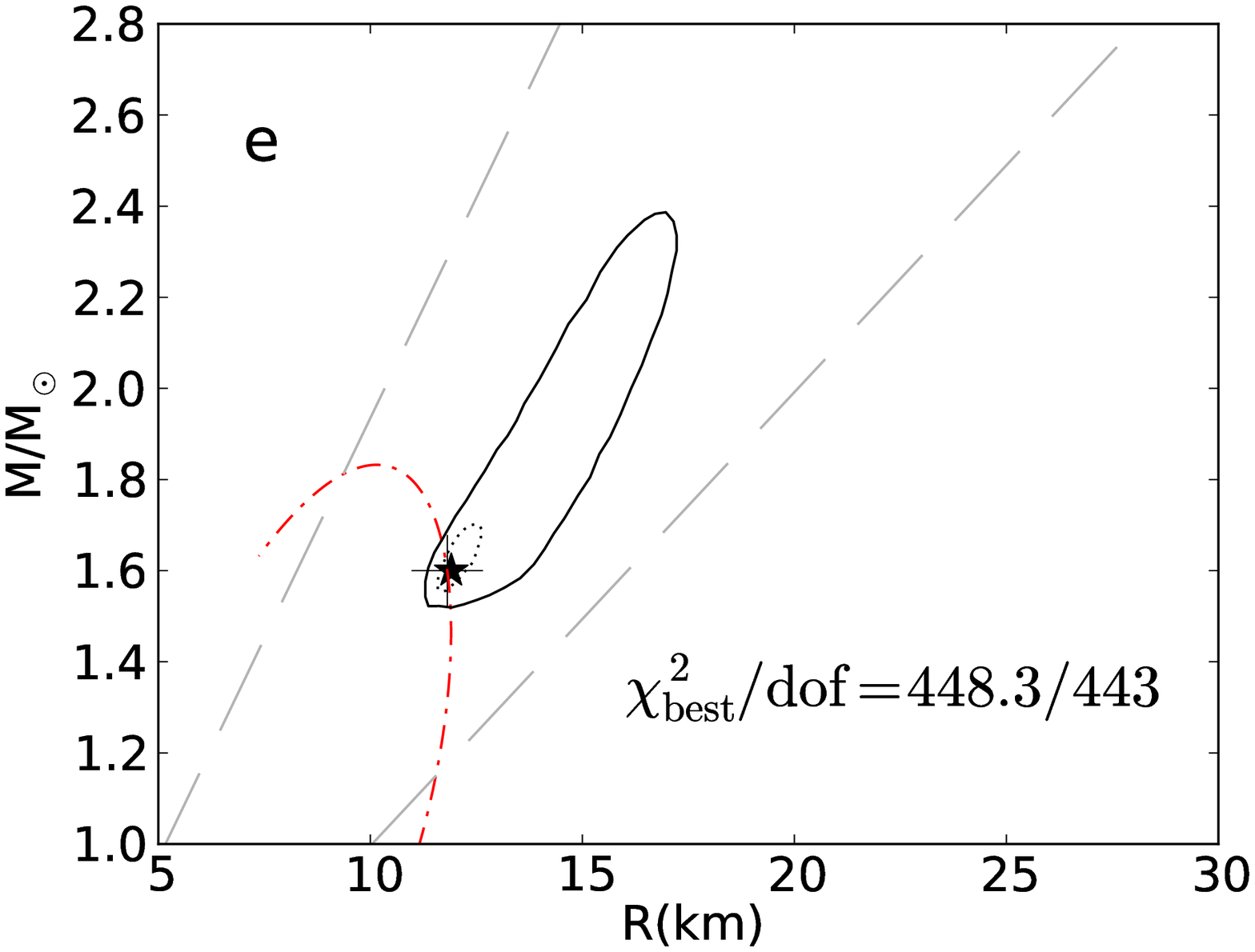}
\includegraphics[height=.26\textheight]{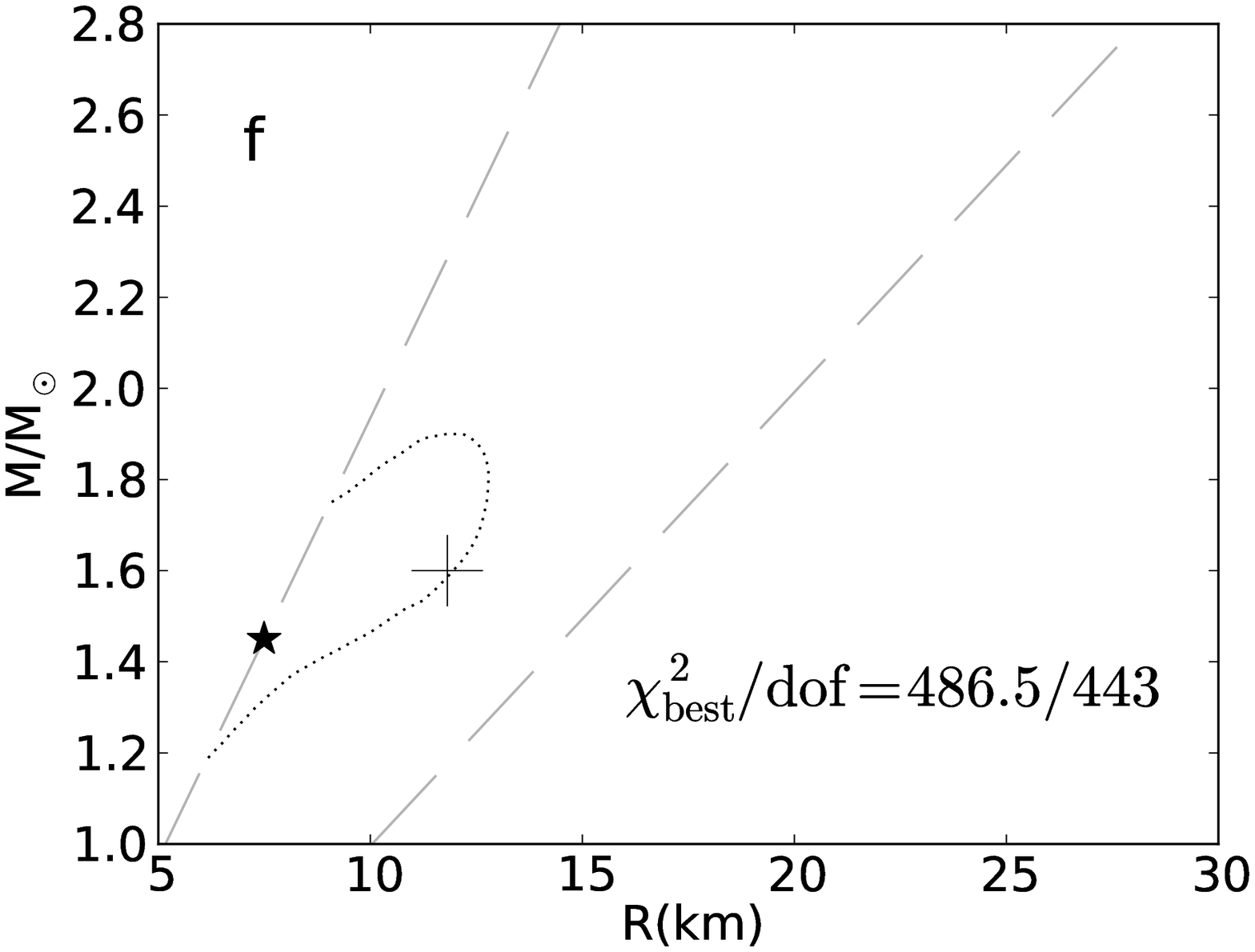}
\end{center}
\vspace{-0.7cm}
\caption{These results show that $M$ and $R$ are tightly constrained when the spot and 
observer inclinations are high and the total number of counts from the background is comparable to or less than the number from the hot spot. The contours in the left column are for the high spot and observer inclination model and (top to bottom) the low-, medium-, and high-background models (for the properties of these models, see Tables~\ref{table:results:reference-cases} 
and~\ref{table:results:summary}). The right column shows the corresponding contours for the low spot and observer inclination model.
The dashed lines show the $R/M=3.5$ and
$R/M=6.8$ boundaries within which the posterior
probability distribution was computed.
The dotted and solid curves
show, respectively, the portions of the $1\sigma$ and $3\sigma$ confidence contours within this domain.
The + symbol indicates the radius and mass that were used in generating the synthetic waveform and the star symbol indicates the values where the marginalized posterior probability density is highest.
The red dashed-dot curve in Fig.~\ref{fig:results:diff_bkg}e shows points of constant $g^{\frac{2}{9}}/(1+z)$, illustrating the complementary constraint that can be obtained by fitting the spectra of X-ray bursts from the same star (see text).
}
\label{fig:results:diff_bkg}
\end{figure*}

\begin{figure*}[!t]
\begin{center}
\includegraphics[height=.26\textheight]{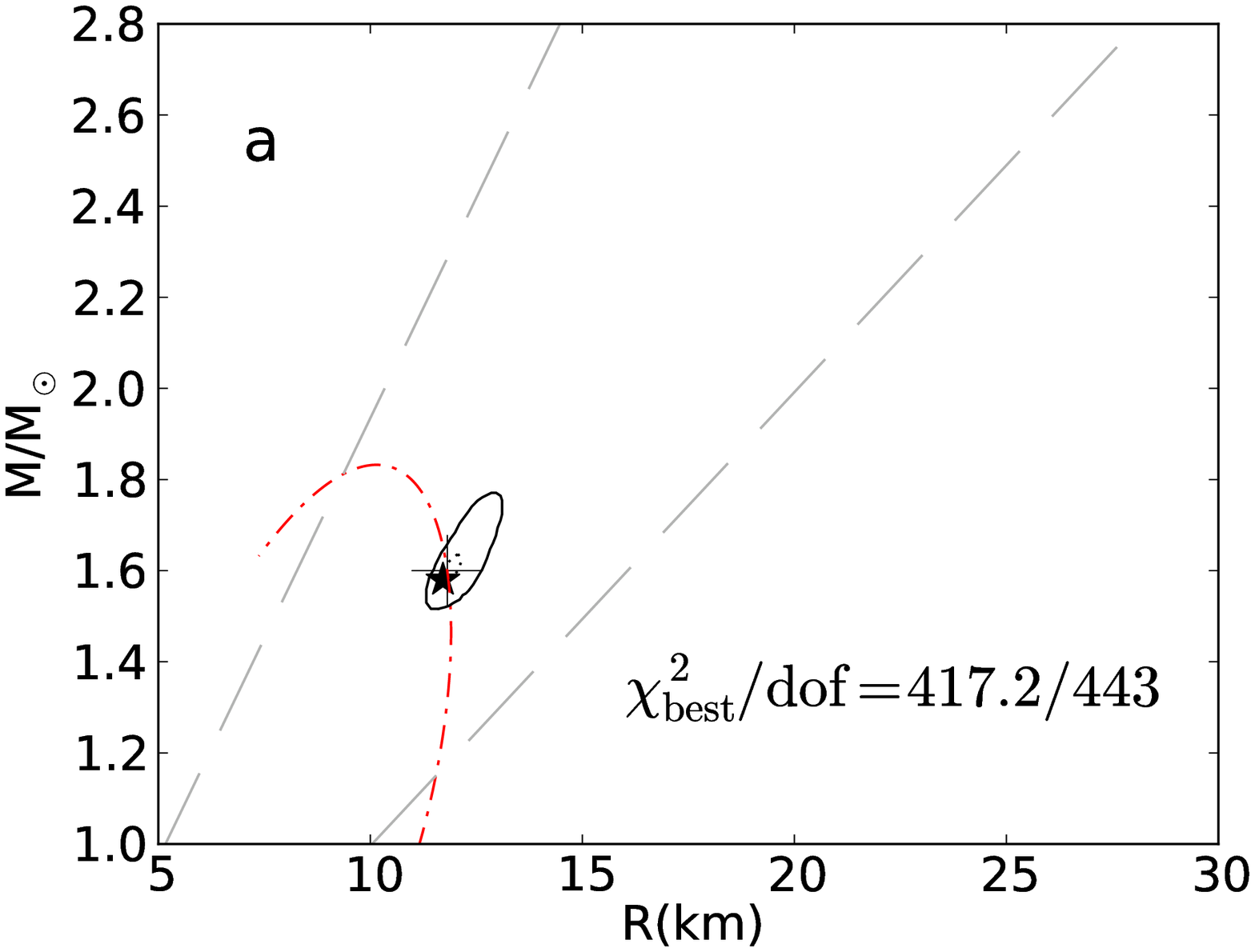}
\includegraphics[height=.26\textheight]{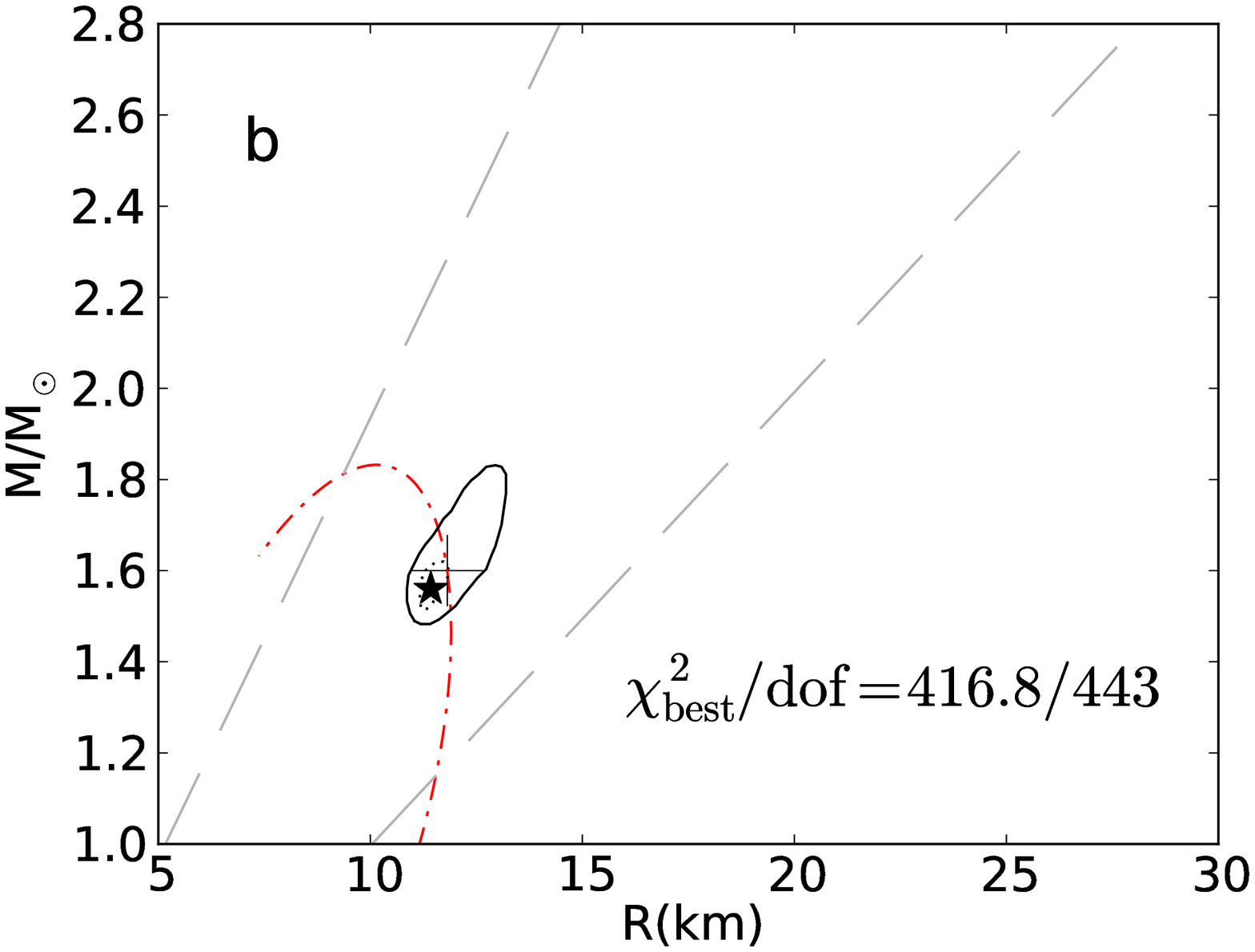}
\end{center}
\vspace{-0.7cm}
\caption{
Comparison of the contours shown here and in 
Fig.~\ref{fig:results:diff_bkg}c demonstrates that the constraints on $M$ and $R$ weaken slowly as the spot inclination decreases from $90^{\circ}$ 
(in Fig.~\ref{fig:results:diff_bkg}c) to $80^{\circ}$ (left panel) and to $60^{\circ}$ (right panel). All three plots assume the observer inclination is $90^{\circ}$ and our medium background ($\sim 10^6$ counts from the hot spot and $\sim 10^6$ background counts).
The red dashed-dot curves show points of constant $g^{\frac{2}{9}}/(1+z)$, illustrating the complementary constraint that can be obtained by fitting the spectra of bursts from the same star (see text).
For the meanings of the line types and symbols, see 
Fig.~\ref{fig:results:diff_bkg}.
}
\label{fig:results:high_incl-ispot80_60}
\end{figure*}

Figure~\ref{fig:results:diff_bkg} (see also Table~\ref{table:results:summary}) illustrates how the constraints on $M$ and $R$ are affected by background counts. The background has very little effect on the uncertainties unless the number of background counts exceeds the number of counts from the hot spot, in which case the uncertainties increase as the background increases, approximately as $1/{\cal R}$, where ${\cal R}\equiv N_{\rm osc}/\sqrt{N_{\rm tot}}$ in terms of the number of counts $N_{\rm osc}$ in the oscillation and the total number of counts $N_{\rm tot}$  (see equation~(\ref{eqn:R-value})).
Thus, the $1\sigma$ and $3\sigma$ uncertainties for our high background, which contributes 9 times as many counts as the hot spot, are several times larger than the uncertainties for our low and medium backgrounds, which contribute, respectively, less than and about the same number of counts as the hot spot.

\begin{deluxetable}{llrl}
\tablewidth{0pt}
\tablecaption{
${\cal R}$ values and rms amplitudes
for selected synthetic waveforms\tablenotemark{a}
}
\tablehead{
\colhead{Inclination} & \colhead{Background} 
& \colhead{~~${\cal R}$}  & \colhead{rms}
}
\startdata
low & low & 340 & 0.21   \\
low & medium & 280 & 0.14   \\
low & high & 120 & 0.028   \\
high & low & 1360 & 0.81   \\
high & medium & 1100 & 0.54   \\
high & high & 490 & 0.11   \\
\enddata
\vskip-10pt
\tablenotetext{a}{
Here ${\cal R}\equiv N_{\rm osc}/\sqrt{N_{\rm tot}}$, where $N_{\rm osc}$ is the number of counts in the oscillating component of the synthetic waveform and $N_{\rm tot}$ is the total number of counts  in the waveform. See equation~(\ref{eqn:R-value}). 
}
\label{table:results:R&rms}
\end{deluxetable}

The values of ${\cal R}$ and the fractional rms amplitudes
of the oscillations in the synthetic waveforms generated in the low- and high-inclination reference cases for our low, medium, and high backgrounds are listed in Table~\ref{table:results:R&rms}.
Scaling from \textit{RXTE} observations suggests ${\cal R} \sim 150$ for a typical burst oscillation observed during a single burst using a 10~m$^2$ detector 
(see Section~\ref{sec:statistical-uncertainties}). ${\cal R}$ scales as the square root of the number of counts, so combining the data from 25 such bursts 
(see Section~\ref{sec:results:constraints:combining-data} and Appendix~\ref{app:jointfits}) could give ${\cal R} \sim 800$, between the values for the synthetic waveforms generated in the high-inclination, medium-background and high-inclination, high-background cases.
The high-inclination, high-background case has a fractional rms amplitude comparable to the $\sim\,$10\% fractional amplitudes typically observed in burst tails using \textit{RXTE}.

\begin{figure*}[!t]
\begin{center}
\includegraphics[height=.26\textheight]{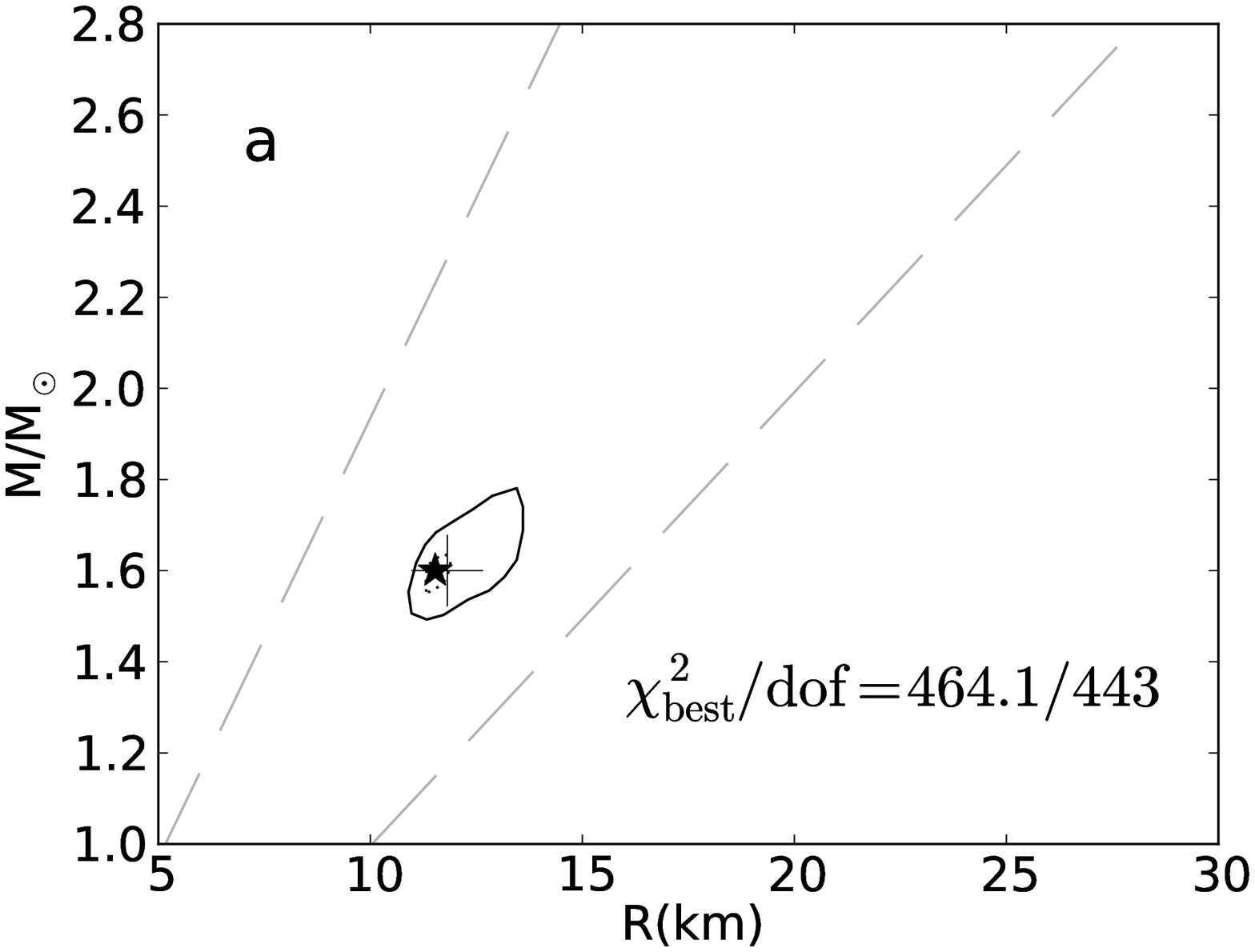}
\includegraphics[height=.26\textheight]{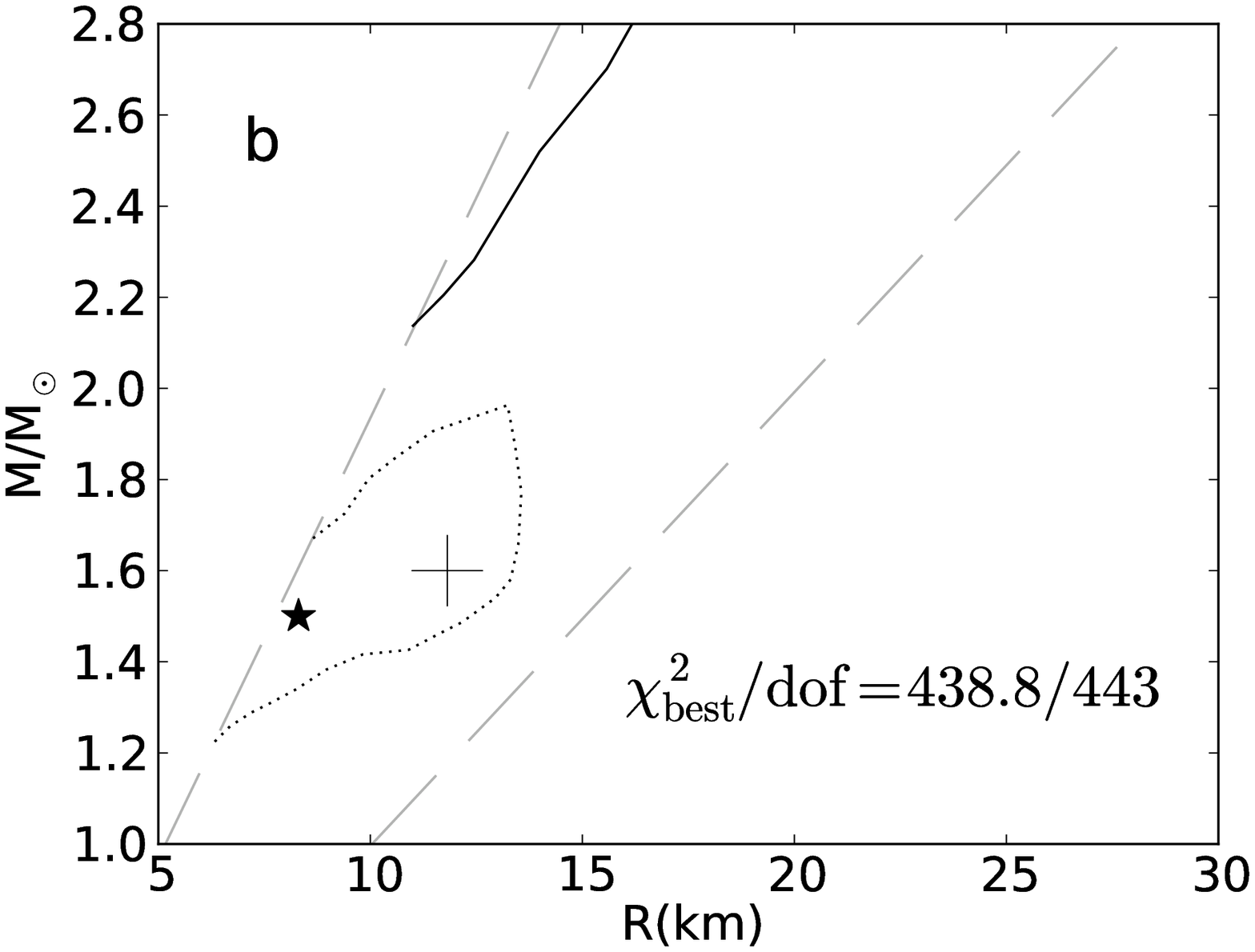}
\vskip-2pt
\includegraphics[height=.26\textheight]{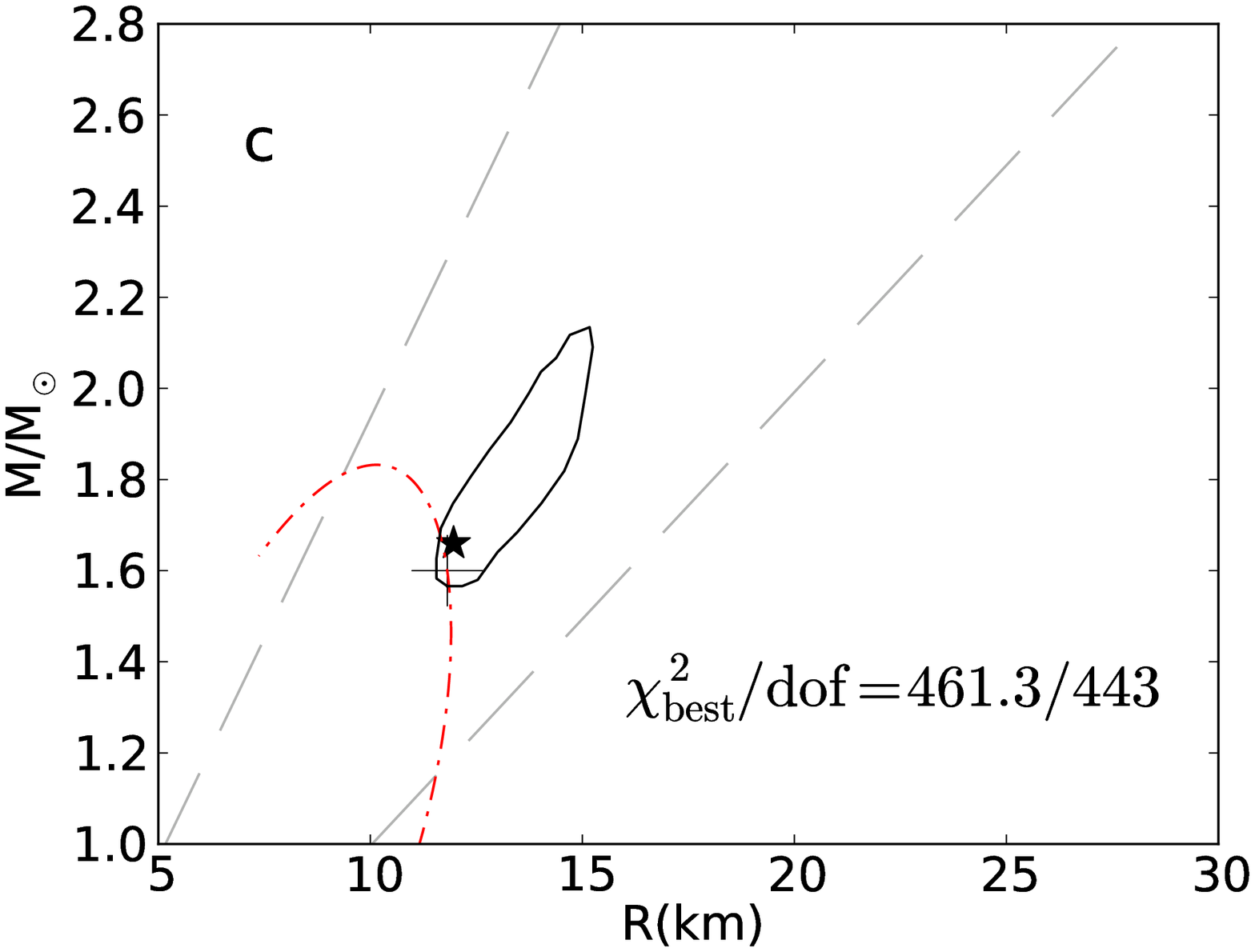}
\includegraphics[height=.26\textheight]{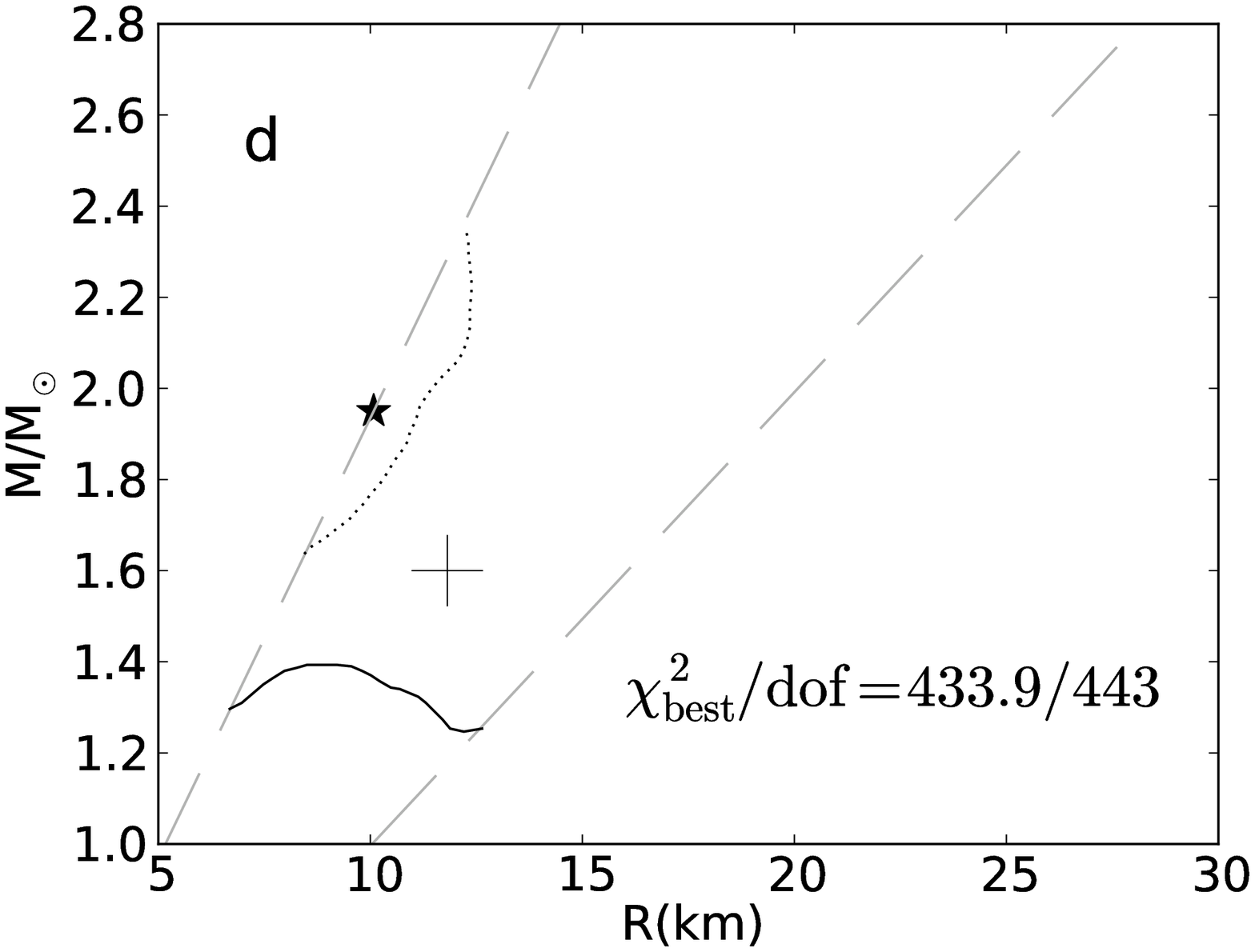}
\vskip-2pt
\includegraphics[height=.26\textheight]{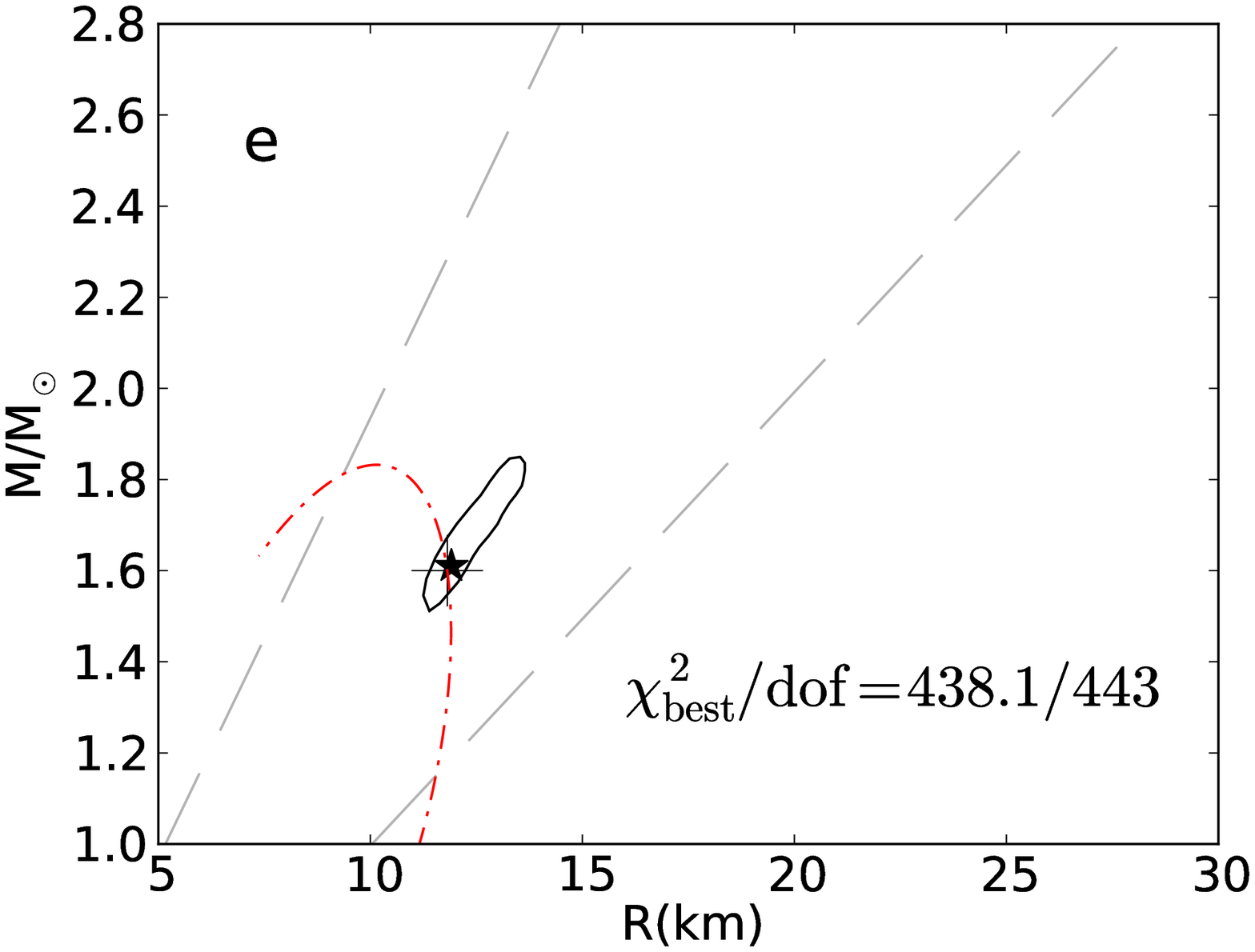}
\includegraphics[height=.26\textheight]{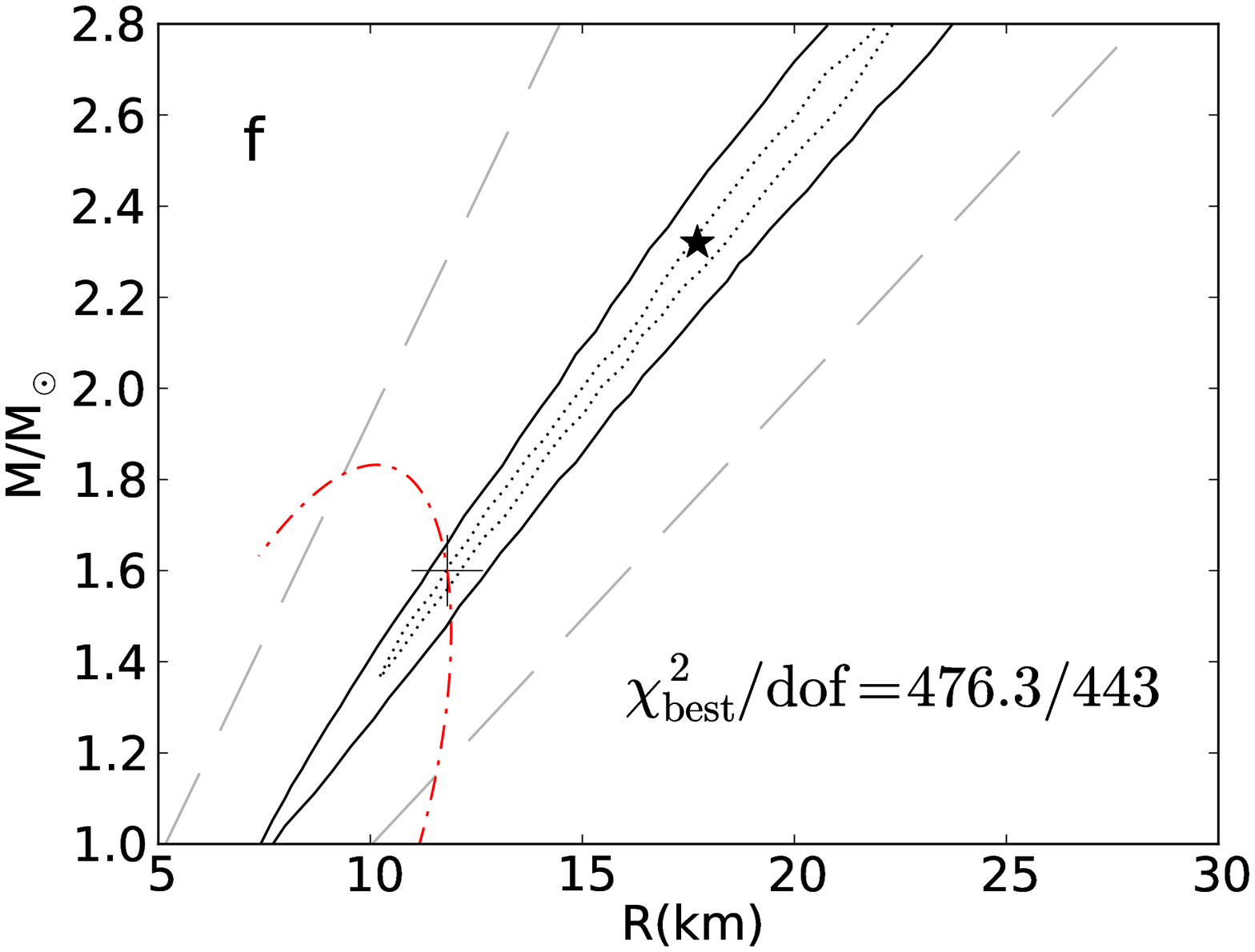}
\end{center}
\vspace{-0.7cm}
\caption{These results show that independent knowledge of some system parameters can significantly tighten the constraints on $M$ and $R$. All results are for our high background. The contours in the left column are for the high spot and observer inclination case and (top to bottom) known observer inclination, known distance, and an identifiable atomic scattering line in the waveform spectrum (see text for details); these contours should be compared with those in Fig.~\ref{fig:results:diff_bkg}e. The right column shows the corresponding contours for the low spot and observer inclination case; these contours should be compared with those in Fig.~\ref{fig:results:diff_bkg}f.
For the meanings of the line types and symbols, see 
Fig.~\ref{fig:results:diff_bkg}.
}
\label{fig:results:diff_knowledge}
\end{figure*}

\begin{figure*}[!t]
\begin{center}
\includegraphics[height=.26\textheight]{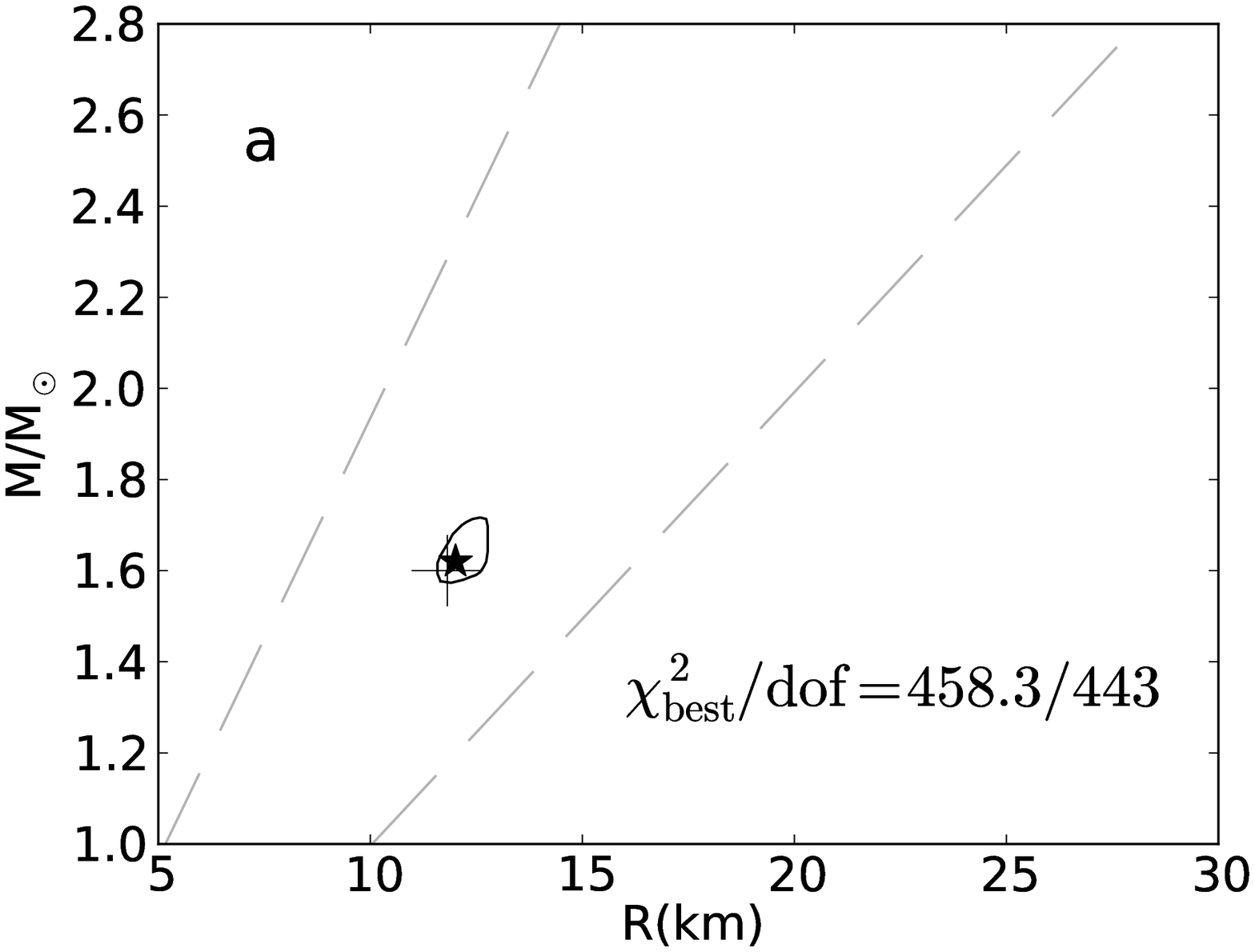}
\includegraphics[height=.26\textheight]{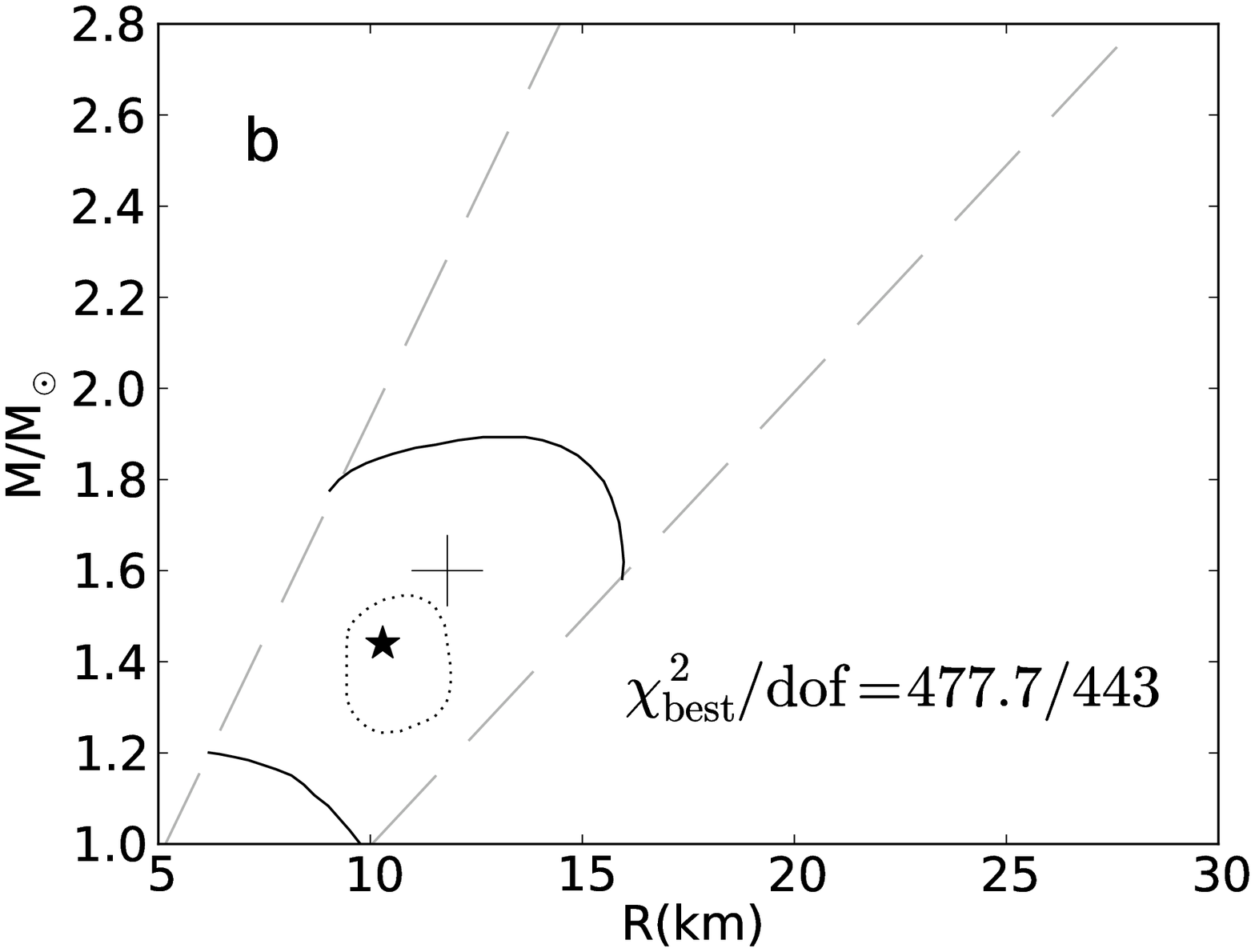}
\end{center}
\vspace{-0.7cm}
\caption{
These results assume that the properties of the background are known.
\textit{Left}: Constraints for the high-inclination reference case (spot and observer inclinations of $90^{\circ}$).
\textit{Right}: Constraints for the low-inclination reference case (spot and observer inclinations of $60^{\circ}$ and $20^{\circ}$, respectively).
Comparison of these contours with those shown in 
Figs.~\ref{fig:results:diff_bkg}c and~\ref{fig:results:diff_bkg}d, which assume the background is unknown, demonstrate that $M$ and $R$ are much more tightly constrained if the background is known.
All these results assume our medium background.
For the meanings of the line types and symbols, see 
Fig.~\ref{fig:results:diff_bkg}.
}
\label{fig:results:known-background}
\end{figure*}

\subsubsection{Effects of independent knowledge of relevant system properties}
\label{sec:results:constraints:independent-info}

The extent to which knowledge that is independent of the waveform fitting process improves the precision of $M$ and $R$ estimates depends strongly on the inclinations of the hot spot and the observer. 
If they are high, the uncertainties in $M$ and $R$ can be reduced by independent knowledge of the observer's inclination or of the distance to the star. In this case observing an atomic scattering line with a known rest energy tightly constrains the stellar compactness $M/R$ and improves the constraints on $M$ and $R$ more than knowing the distance and approximately as much as knowing the observer's inclination.
If instead the spot and observer inclinations are low, observing a scattering line with a known rest energy still tightly constrains the stellar compactness but improves the constraints on $M$ and $R$ very little.
The uncertainties in $M$ and $R$ can be reduced substantially by independent knowledge of the size and spectrum of the background, especially if the spot and observer inclinations are high. 
Table~\ref{table:results:summary} lists the $1\sigma$ fractional uncertainties for each of these cases.
Analysis of high-quality observations of the spectra of the bursts using high-precision spectral models can help to further constrain $M$ and $R$.

Figure~\ref{fig:results:diff_knowledge}a 
shows that if the spot and observer inclinations are both $90^{\circ}$, knowing the inclination of the observer reduces the $1\sigma$ uncertainty from $\sim\,$9\% to $\sim\,$5\% and the $3\sigma$ uncertainty from $\sim\,$50\% to $\sim\,$20\% for our high background (compare this figure with Figure~\ref{fig:results:diff_bkg}e). 
Figure~\ref{fig:results:diff_knowledge}c shows that knowing the distance to the star reduces the $1\sigma$ uncertainty to $\sim\,$7\% and the $3\sigma$ uncertainty to $\sim\,$30\% (again compare 
this figure with 
Figure~\ref{fig:results:diff_bkg}e). 
Additional computations show that if the spot is at $80^{\circ}$ and the observer's inclination is known, the uncertainties in $M$ and $R$ are similar to the uncertainties obtained if the spot is at $90^{\circ}$.
If the spot and observer inclinations are high and the observer inclination and the distance are both known, the $1\sigma$ uncertainties in $M$ and $R$ are similar to the uncertainties obtained if only the observer's inclination is known. The similarity of the $1\sigma$ uncertainties in these two cases is not surprising, because independent knowledge of the observer's inclination is usually more helpful than independent knowledge of the distance.
If instead the spot and observer inclinations are low, knowing the inclination of the observer or the distance to the star reduces the uncertainties only modestly (compare 
Figures~\ref{fig:results:diff_knowledge}b and~\ref{fig:results:diff_knowledge}d with 
Figure~\ref{fig:results:diff_bkg}f).

Figures~\ref{fig:results:diff_knowledge}e 
and~\ref{fig:results:diff_knowledge}f illustrate the effect on the precision of $M$ and $R$ estimates if an atomic scattering line with a known rest energy is observed in the emission from the hot spot. The results shown here are for a line at 6.4~keV with a FWHM of 0.2~keV and an optical depth at the line center of 0.24. 
If the spot and observer inclinations are high, observing this line   tightly constrains the stellar compactness and tightens the constraints on $M$ and $R$, decreasing the $1\sigma$ uncertainty from $\sim\,$9\% to $\sim\,$4\% and the $3\sigma$ uncertainty from $\sim\,$50\% to $\sim\,$20\% for our high background, a greater improvement than would be achieved by independently determining the distance and approximately as much as would be produced by independently determining the observer's inclination (compare Figure~\ref{fig:results:diff_knowledge}e with Figure~\ref{fig:results:diff_bkg}e). If instead the spot and observer inclinations are low, observing this scattering line still tightly constrains the stellar compactness but improves the constraints on $M$ and $R$ very little (compare Figure~\ref{fig:results:diff_knowledge}f with 
Figure~\ref{fig:results:diff_bkg}f; for the particular realization shown in Figure~\ref{fig:results:diff_knowledge}f, the constraints on $M$ and $R$ are weaker with a scattering line present, probably due to a sampling fluctuation).

Figure~\ref{fig:results:known-background} shows that the knowing the size and spectrum of the background greatly improves the constraints on $M$ and $R$. For our high-inclination case, knowing the background decreases the $1\sigma$ uncertainties in $M$ and $R$ from $\sim\,$9\% to $\sim\,$4\% and the $3\sigma$ uncertainties from $\sim\,$50\% to $\sim\,$9\% (compare Figure~\ref{fig:results:known-background}a with
Figure~\ref{fig:results:diff_bkg}e). For our low-inclination case, knowing the background leads to a $1\sigma$ uncertainty of $\sim\,$20\%, whereas without this knowledge one obtains no useful constraints on $M$ and $R$ (compare Figure~\ref{fig:results:known-background}b with Figure~\ref{fig:results:diff_bkg}f).

\begin{figure*}[!t]
\begin{center}
\includegraphics[height=.26\textheight]{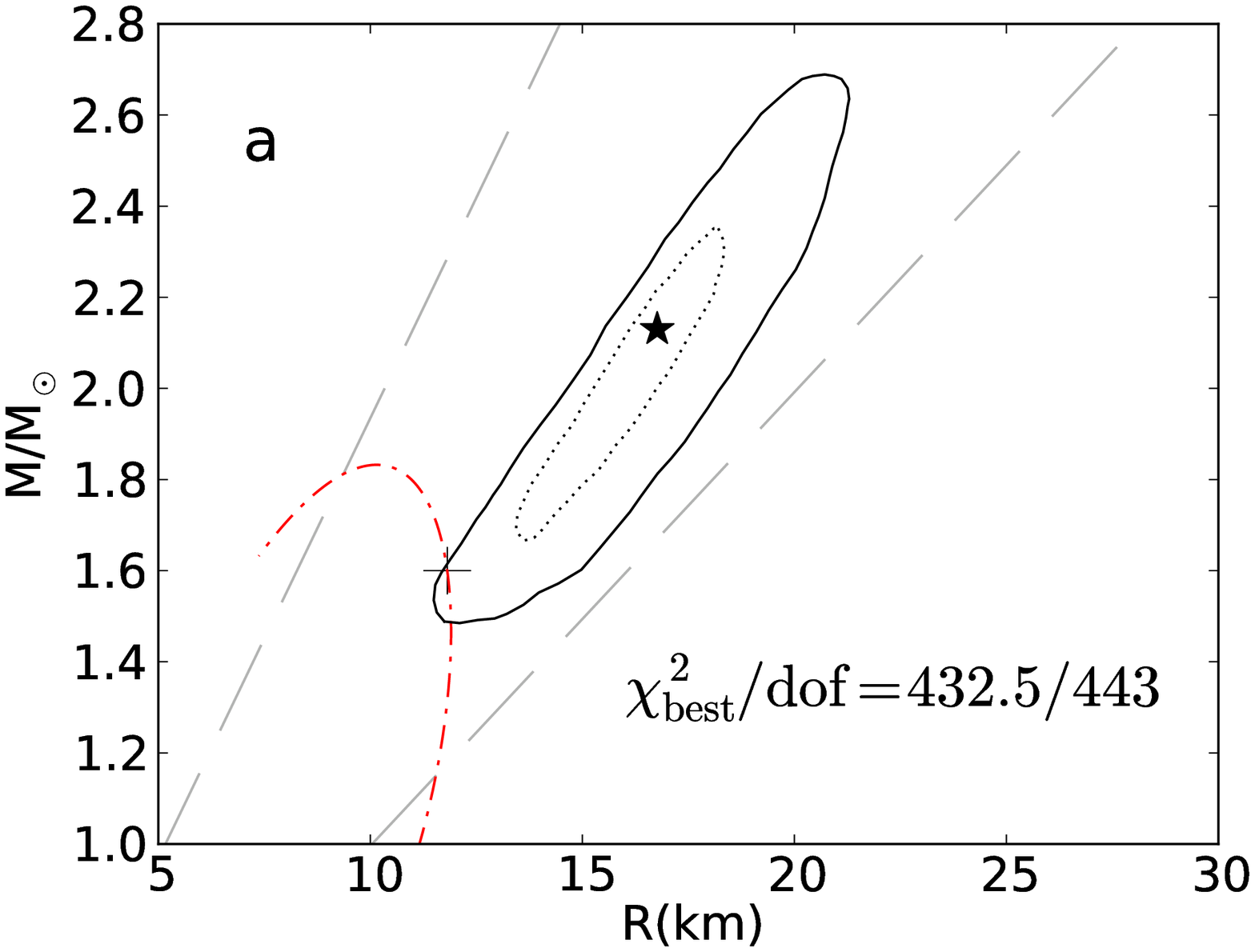}
\includegraphics[height=.26\textheight]{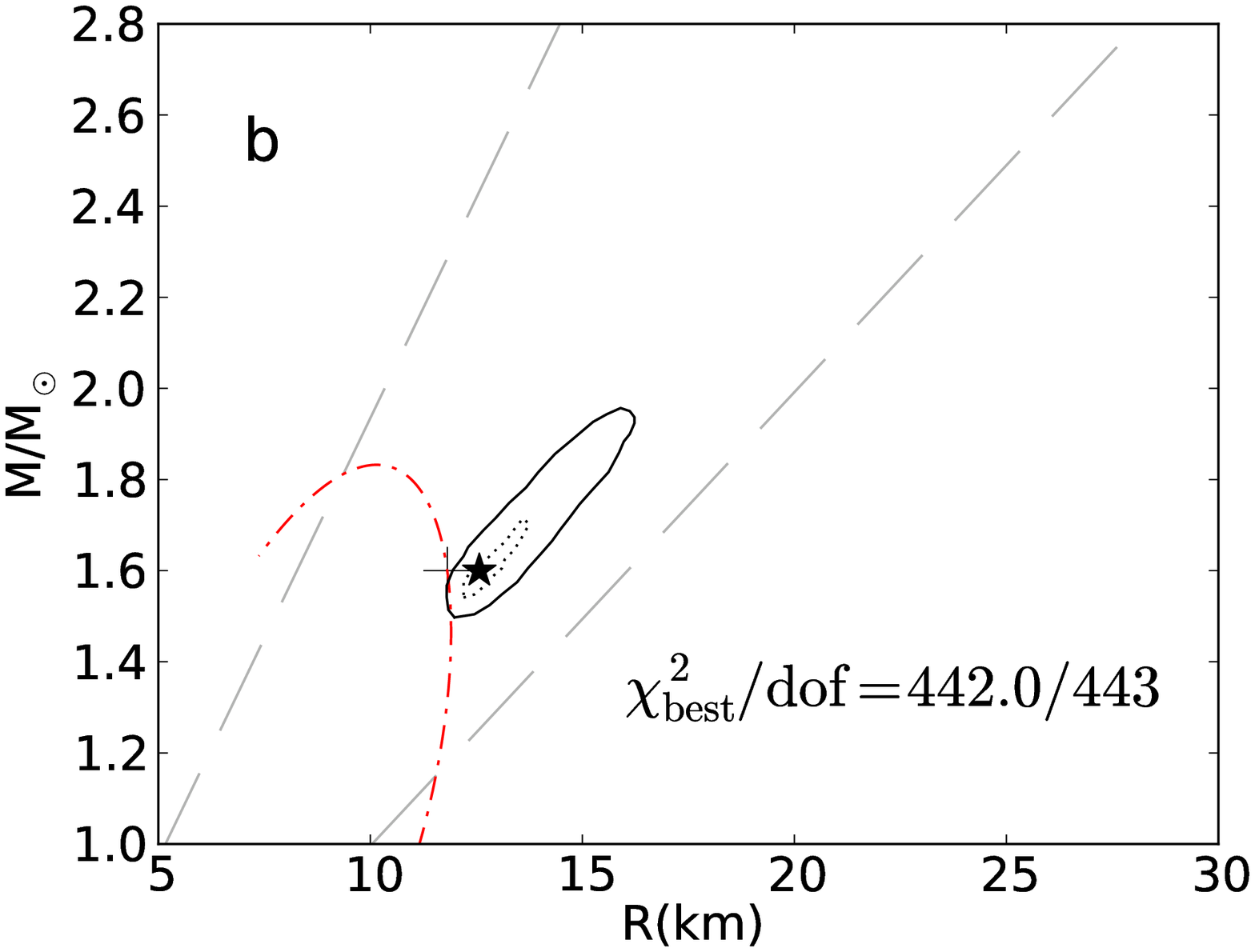}
\end{center}
\vspace{-0.7cm}
\caption{
These results show that fitting the observed waveform using a model with an incorrect spot shape increases the uncertainty of the $M$ and $R$ estimates and can sometimes bias them. 
\textit{Left}: Constraints obtained when the actual emitting area is elongated in the north-south direction, represented here by two uniform, circular spots of angular radius $25^{\circ}$, centered at the same longitude, with inclinations of $40^{\circ}$ and $90^{\circ}$, but the waveform is fit assuming a circular spot.
\textit{Right}: Constraints obtained when the actual emitting area is rotational equator and elongated in the east-west direction, represented here by three uniform, circular spots of angular radius $25^{\circ}$, with centers at the same $90^{\circ}$ inclination and spaced $22.5^{\circ}$ apart in the azimuthal direction, but the waveform is fit assuming a circular spot.
Both results are for our medium background.
For the meanings of the line types and symbols, see 
Fig.~\ref{fig:results:diff_bkg}.
}
\label{fig:results:high_incl-NS_EW}
\end{figure*}

Fitting high-quality observations of the X-ray spectra of many bursts from the same star using detailed, high-precision model atmosphere calculations can produce a constraint on $M$ and $R$ that is complementary to that obtained by fitting burst oscillation waveforms. This is illustrated by the dash-dotted red curves in Figures~\ref{fig:results:diff_bkg}e,
\ref{fig:results:high_incl-ispot80_60}a,
\ref{fig:results:high_incl-ispot80_60}b,
\ref{fig:results:diff_knowledge}c,
\ref{fig:results:diff_knowledge}e, 
\ref{fig:results:diff_knowledge}f,
\ref{fig:results:high_incl-NS_EW}a,
and~\ref{fig:results:high_incl-NS_EW}b, which show the relation $(1+z)/g^{\frac{2}{9}} = {\rm const.}$ for the values of the redshift $z$ and surface gravity $g$ that correspond to the $M$ and $R$ values used in generating these synthetic waveforms. As discussed in 
Section~\ref{sec:results:waveform-analysis:additional-info}, this combination of $g$ and $z$ can be determined by accurately measuring the spectra of the X-ray bursts. The red dashed-dot curve intersects the confidence region derived by fitting burst oscillation waveforms at a high angle. Combining these two complementary methods is therefore a promising way to further constrain $M$ and $R$.

\begin{figure*}[!t]
\begin{center}
\includegraphics[height=.26\textheight]{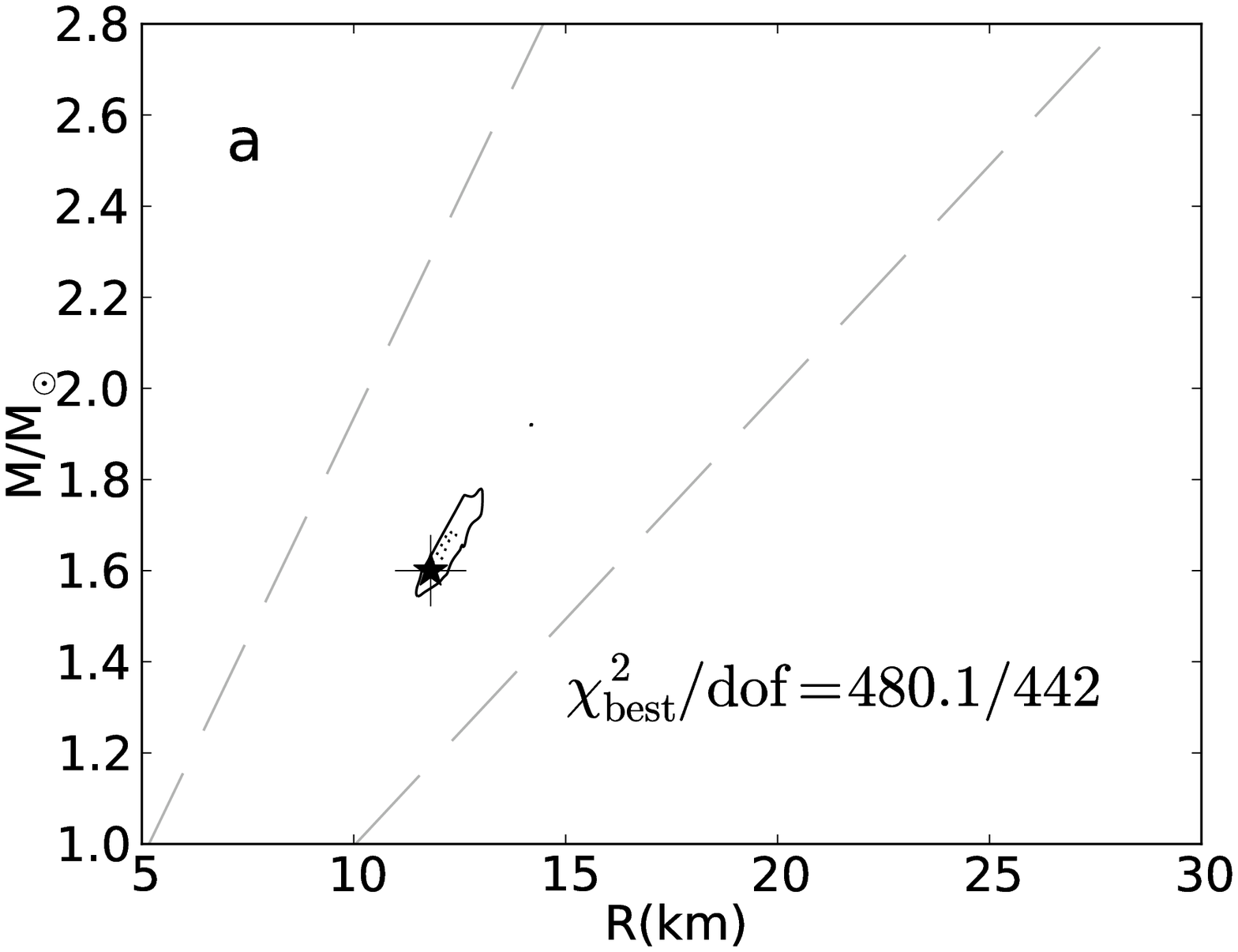}
\includegraphics[height=.26\textheight]{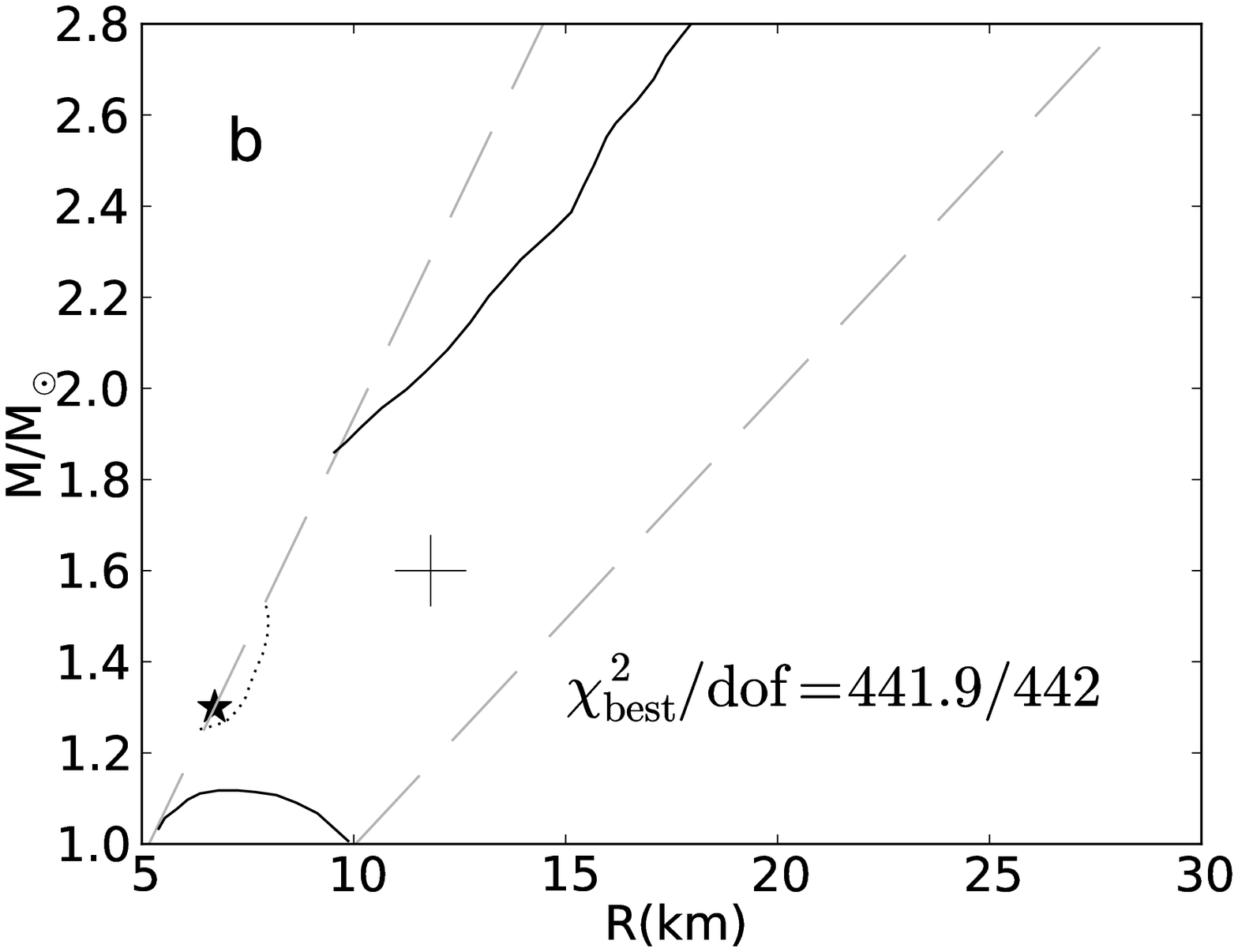}
\vskip-2pt
\includegraphics[height=.26\textheight]{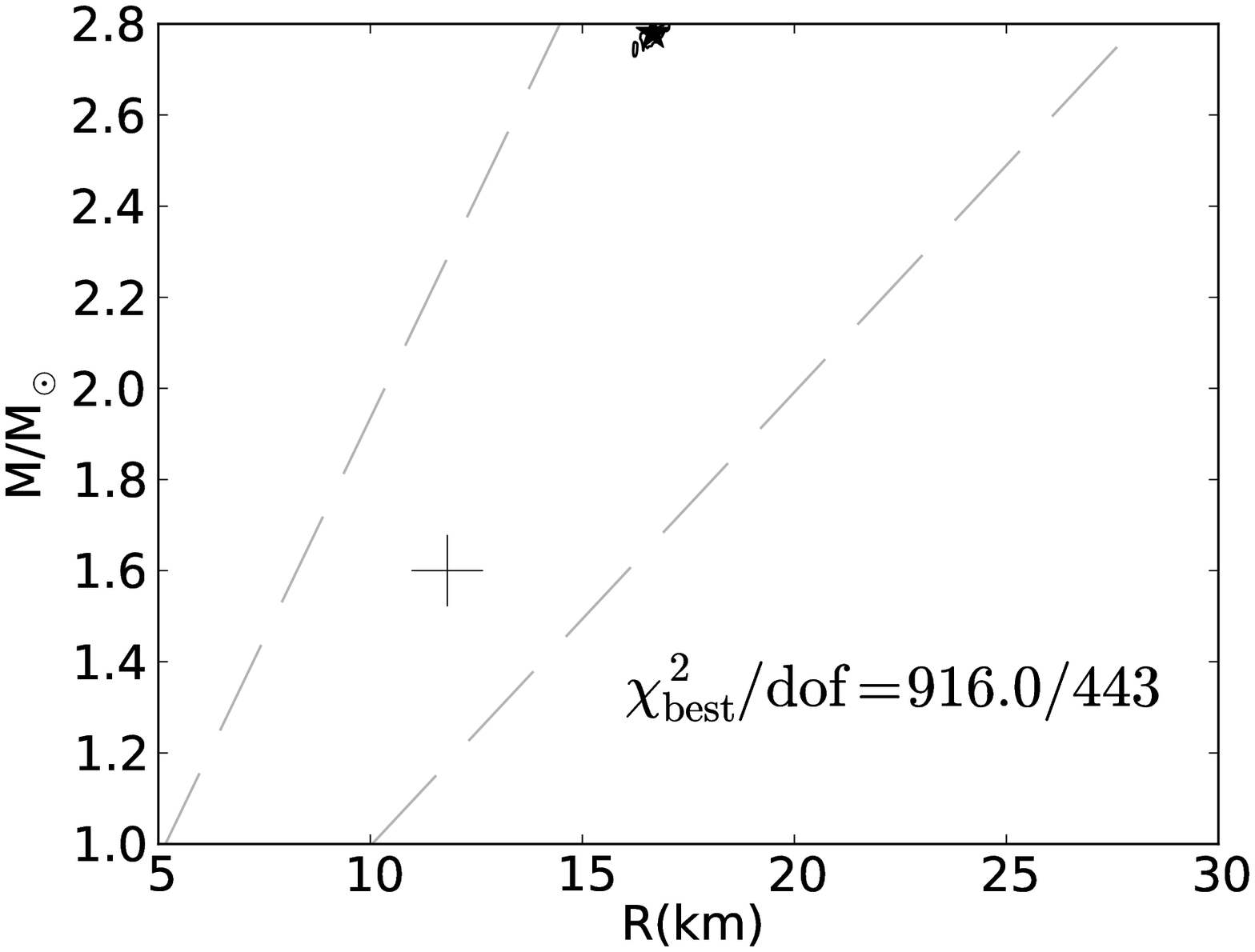}
\includegraphics[height=.26\textheight]{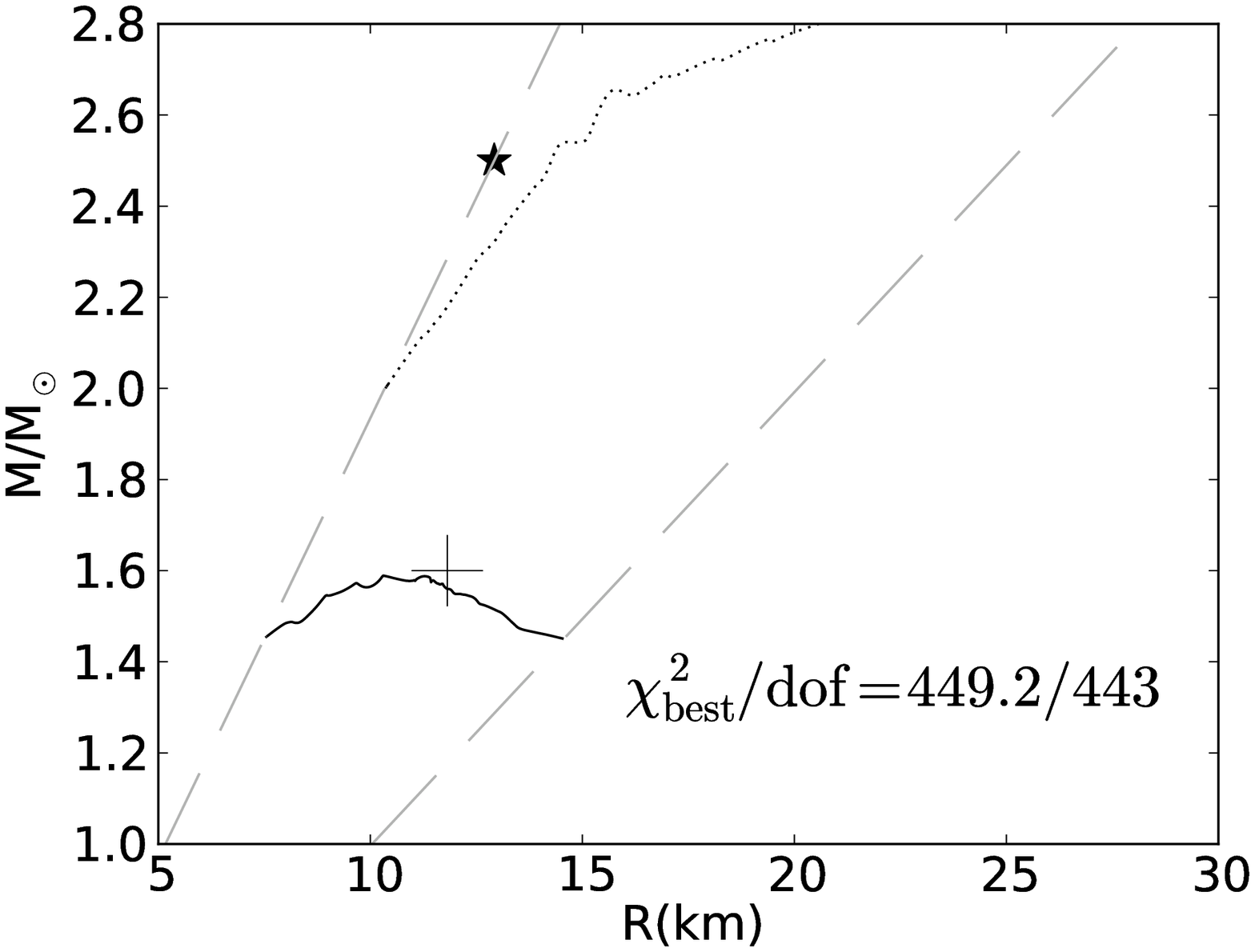}
\end{center}
\vspace{-0.7cm}
\caption{
The top panels show that fitting the observed waveform using a model in which the emission spectrum differs by a modest amount from the actual emission spectrum has little effect on the estimated values of $M$ and $R$ or their uncertainties (compare 
Figs.~\ref{fig:results:diff_bkg}a 
and~\ref{fig:results:diff_bkg}b).
The bottom panels show that fitting the observed waveform using a model in which the beaming pattern differs substantially from the actual beaming pattern increases the uncertainty of $M$ and $R$ estimates and can sometimes bias them a moderate amount. The contours in the left column are for high spot and observer inclinations, whereas the contours in the right column are for low spot and observer inclinations. All results assume our medium background. The top row shows the constraints obtained when the actual emission spectrum has the shape of a Bose-Einstein spectrum with a chemical potential $\mu = - kT$ but the waveform model assumes the emission spectrum has the shape of a Planck spectrum. The bottom row shows the constraints obtained when the actual emission is isotropic but the waveform model assumes the emission has the beaming pattern described by the Hopf function. For the meanings of the line types and symbols, see Fig.~\ref{fig:results:diff_bkg}.
}
\label{fig:results:wrong_spectrum_or_beaming}
\end{figure*}

\subsubsection{Effects of errors in the assumed properties of the hot spot}
\label{sec:results:constraints:model-errors}

In the results discussed previously, we assumed that we knew the shape of the hot spot and the spectrum and beaming pattern of the radiation from the spot. In practice, we may not know these properties. Any discrepancies between the actual properties of the hot spot and the properties assumed in the waveform model will produce systematic errors in the constraints on $M$ and $R$, in addition to the statistical uncertainties, and these errors should be quantified. A full analysis of possible systematic errors is beyond the scope of this work, but our initial exploration shows that inaccuracies in modeling the waveform can sometimes increase the uncertainties in measurements of $M$ and $R$.

\textit{Errors in the assumed spot shape}.  The effects of errors in the assumed shape of the hot spot are illustrated in Figure~\ref{fig:results:high_incl-NS_EW}. For these examples, we generated synthetic waveforms using hot spots elongated in the latitudinal (east-west) and longitudinal (north-south) directions by superimposing two or three circular emitting areas, as described in the figure caption, and then fitting the resulting waveforms using our model, which assumes a circular hot spot.

Figure~\ref{fig:results:high_incl-NS_EW}a shows the results for the moderately high-inclination spot elongated in the north-south direction by $50^{\circ}$ and our medium background. The $1\sigma$ and $3\sigma$ uncertainties in $M$ and $R$ are, respectively, $\sim\,$40\% and $\sim\,$75\%, much larger than the $\sim\,$4\% and $\sim\,$15\% uncertainties for the waveform produced by a circular spot in the rotational equator and the same background 
(see Figure~\ref{fig:results:diff_bkg}c).
The fit appears excellent ($\chi^2/{\rm dof} = 432.5/443$), providing no warning that the model is incorrect.
Although the best-fit values of $M$ and $R$ differ substantially from the values assumed in generating the waveform, the differences are only marginally significant.

Figure~\ref{fig:results:high_incl-NS_EW}b shows the results for the spot in the rotational equator elongated in the east-west direction by $45^{\circ}$ and our medium background. The $1\sigma$ uncertainties in $M$ and $R$ are, respectively, $\sim\,$8\% and $\sim\,$17\%, substantially larger than the $\sim\,$4\% uncertainties for the waveform produced by a circular spot in the rotational equator and the same background 
(see Figure~\ref{fig:results:diff_bkg}c). 
The $3\sigma$ uncertainties in $M$ and $R$ are, respectively, $\sim\,$30\% and $\sim\,$40\%, much larger than the $\sim\,$15\% uncertainties that were achieved for a circular spot
(again see Figure~\ref{fig:results:diff_bkg}c). 
The fit is good ($\chi^2/{\rm dof} = 442/443$), providing no warning that the model is incorrect.
However, in this case the best-fit values of $M$ and $R$ are very close to the values assumed in generating the waveform.

\textit{Errors in the assumed spot spectrum}.  The effects of errors in the assumed spectrum of the emission from the hot spot are illustrated in 
Figures~\ref{fig:results:wrong_spectrum_or_beaming}a 
and~\ref{fig:results:wrong_spectrum_or_beaming}b. For these examples, we generated synthetic waveforms assuming the spectrum of the radiation from the surface of the hot spot has the shape of a Bose-Einstein spectrum and then fitted the waveforms using our model, which assumes that the radiation has the shape of a Planck spectrum.

For the high-inclination reference case 
(Figure~\ref{fig:results:wrong_spectrum_or_beaming}a), the 
best-fit values of $M$ and $R$ are very close to their actual values and the quality of the fit is acceptable (there would be a 10\% probability of a reduced $\chi^2$ this high if the model were correct).
The $1\sigma$ uncertainties in $M$ and $R$ are both about 6\%, whereas the $3\sigma$ uncertainties in $M$ and $R$ are 15\% and 12\%, respectively. These are comparable to the 3\%--4\%  $1\sigma$ and about 15\% $3\sigma$ uncertainties achieved by fitting a model with the correct spectral shape to the synthetic waveform 
(see Section~\ref{sec:results:constraints:inclination-effects}).

For the low-inclination reference case 
(Figure~\ref{fig:results:wrong_spectrum_or_beaming}b),
the confidence regions are large, the $1\sigma$ region extends beyond the upper boundary of the search domain, and the best-fit $M$-$R$ pair is on the boundary, all indicating that the most probable solution has not been determined. The reduced $\chi^2$ value is close to unity and hence does not itself show that the waveform model is wrong, but this is of little consequence because the constraints on $M$ and $R$ are very weak.

\textit{Errors in the assumed spot beaming function}.  The effects of errors in the assumed beaming pattern of the radiation from the hot spot are illustrated in
Figures~\ref{fig:results:wrong_spectrum_or_beaming}c 
and~\ref{fig:results:wrong_spectrum_or_beaming}d.
For these examples, we generated synthetic waveforms assuming that the surface of the hot spot radiates isotropically, but then fit these waveforms using our standard model, which assumes the beaming pattern from the surface of the hot spot is described by the Hopf function (the beaming pattern for radiation from a Thomson scattering atmosphere).

For the high-inclination reference case (see Figure~\ref{fig:results:wrong_spectrum_or_beaming}c), the $1\sigma$ and $3\sigma$ contours are very small and far from the $M$-$R$ pair used in generating the synthetic waveform, but the reduced $\chi^2$ for this fit is extremely large, showing that the model being fit is wrong.

For the low-inclination reference case (see Figure~\ref{fig:results:wrong_spectrum_or_beaming}d), the best-fit $M$-$R$ pair is again far from the pair used in generating the synthetic waveform. Although the reduced $\chi^2$ value is close to unity and hence does not itself show that the waveform model is wrong, this is of little consequence, because the constraints on $M$ and $R$ are very weak, the $1\sigma$ and $3\sigma$ confidence regions extend beyond one or both of the boundaries of the search domain, and the best-fit $M$-$R$ pair is on the high-compactness boundary, all indicating that the most probable solution has not been determined.

\subsubsection{Combining different segments of data}
\label{sec:results:constraints:combining-data}

We have seen that for favorable system properties, obtaining strong constraints on $M$ and $R$ usually requires $\sim\,$$10^6$ counts from the hot spot.
Acquiring this many counts will typically require collection of hundreds to thousands of seconds of data, even using an instrument with an effective area $\sim 10$~m$^2$. Thus, unless a superburst is observed, the required data must be accumulated over many bursts. 
We therefore analyzed how the constraints on $M$ and $R$ reported here would be affected if $\sim\,$$10^6$ hot spot counts were collected by combining data from multiple bursts from the same neutron star. The same considerations apply if the data from a single burst are divided into several time segments that are then analyzed jointly.

We find that the constraints on interesting parameters obtained by jointly fitting many data sets are usually comparable to the constraints that could be obtained by fitting a single set of waveform data with the same average profile and the same total number of counts as the waveform obtained by combining the data sets (see Appendix~\ref{app:jointfits}). Thus, if $\sim\,$$10^6$ counts can be accumulated from the hot spots of many bursts produced by a given star, the prospects are good that constraints on $M$ and $R$ similar to those reported here can be obtained. This is true even though one might intuitively expect that the extra parameters involved in fitting many data sets with changing waveforms would compromise the constraints on the model parameters that are fixed ($M$, $R$, and $\theta_{\rm obs}$). 

The effects on the waveform of the special relativistic Doppler boost and aberration play an important role in constraining $M$ and $R$. These effects can be accurately captured even if the hot spot rotation frequency is changing, provided it can be tracked accurately enough to maintain the correct oscillation phase when folding successive periods of the oscillation. If this can be done, the constraints on $M$ and $R$ that can be obtained by analyzing the resulting folded waveform are often nearly the same as those that could be obtained by analyzing a similar waveform with a fixed oscillation frequency and the same number of counts; moderate changes in the values of nuisance parameters other than the spot rotation frequency often have only a small effect.

If the oscillation frequency varies so rapidly or erratically during a burst that it cannot be tracked or other important parameters (such as the spot inclination and spot radius) vary sufficiently during a single burst or from burst to burst that this has a substantial effect on the waveform, the full data set may have to be divided into a series of shorter segments and each segment analyzed separately. We find that the resulting constraints on $M$ and $R$ will usually be similar to the constraints obtained by analyzing a single data set in which these nuisance parameters do not vary and that has the same number of counts. Analyzing a sequence of data segments increases the computational burden only linearly with the number of segments. If the segments are properly analyzed using a Bayesian approach, the order in which they are analyzed does not matter.

In summary, constraints similar to those reported here can usually be obtained if $10^6$ hot spot counts are acquired by combining data from different segments of a single burst or from multiple bursts from the same neutron star, even if the oscillation frequency and other physical parameters vary during the burst or from burst to burst.

\clearpage
\newpage


\section{SUMMARY AND CONCLUSIONS}
\label{sec:conclusions}

The goal of this paper was to explore the constraints on the masses and radii of neutron stars that could be inferred by analyzing observations of X-ray burst oscillations made using a future, satellite-borne detector with 2--30~keV energy coverage and an effective area 10 to 20 times larger than the \textit{RXTE PCA}. This research was motivated by the new concepts for large-area X-ray timing space missions that are now being proposed. These include \textit{LOFT} \citep{mign12, delm12} and \textit{AXTAR} \citep{chak08, ray11}. Although we focused here on exploring the constraints on $M$ and $R$ that can be obtained by analyzing the waveforms of X-ray burst oscillations, our results are equally relevant to plans for analyzing the oscillations produced by X-ray emission from the heated polar caps of isolated rotation-powered millisecond pulsars, the goal of the proposed NICER mission \citep{gend12}.

\textit{Approach to the problem}.  We described and explained our approach in Section~\ref{sec:approach}. There we considered the relative merits of using observations of oscillations during burst rises and burst tails (see Section~\ref{sec:approach:rise-or-tail}). We find that the uncertainties in $M$ and $R$ estimates obtained by analyzing oscillations during the tails of bursts are likely to be smaller than the uncertainties in estimates derived by analyzing the oscillations observed during the rises of bursts. Hence analyzing tail oscillations is likely to be the better approach.

\textit{Computational methods}.  We described our computational methods in Section~\ref{sec:methods}. 
We explored the constraints on $M$ and $R$ that could be derived from X-ray burst oscillations by first generating energy-dependent synthetic observed waveforms for a variety of neutron star and hot spot properties. We then used a Bayesian approach and MCMC sampling methods to determine the constraints on $M$ and $R$ that can be obtained by analyzing these synthetic observed waveforms. We computed the joint posterior probability distribution of the parameters in our waveform model, marginalized the parameters other than $M$ and $R$, and then determined the most probable values of $M$ and $R$ and their confidence intervals. 

The purpose of the analysis presented here was not to reproduce all the steps that would be needed for a full Bayesian analysis of a real observed waveform, but rather to determine the precision and accuracy with which such a full analysis could determine $M$ and $R$. We therefore made use of several shortcuts that reduced the computational burden substantially without significantly altering the results. Using these shortcuts, we could compute the most probable values of $M$ and $R$ and their confidence intervals for a single synthetic observed waveform in 50--100 clock hours, running the parallel version of our code on 150 nodes of a fast CPU cluster.

\textit{Assumed counts from the hot spot and background}.  We assumed that about $10^6$ hot spot counts are available from the star. A 10~m$^2$ detector could collect this many hot spot counts by observing 20--25 bright X-ray bursts (see Section~\ref{sec:approach:rise-or-tail}). We find that the constraints on $M$ and $R$ obtained by combining data from multiple segments of a single burst or from multiple bursts from the same star are usually similar to the constraints that would be obtained by analyzing an observation of a single, time-independent waveform that has the same number of counts and the same shape as the average of the waveform over the multiple data sets (see below, and 
Section~\ref{sec:results:constraints:combining-data} and Appendix~\ref{app:jointfits}).

We assumed that all sources of background, when combined, contribute about $0.3\times10^6$, 10$^6$, or $9\times10^6$ counts. Our treatment of the background was very conservative, in the sense that we usually made no assumptions about the magnitude or spectrum of the background. We did not even assume that the background is constant, only that it does not vary at frequencies commensurate with the burst oscillation frequency.

\textit{Our main results}.  We described our results in detail in Section~\ref{sec:results}. There we showed that if $\sim\,$$10^6$ counts were collected from the hot spot of a single bursting neutron star and the values of all the parameters in the waveform model other than $M$ and $R$ were known independently of the waveform analysis,
the uncertainties in $M$ and $R$ estimates produced by statistical fluctuations in the waveform would typically be only a percent or two \citep[compare the analysis of a similar situation in][]{stro04}. For realistic situations in which some of the other parameters in the model must be determined from the waveform, the statistical uncertainties are usually much larger. The reason for this is that the effects on the waveform of changing different parameters in the waveform model are often very similar. These degeneracies of the waveform with respect to changes in the values of the model parameters are an inherent property of any physical model based on a rotating hot spot and cannot be removed by ``improving'' the model. These degeneracies inflate the uncertainties in $M$ and $R$ estimates produced by the statistical fluctuations in the waveform.

We first explored how the uncertainties in the measured values of $M$ and $R$ depend on the inclinations of the hot spot and observer and the background count rate. We find that the uncertainties depend strongly on the inclination of the hot spot and observer relative to the stellar spin axis. Specifically:
\begin{itemize}\itemsep0pt
\item If the hot spot is within 10$^\circ$ of the rotation equator, both $M$ and $R$ can usually be determined with an uncertainty of about 10\%.
\item If instead the spot is within 20$^\circ$ of the rotation pole, the uncertainties in $M$ and $R$ are so large that waveform measurements alone provide no useful constraints.
\item The uncertainties in $M$ and $R$ are affected little by background count rates less than or comparable to the count rate from the hot spot, but become significantly larger for higher background count rates.
\end{itemize}

Next, we explored the effects on the uncertainties in $M$ and $R$ if the distance to the star or the inclination of the observer are known from other measurements, if a resonance scattering line is observed in the burst oscillation spectrum, or if the properties of the background are independently known. We find that:
\begin{itemize}\itemsep0pt
\item Independent information about the background can greatly reduce the uncertainties in estimates of $M$ and $R$.
\item Independent knowledge of the observer's inclination can also greatly reduce the uncertainties.
\item Knowledge of the star's distance can help, but not as much as knowledge of the background or the observer's inclination.
\item Observation of an identifiable atomic line in the hot-spot emission always tightly constrains $M/R$; it can also tightly constrain $M$ and $R$ individually, if the hot spot is within about 30$^\circ$ of the rotation equator.
\end{itemize}

Finally, we explored the effects of deviations in the actual shape of the hot spot, radiation beaming pattern, and spectrum from those assumed in the our waveform model. We find that:
\begin{itemize}\itemsep0pt
\item Modest deviations of the actual spectrum from that assumed in the waveform model have little effect on the accuracy or uncertainty of $M$ and $R$ estimates.
\item  Large deviations of the actual radiation beaming pattern from the pattern assumed in the waveform model can increase the uncertainties of $M$ and $R$ estimates substantially.
\item  In some cases, but not always, large deviations of the actual shape of the hot spot from the circular shape assumed in the waveform model can increase the uncertainties of $M$ and $R$ estimates and bias them by moderate amounts. The physical conditions that produce tight constraints on $M$ and $R$ (relatively small spots far from the rotation axis) are the conditions in which the shape of the spot is unimportant \citep[see][]{lamb09a,lamb09b}.
\end{itemize}

\textit{Combining data from different bursts}.  
Our results show that for favorable system properties, strong constraints on $M$ and $R$ can be obtained if one can analyze $\sim\,$$10^6$ counts from the hot spot. This will typically require analyzing hundreds or thousands of seconds of data, even if the detector has an effective area $\sim 10$~m$^2$. Thus, unless a superburst is observed, the data required must be accumulated from many bursts. 
We therefore analyzed how the constraints on $M$ and $R$ reported here would be affected if the $\sim\,$$10^6$ hot spot counts were obtained by combining data from multiple bursts from the same star. The same considerations apply if the data from a single burst are divided into several time segments and then analyzed jointly. We find that the constraints on $M$ and $R$ obtained by jointly fitting many data sets are usually comparable to the constraints that could be obtained by fitting a single segment of waveform data having the same profile and as the average profile of the combined data sets and the same total number of counts (see 
Section~\ref{sec:results:constraints:combining-data} and~Appendix~\ref{app:jointfits}).

\textit{Time varying waveforms}.  
We also investigated the problem of analyzing burst oscillation waveforms that change with time. If the oscillation frequency changes during a burst but the change can be modeled accurately enough to maintain the correct oscillation phase when folding successive periods of the oscillation, the constraints on $M$ and $R$ obtained by analyzing the resulting folded waveform will be nearly the same as those that could be obtained by analyzing a similar waveform with a fixed oscillation frequency and the same number of counts.
Even if the oscillation frequency varies too rapidly or irregularly during the burst rise or tail to be described accurately by a simple frequency model, the full burst oscillation data set can be divided into smaller time segments and analyzed using standard Bayesian techniques. This approach can also be used if other physical properties of the system, such as the size and inclination of the emitting region, vary significantly. The computational burden of this kind of analysis increases only linearly with the number of segments. Consequently, variations of the burst oscillation waveform on timescales shorter than the burst rise or burst tail (but substantially longer than the burst oscillation period) do not appear to pose an insurmountable analysis problem (again see Section~\ref{sec:results:constraints:combining-data} and~Appendix~\ref{app:jointfits}).

\vskip-5pt
\acknowledgments
We thank Anthony Chan, Stephen Drake, Jos{\'e} Garmilla, Aaron Hanks, John Hoffman, Miaotianzi Jin, David Kolschowsky, and Dave Kotan for assistance in developing and testing the ray-tracing code and Novarah Kazmi for help in assembling the references. We also thank Stratos Boutloukos and Juri Poutanen for useful discussions. This research was supported in part by a grant from the Simons Foundation (grant number 230349) to MCM and NSF grant AST0708424 at Maryland, and by NSF grant AST0709015 and funds of the Fortner Endowed Chair at Illinois. MCM gratefully acknowledges the hospitality of the Aspen Center for Physics, where part of this work was completed. MCM also thanks the Department of Physics and Astronomy at Johns Hopkins University for their hospitality during his sabbatical.

\clearpage
\newpage

\appendix


\section{CODE TESTS}
\label{app:code-validation}

\setcounter{table}{0}
\renewcommand\thetable{\Alph{section}\arabic{table}}

\setcounter{figure}{0}
\renewcommand\thefigure{\Alph{section}\arabic{figure}}

In this appendix, we summarize the suite of test problems we routinely solve to validate both our waveform code and our MCMC code (of which the waveform code is an essential part). The waveform code we use here is based on the code developed and validated by \citet{lamb09a}. We described some tests of this code there. The code validation tests we describe here are designed to check many of the physical effects included in our codes, by using them to solve problems where the answer is known analytically. This suite of test problems includes the following problems:

\begin{itemize}\itemsep0pt

\item computing the deflection angle as a function of the star's compactness and the direction of the ray relative to the normal to the stellar surface (Section~\ref{sec:tests:WF:deflection})

\item computing the light travel time as a function of the star's compactness and the direction of the ray relative to the normal to the stellar surface (\ref{sec:tests:WF:dtime}) 

\item computing the bolometric flux seen by a distant observer located directly above the center of an emitting spot on a non-rotating star (Section~\ref{sec:tests:WF:absolute_flux})

\item computing the radiation pattern produced by a small emitting spot on a non-rotating star in flat spacetime (Section~\ref{sec:tests:WF:nonrot-flat}) 

\item computing the waveform produced by a small emitting spot on a rotating star in flat spacetime (Section~\ref{sec:tests:WF:rot-flat})

\item computing the angular deflection of a pencil beam from a small emitting spot on a rotating star (Section~\ref{sec:tests:WF:pencil}) 

\item computing the observed temperature of thermal emission from a small emitting spot on a rotating star in flat spacetime, as a function of rotational phase (Section~\ref{sec:tests:WF:doppler-temperature})  
 
\item determining the stellar compactness, spot temperature, and distance, given uniform thermal emission from the entire surface of a non-rotating star with no background  
(Section~\ref{sec:tests:fitting:nonrot-nobkg-R_T}) 

\item determining the spot inclination, observer inclination, and distance, given thermal emission from a small hot spot on a non-rotating star in flat spacetime with no background 
(Section~\ref{sec:tests:fitting:nonrot-nobkg-ispot_iobs}) 

\item determining the stellar compactness, spot temperature, and distance, given thermal emission from a small spot on a non-rotating star in flat spacetime with a background 
(Section~\ref{sec:tests:fitting:nonrot-withbkg-R_T}).

\end{itemize}

The values of the waveform code integration, hot spot, and angular resolution parameters that we use for the code tests reported here are the values listed in 
Table~\ref{table:tests:WF:resolution-parameters}, unless otherwise noted. We use these same parameter values when computing $M$ and $R$ confidence regions (see 
Section~\ref{sec:results:constraints}), except that for that purpose we set {\tt Nphase} to 16, which speeds up the computations while providing accuracy sufficient for the waveform-fitting process. As discussed in 
Section~\ref{sec:tests:convtests}, we have carefully chosen the values of these code parameters to provide a resolution fine enough to meet our accuracy requirements, but no finer, so that our code runs as fast as possible.

We now discuss the suite of validation tests in detail.

\begin{deluxetable}{ l r }
\tablewidth{0pt}
\tablecaption{
Values of the waveform code resolution parameters
\label{table:tests:WF:resolution-parameters}
}
\tablehead{
\colhead{Parameter} & \colhead{Value}
}
\startdata
{\tt Nlat} (spot grid points in stellar latitude) & 100 \\
{\tt Nlong} (spot grid points in stellar longitude) & 100 \\
{\tt Nalpha} (grid points in $\alpha$) & 1000 \\
{\tt Nphase} (grid points in phase w. time delays included) & 1000 \\
{\tt Nloc} (grid points in phase w.o. time delays included) & 1001 \\
$n_d$ (grid points used in computing the deflection angle) & 100 \\
$\epsilon_d$ (deflection angle integration parameter) & 0.1 \\
$n_t$ (grid points used in computing the time delay) & 100 \\
$\epsilon_0$ (first time delay integration parameter) & 0.01 \\
$\epsilon_1$ (second time delay integration parameter) & 0.01 \\
\enddata
\end{deluxetable}

\subsection{Tests of the Waveform Code}
\label{sec:tests:WF}
\subsubsection{Deflection angle as a function of compactness and direction}
\label{sec:tests:WF:deflection}

In the Schwarzschild spacetime, the total angle $\psi$ by which a light ray from the stellar surface is eventually deflected is a function only of the initial angle $\alpha$ of the ray relative to the normal to the stellar surface (as measured in the static frame) and the compactness $M/R$ of the star, and is given by the expression \citep[see][equation~(2.13) for a similar expression]{pech83}
\begin{equation}
\psi\left(\alpha,\frac{M}{R}\right) = \int^1_0 \frac{\sin\alpha~\dd x}{\sqrt{\left(1-2M/R\right)-\left(1-2Mx/R\right)x^2\sin^2\alpha}} \;,
\label{eqn:tests:WF:psi}
\end{equation}
The integrand in equation~(\ref{eqn:tests:WF:psi}) diverges as $x \rightarrow 1$ and $\sin\alpha \rightarrow 1$; numerical integration near this limit therefore requires special treatment. In the waveform code we break the integral into two pieces, from $x=0$ to $x=1-\epsilon_d$ and from $x=1-\epsilon_d$ to $x=1$, with $\epsilon_d \ll 1$. Our waveform code computes the first piece numerically, using Simpson's rule with $n_d$ divisions, and computes the second piece using an approximate analytical expression, derived as follows. 

Setting $x=1-\epsilon_d$, the square root in the integrand becomes
\begin{align}
\sqrt{~\ldots~} &= \sqrt{\left(1-\frac{2M}{R}\right) - \left(1-\frac{2M(1-\epsilon_d)}{R}\right)(1-\epsilon_d)^2\sin^2\alpha} \\ 
&= \sqrt{\left(1-\frac{2M}{R}\right)\cos^2\alpha + 2\sin^2\alpha\left(1-\frac{3M}{R}\right)\epsilon_d + \sin^2\alpha\left(\frac{6M}{R}-1\right)\epsilon_d^2 + \mathcal{O}(\epsilon_d^3)} \quad.
\label{approxmatedenominator}
\end{align}
Dropping the $\mathcal{O}(\epsilon_d^3)$ terms inside the square root, the second piece of the integral becomes
\begin{align}
\int^1_{1-\epsilon_d} \frac{\dd x}{\sqrt{\left(1-2M/R\right)-\left(1-2Mx/R\right)x^2\sin^2\alpha}} &\simeq \int^{\epsilon_d}_0 \frac{\dd\epsilon_d}{\sqrt{a+b\epsilon_d+c\epsilon_d^2}} \\ \nonumber
&= \left\{\begin{array}{cc} \sqrt{c}~\ln\left[2\sqrt{c(a+b\epsilon_d+c\epsilon_d^2)} + 2cx + b\right]\;, & c>0 \\
	-(-c)^{\frac{1}{2}}\sin^{-1}\left[\frac{2c\epsilon_d+b}{\sqrt{b^2-4ac}}\right]\;, & c<0 \\
	\frac{2}{b}\sqrt{a+b\epsilon_d}\;, & c=0 \end{array}\right.
\end{align}
where $a$, $b$, and $c$ are the coefficients of 1, $\epsilon_d$, and $\epsilon_d^2$ in the square root~(\ref{approxmatedenominator}). Because $c=\sin^2\alpha\left({6M}/{R}-1\right)$, different integrals are performed for $R<6M$ and $R>6M$. Note also that for $R=3M$ and $\sin\alpha \rightarrow 1$, $a=b=0$, causing the integral to diverge, as it must, because $r=3M$ is the photon orbit in the Schwarzschild geometry.

\begin{figure*}[!t]
\begin{center}
\includegraphics[height=.26\textheight]{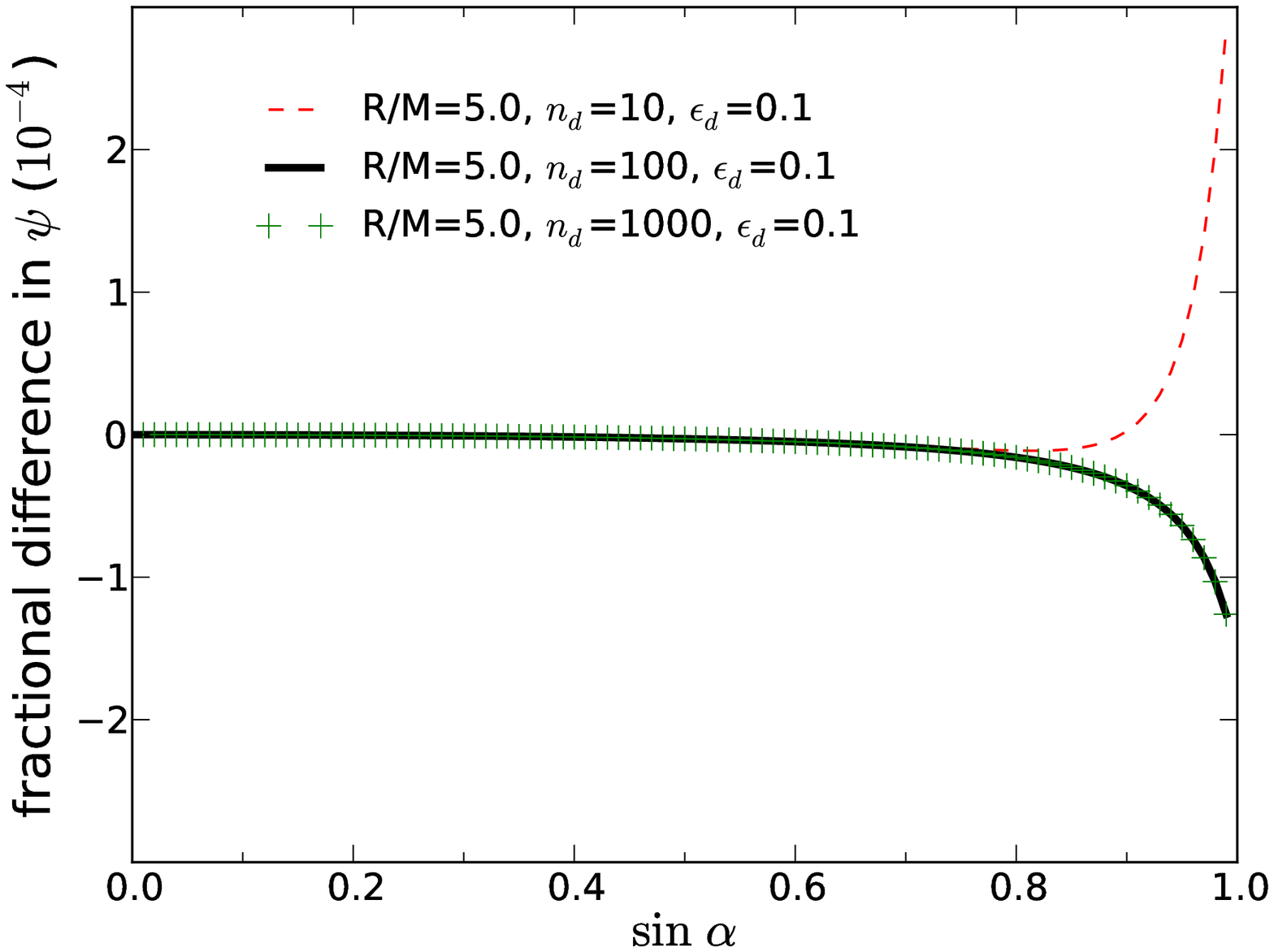}
\includegraphics[height=.26\textheight]{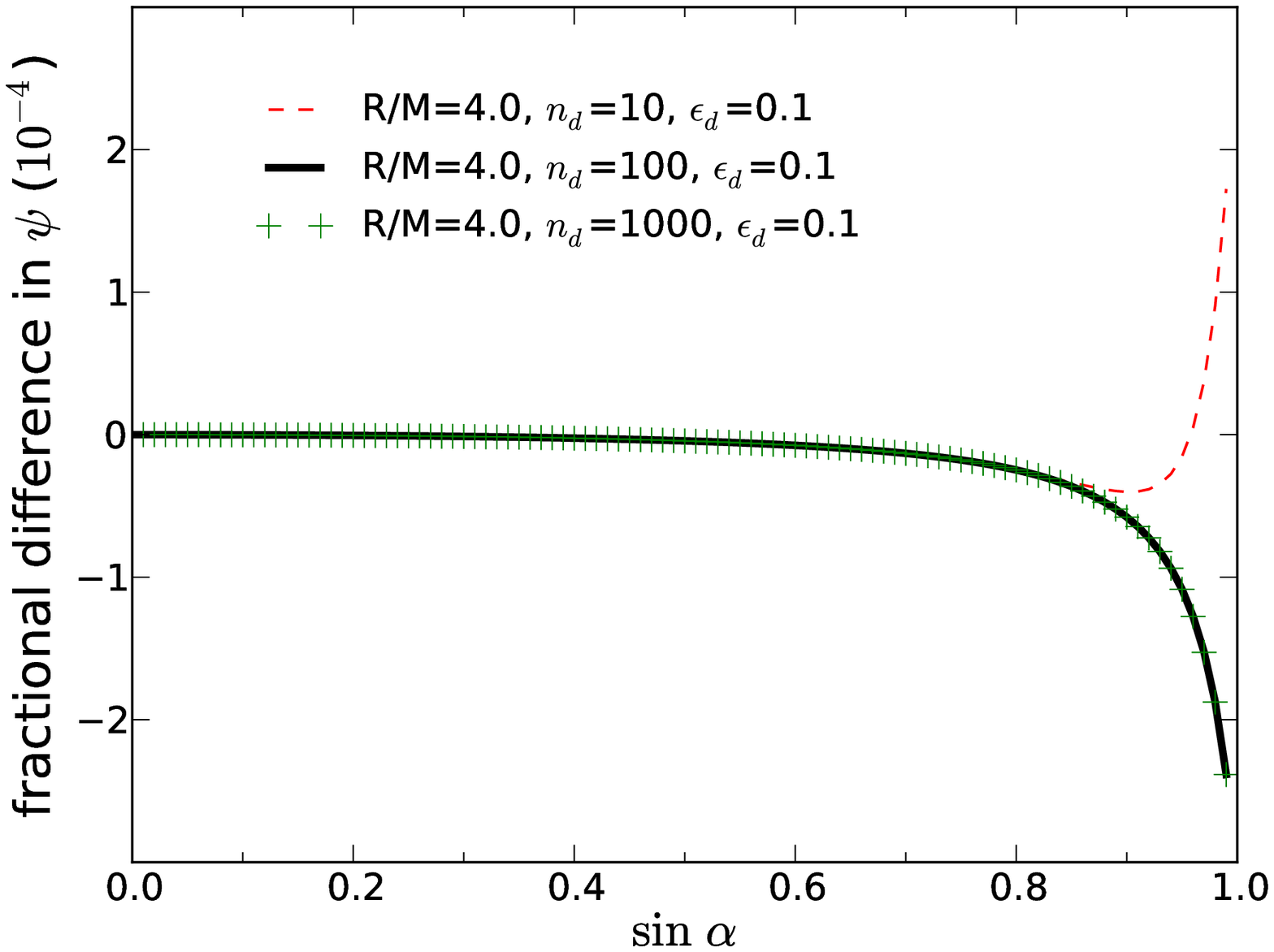}
\end{center}
\vspace{-0.5cm}
\caption{
Fractional difference between the deflection angle $\psi$ as a function of the initial ray direction $\alpha$ computed using our code and using Mathematica~8, for three different numbers $n_d$ of integration steps and two values of the stellar compactness (left: $R/M=5.0$; right: $R/M=4.0$). This figure shows that the value $n_d=100$ used in our waveform fitting is adequate.
See Section~\ref{sec:tests:WF:deflection} for further details.
\label{fig:tests:WF:deflection:compare_n}
}
\end{figure*}

\begin{figure*}[!t]
\begin{center}
\includegraphics[height=.26\textheight]{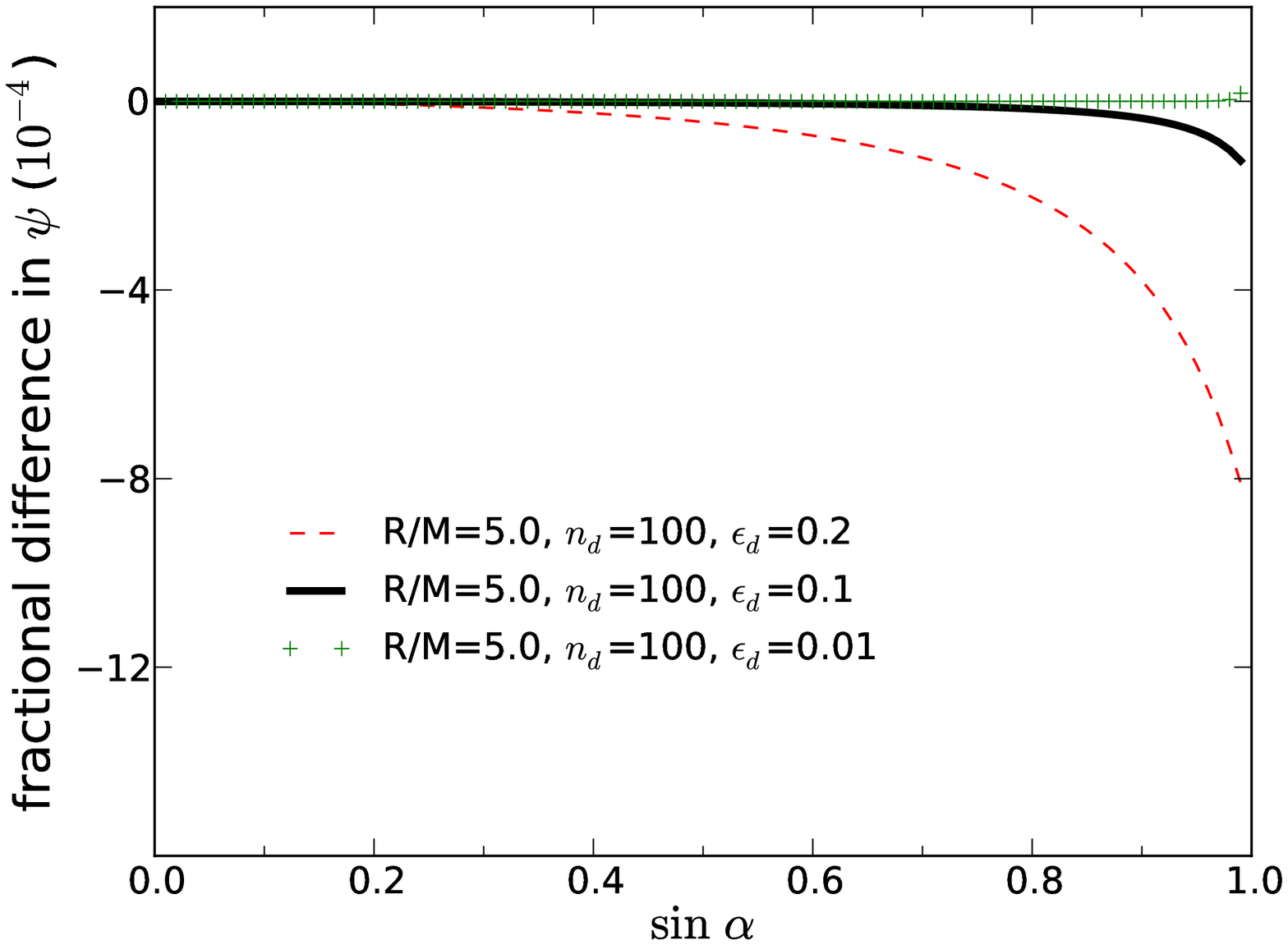}
\includegraphics[height=.26\textheight]{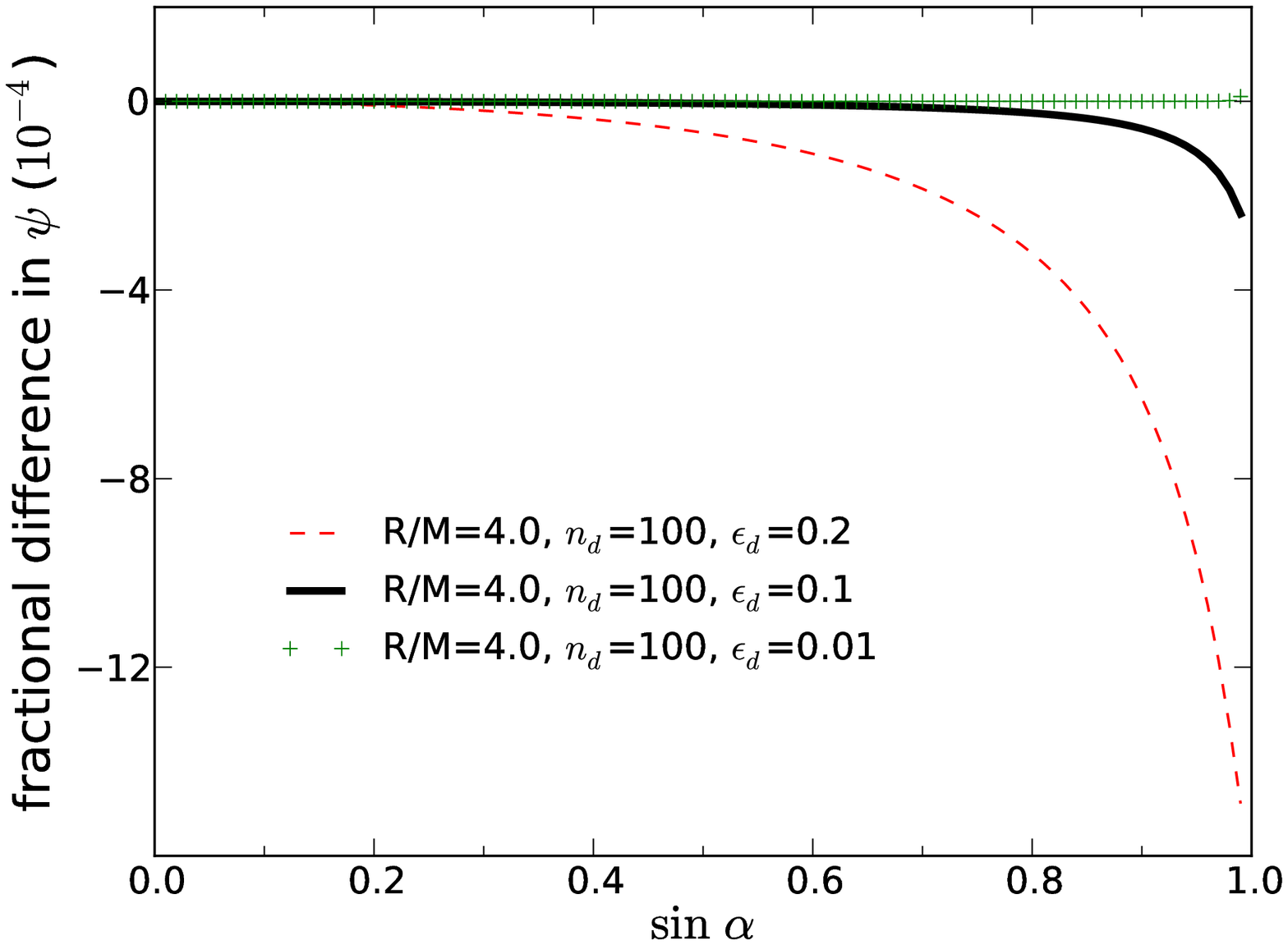}
\end{center}
\vspace{-0.5cm}
\caption{
Fractional difference between the deflection angle $\psi$ as a function of the initial ray direction $\alpha$ computed using our code and using Mathematica~8, for three values of the integration parameter $\epsilon_d$ and two values of the stellar compactness (left: $R/M=5.0$; right: $R/M=4.0$). This figure shows that the value $\epsilon_d=0.1$ used in our waveform fitting is adequate.
See Section~\ref{sec:tests:WF:deflection} for further details.
\label{fig:tests:WF:deflection:compare_eps}
}
\end{figure*}

\begin{figure*}[!t]
\begin{center}
\includegraphics*[height=.26\textheight]{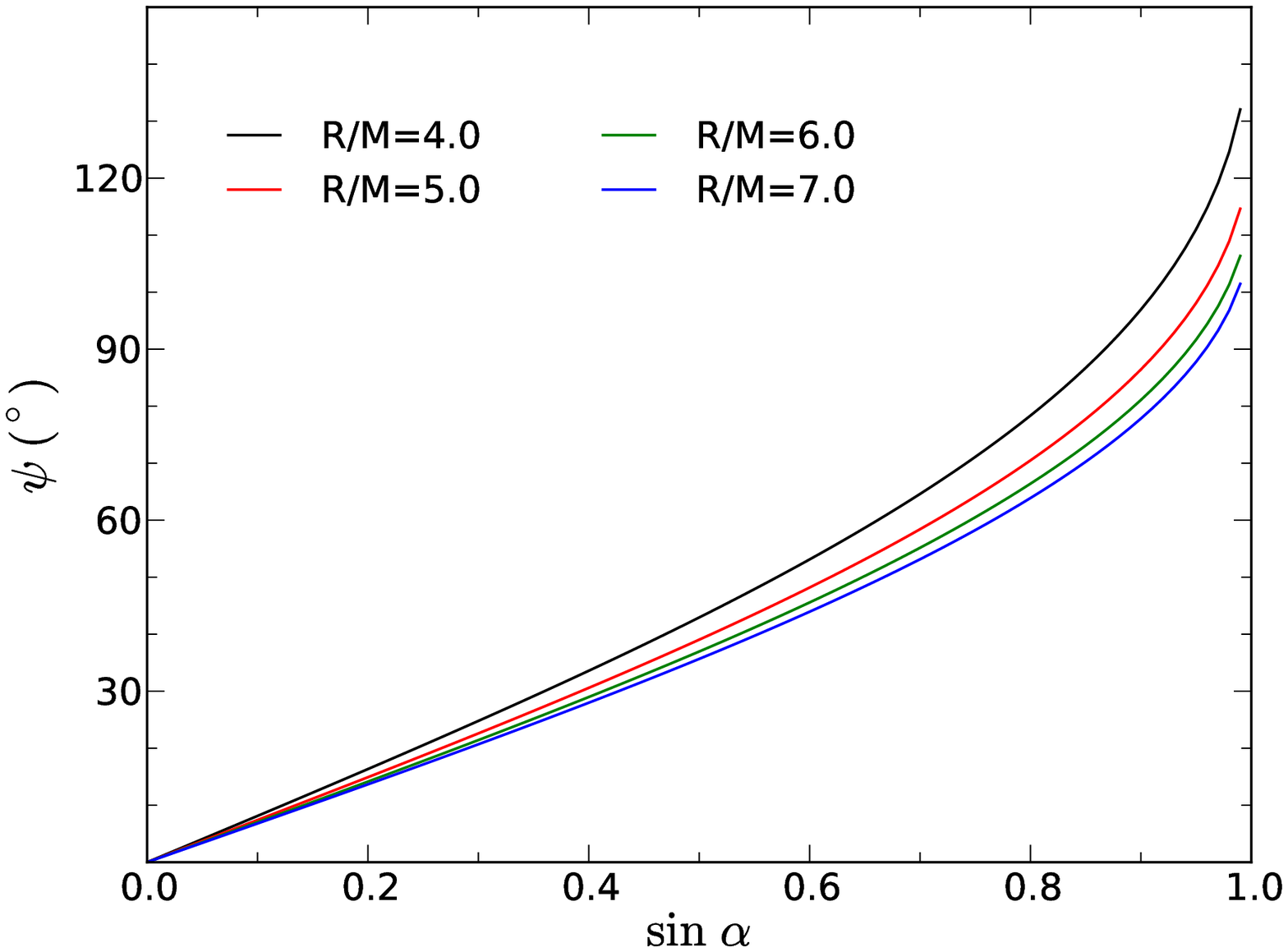}
\includegraphics*[height=.26\textheight]{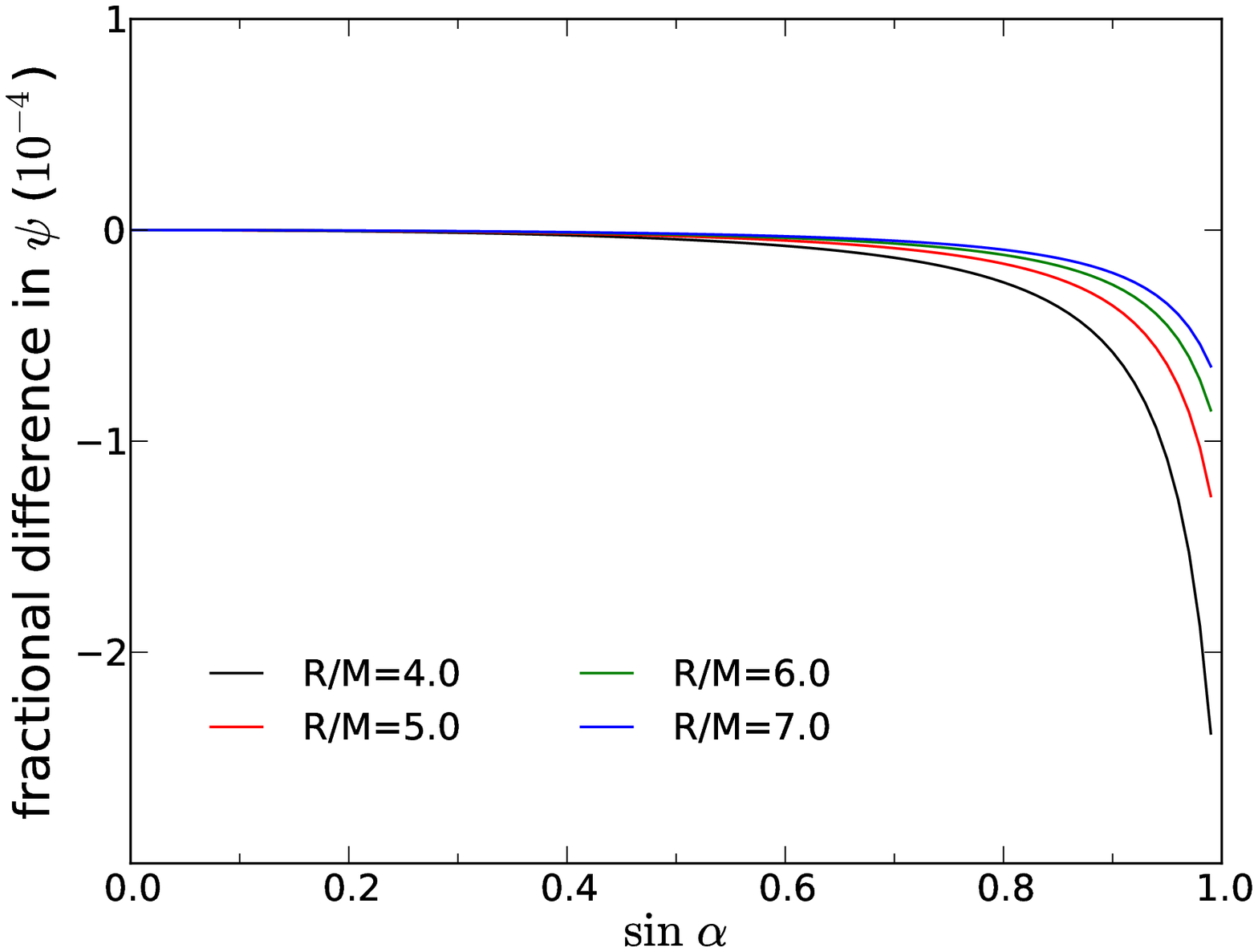}
\end{center}
\vspace{-0.5cm}
\caption{
\textit{Left}: Deflection angle $\psi$ as a function of the initial ray direction $\alpha$ computed using our code, for four values of the stellar compactness. The deflection angle is larger for smaller values of $R/M$, i.e., for more compact stars.
\textit{Right}: Fractional difference between the deflection angle $\psi$ computed using our code and using Mathematica~8, as a function of $\alpha$, for the same four values of the stellar compactness. The fractional difference is larger for smaller values of $R/M$, i.e., for more compact stars, but is smaller in magnitude than $2.5\times10^{-4}$ for all these values of $R/M$.
This figure shows that the resolution parameter values $n_d=100$ and $\epsilon_d=0.1$ used here provide sufficient accuracy for our waveform fitting.
See Section~\ref{sec:tests:WF:deflection} for further details.
\label{fig:tests:WF:deflection:mathematica}
}
\end{figure*}

We test the deflection angle $\psi$ given by our waveform code by comparing it with the value of $\psi$ given by evaluating integral~(\ref{eqn:tests:WF:psi}) using Mathematica~8. 
Figure~\ref{fig:tests:WF:deflection:compare_n} shows the fractional difference $[(\psi_{\rm code} - \psi_{\rm mathematica}) / \psi_{\rm mathematica}]$ for three values of $n_d$ and two values of the stellar compactness. 
Figure~\ref{fig:tests:WF:deflection:compare_eps} shows the fractional difference for three values of $\epsilon_d$ and two values of the stellar compactness.
The left and right panels of 
Figure~\ref{fig:tests:WF:deflection:mathematica} show, respectively, the relation $\psi(\alpha)$ and the fractional difference in $\psi$ given by our code using $n_d=100$ and $\epsilon_d=0.1$, for four values of the stellar compactness that span the range considered in this work.
Figures~\ref{fig:tests:WF:deflection:compare_n}--\ref{fig:tests:WF:deflection:mathematica} show that the values $n_d=100$ and $\epsilon_d=0.1$ of these resolution parameters that we use in our waveform code when fitting synthetic waveform data provide sufficient accuracy for this purpose.

\subsubsection{Light travel time as a function of compactness and direction}
\label{sec:tests:WF:dtime}

The time it takes for a light ray emitted at an angle $\alpha$ relative to the surface normal (in the static frame) to travel from the surface to infinity, as measured at infinity, is given by
\begin{equation}
t = \int^1_0 \dd y~\frac{R/c}{y^2\left(1-2My/R\right)}\left[\frac{1}{\sqrt{1-\sin^2\alpha\left(1-2M/y\right)^{-1}\left(1-2My/R\right)y^2}}\right] \;.
\label{sec:tests:WF:travel-time}
\end{equation}
This expression accounts for the gravitational time delay in the Schwarzschild spacetime as well as the purely geometrical effect that the travel time is different for rays emitted from the stellar surface at different angles. The travel time given by 
equation~(\ref{sec:tests:WF:travel-time}) is formally infinite, because the ray is traced to infinity. However, what we are interested in is not the total travel time of each ray, but rather the \textit{differential} travel times of different rays. 

To be able to compute the differential travel times of any two rays, we calculate the differential travel time of every ray relative to the travel time of a reference ray. We choose as our reference ray the radial ray coming from the point on the stellar surface directly under the observer. The differential light travel time of an arbitrary ray compared to the reference ray is then
\begin{equation}
\Delta t \equiv t-[t]_{\alpha=0} = \int^1_0 \dd y~\frac{R/c}{y^2\left(1-2My/R\right)}\left[\frac{1}{\sqrt{1-\sin^2\alpha\left(1-2M/y\right)^{-1}\left(1-2My/R\right)y^2}}-1\right] \;.
\label{eqn:tests:WF:differential-travel-time}
\end{equation}
The integrand in this expression can diverge at the two limits of the integral. Our waveform code therefore calculates this integral in three pieces. Near the upper endpoint, we set $y=1-\epsilon_1$, with $\epsilon_1 \ll 1$, expand the expression inside the square root in powers of $\epsilon_1$, and drop the terms that are $\mathcal{O}(\epsilon_1^2)$. The result is
\begin{equation}
\int^1_{z_0} \dd z~\frac{\dd\Delta t}{\dd z} \simeq \int^1_{z_0} \dd z~\frac{R/c}{2\left(1-3M/R\right)}\frac{1}{z}\left[\frac{1}{\sqrt{1-z\sin^2\alpha}}-1\right]
\end{equation}
\begin{equation}
 = \left.\frac{R/c}{2(1-3M/R)}\left[-2\ln\left(1+\sqrt{1-z\sin^2\alpha}\right)\right]\right|^1_{z_1} \;,
\end{equation}
where $z \equiv 1-2\epsilon_1+\epsilon_1(2M/R)/(1-2M/R)$ and $z_1 \equiv 1-2\epsilon_1(1-3M/R)/(1-2M/R)$. Near the lower endpoint of the integral, we set $y=\epsilon_0$, with $\epsilon_0 \ll 1$, expand the integrand in powers of $\epsilon_0$, and thereby obtain the expression
\begin{equation}
\int^{\epsilon_0}_0 \dd z~\frac{\dd\Delta t}{\dd z} \simeq \frac{R/c}{2}\left[\frac{\sin^2\alpha}{1-2M/R}\epsilon_0 + \frac{3}{4}\frac{\sin^4\alpha}{(1-2M/R)^2}\left(\frac{\epsilon_0^3}{3}-\frac{M}{2R}\epsilon_0^4\right)\right] \;.
\end{equation} 
Our waveform code computes the middle piece of the integral, from $y=\epsilon_0$ to $y=1-\epsilon_1$, numerically, using Simpson's rule with $n_t$ divisions in this interval.

\begin{figure*}[!t]
\begin{center}
\includegraphics[height=.26\textheight]{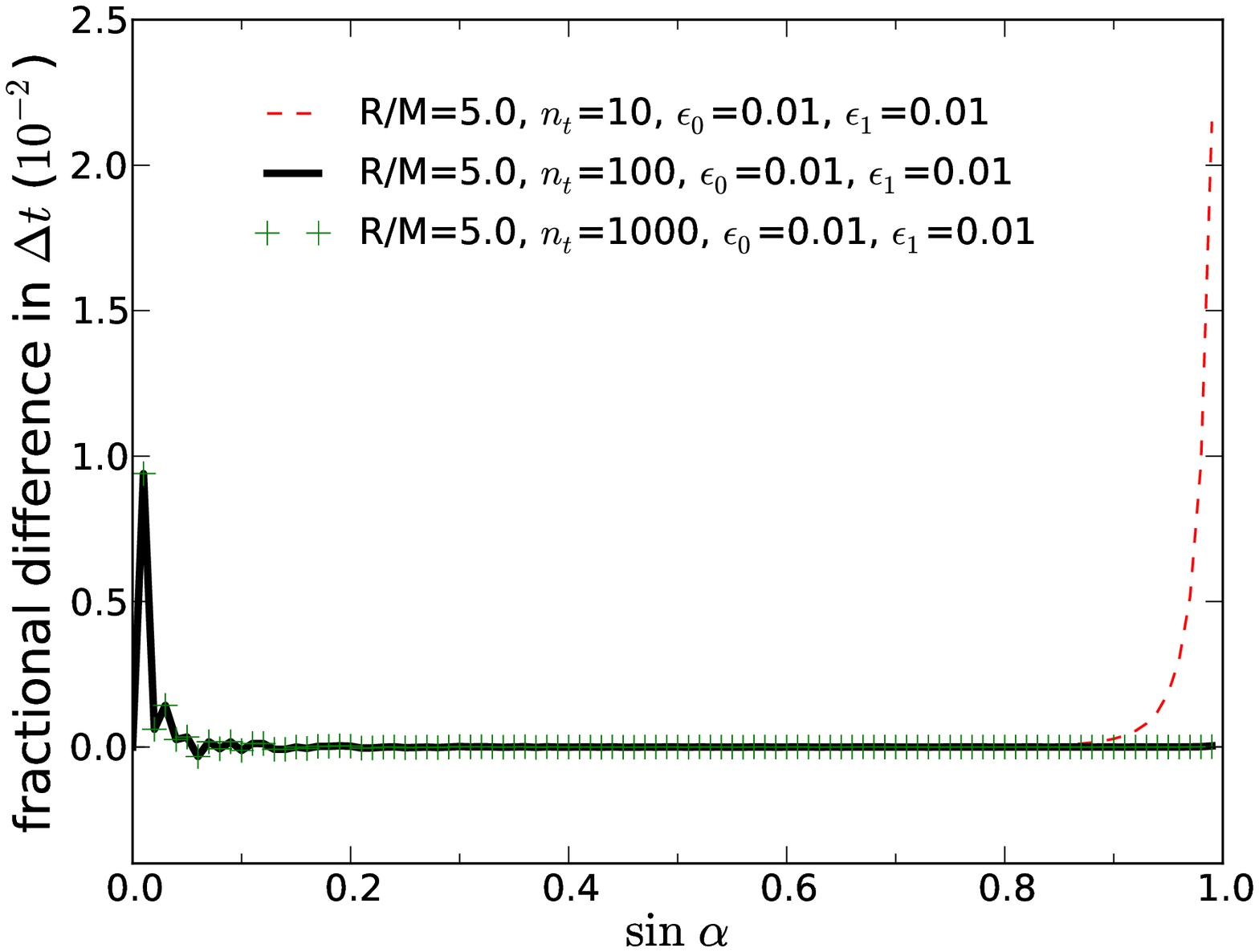}
\includegraphics[height=.26\textheight]{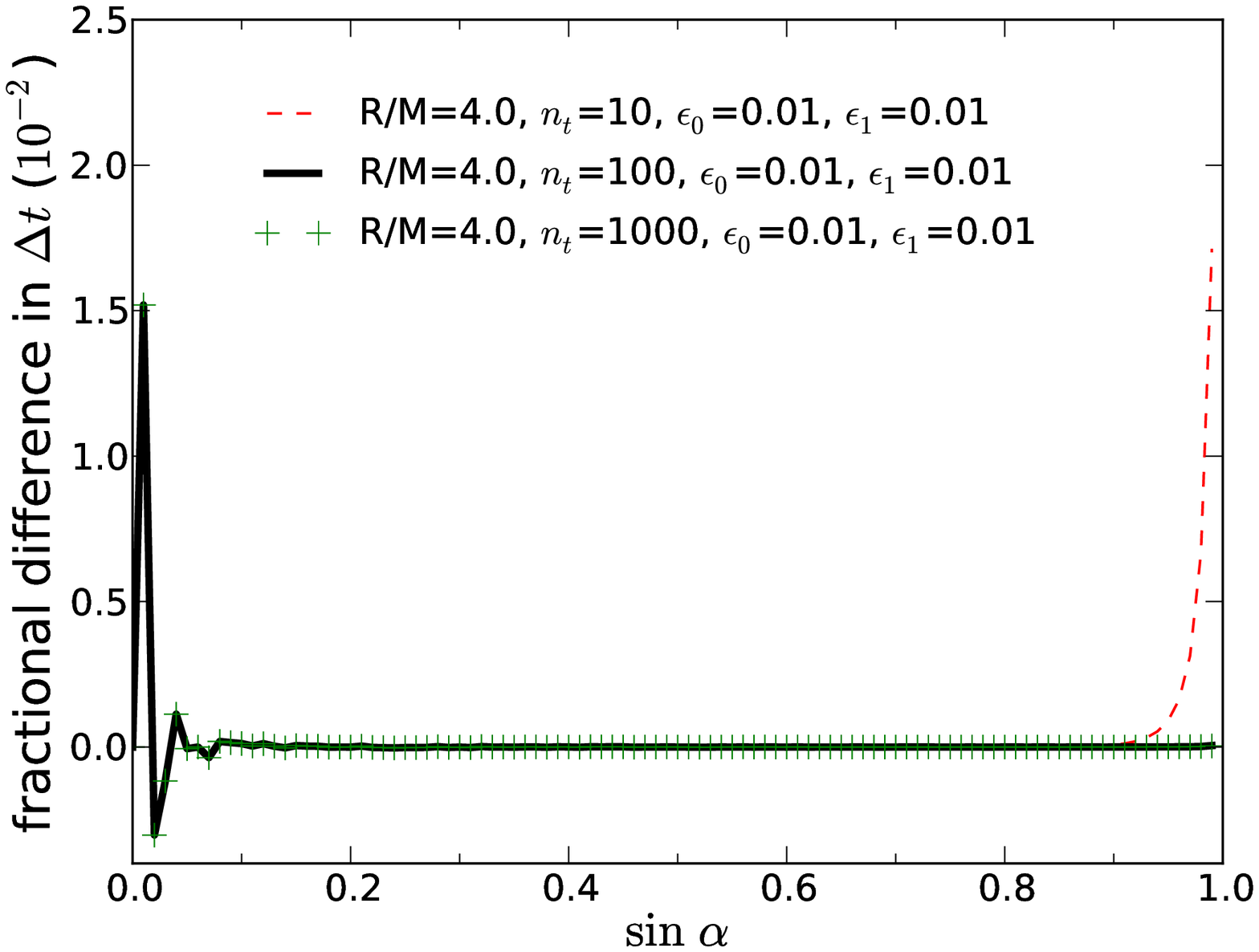}
\end{center}
\vspace{-0.5cm}
\caption{
Fractional difference between the differential light travel time ${\Delta t}$ as a function of the initial ray direction $\alpha$ computed using our code and using Mathematica~8, for three different numbers $n_t$ of integration steps and two values of the stellar compactness (left: $R/M=5.0$; right: $R/M=4.0$). This figure shows that the resolution $n_t=100$ used in our waveform fitting is adequate.
See Section~\ref{sec:tests:WF:dtime} for further details.
\label{fig:tests:WF:dtime:compare_n}
}
\end{figure*}

\begin{figure*}[!t]
\begin{center}
\includegraphics[height=.26\textheight]{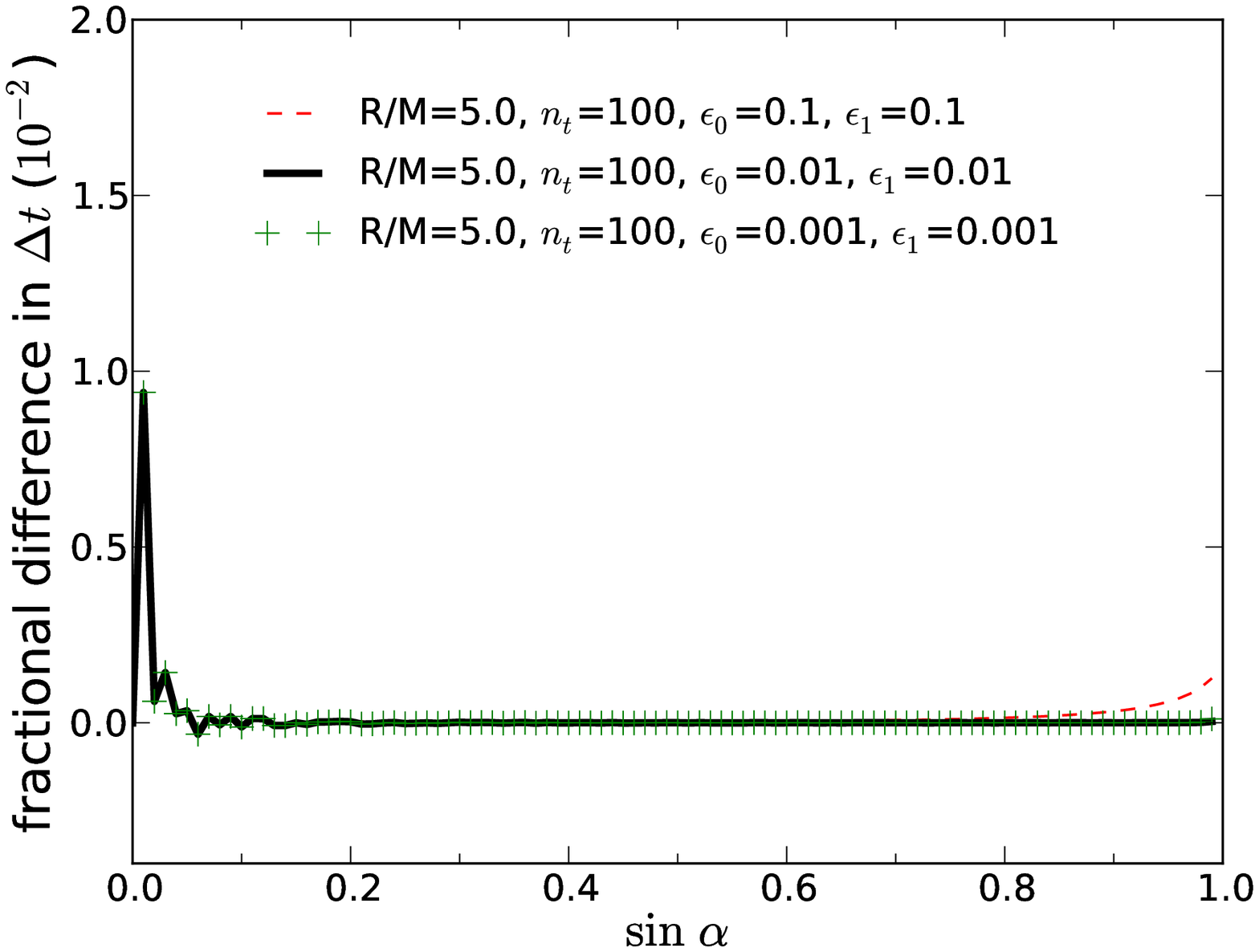}
\includegraphics[height=.26\textheight]{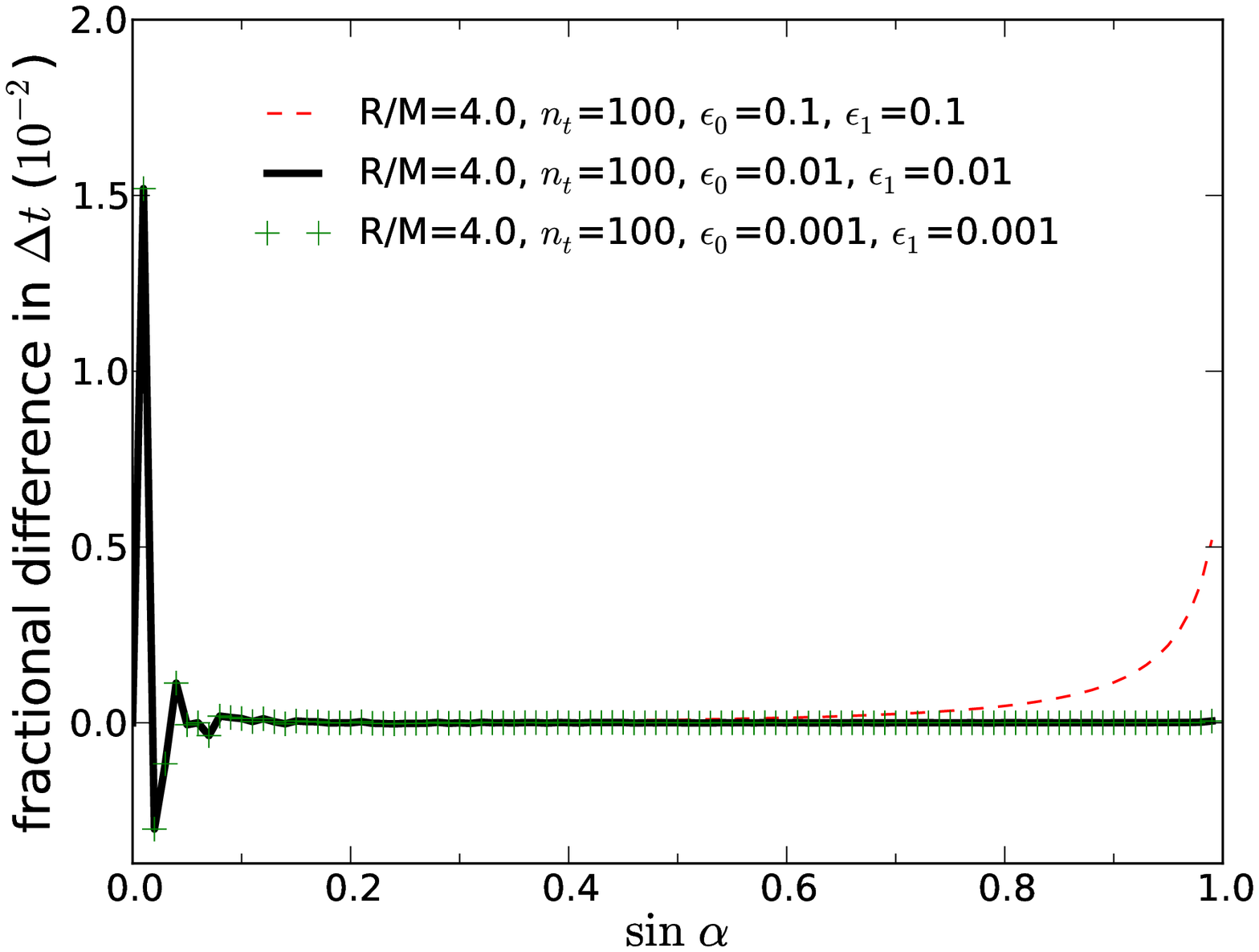}
\end{center}
\vspace{-0.5cm}
\caption{
Fractional difference between the differential light travel time ${\Delta t}$ as a function of the initial ray direction $\alpha$ computed using our code and using Mathematica~8, for three different values of $\epsilon_0$ and $\epsilon_1$ and two values of the stellar compactness (left: $R/M=5.0$; right: $R/M=4.0$). This figure shows that the values $\epsilon_0=\epsilon_1=0.01$ used in our waveform fitting are adequate.
See Section~\ref{sec:tests:WF:dtime} for further details.
\label{fig:tests:WF:dtime:compare_eps_delt}
}
\end{figure*}

\begin{figure*}[!t]
\begin{center}
\includegraphics[height=.26\textheight]{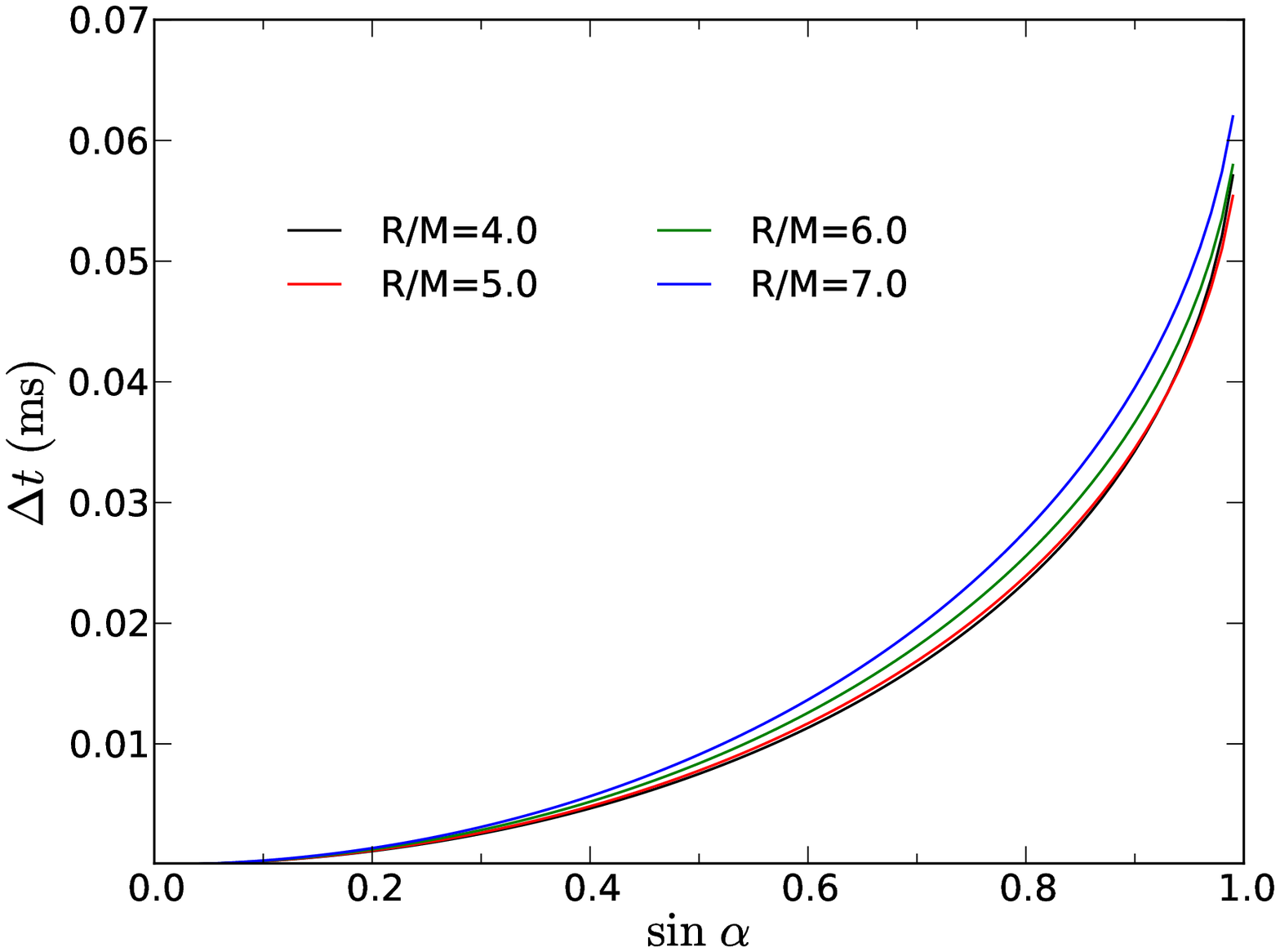}
\includegraphics[height=.26\textheight]{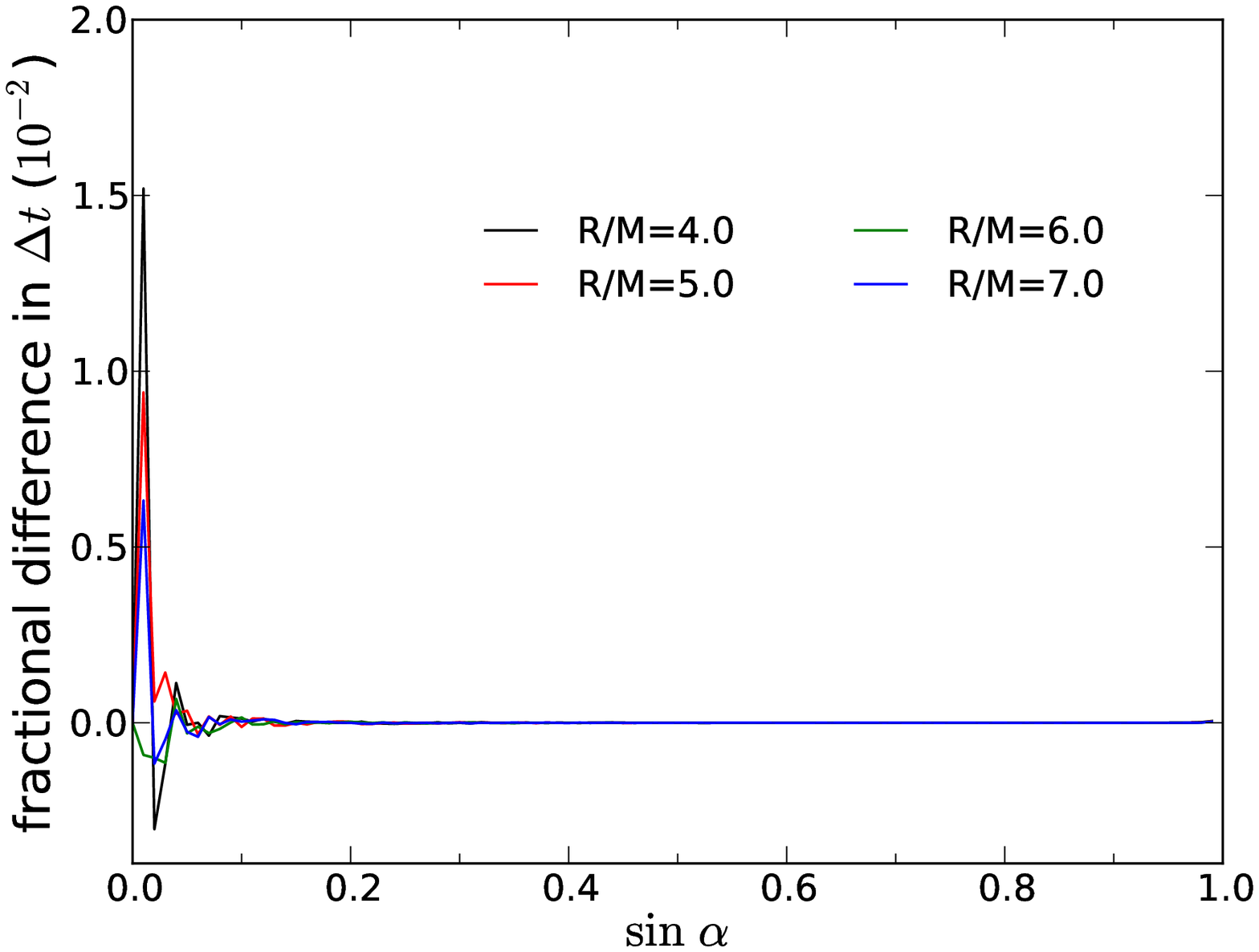}
\end{center}
\vspace{-0.5cm}
\caption{
\textit{Left}: Differential light travel time ${\Delta t}$ as a function of the initial ray direction $\alpha$ computed using our code, for four values of the stellar compactness. The differential light travel time is smaller for smaller values of $R/M$, i.e., for more compact stars.
\textit{Right}: Fractional difference between the light travel time ${\Delta t}$ as a function of $\alpha$ computed using our code and using Mathematica~8, for the same four values of the stellar compactness. The fractional difference is larger for smaller values of $R/M$, i.e., for more compact stars, but is less than $1.5\times10^{-2}$ for all these values of $R/M$.
This figure shows that the resolution parameter values $n_t=100$ and $\epsilon_0=\epsilon_1=0.01$ used in our waveform fitting are adequate. We have assumed $M=1.4M_{\odot}$ to obtain $R$ from $R/M$.
See Section~\ref{sec:tests:WF:dtime} for further details.
\label{fig:tests:WF:dtime:mathematica}
}
\end{figure*}

We test the differential light travel time ${\Delta t}$ given by our waveform code by comparing it with the value of ${\Delta t}$ given by evaluating 
integral~(\ref{eqn:tests:WF:differential-travel-time}) using Mathematica~8. 
Figure~\ref{fig:tests:WF:dtime:compare_n} shows the fractional difference $[({\Delta t}_{\rm code} - {\Delta t}_{\rm mathematica}) / {\Delta t}_{\rm mathematica}] \times 100\%$ for three values of $n_t$ and two values of the stellar compactness.
Figure~\ref{fig:tests:WF:dtime:compare_eps_delt} shows the fractional difference for three values of $\epsilon_0$ and $\epsilon_1$ and two values of the stellar compactness.
The left and right panels of 
Figure~\ref{fig:tests:WF:dtime:mathematica} show, respectively, the relation ${\Delta t}(\alpha)$ and the fractional difference given by our code using $n_t=100$ and $\epsilon_0=\epsilon_1=0.01$, for four values of the stellar compactness that span the range considered in this work.
Figures~\ref{fig:tests:WF:dtime:compare_n}--\ref{fig:tests:WF:dtime:mathematica} show that the values $n_t=100$ and $\epsilon_0=\epsilon_1=0.01$ of these resolution parameters that we use in our waveform code when fitting synthetic data provide sufficient accuracy for this purpose.

\subsubsection{Bolometric flux above the center of an emitting spot on a non-rotating star}
\label{sec:tests:WF:absolute_flux}

We test the accuracy of our waveform code's computation of the gravitational redshift and deflection angle and its integration over the area of the hot spot by comparing the \textit{bolometric} flux that it predicts will be seen by an observer positioned on the radial vector from the center of a non-rotating star through the center of the spot, as a function of the compactness of the star and the angular radius of the spot, with the analytic result for this flux.

Consider a uniform circular spot of angular radius $\Delta\theta$ on the surface of a non-rotating star with compactness $M/R$, emitting radiation with an isotropic beaming pattern. In the Schwarzschild spacetime, the bolometric flux seen by a distant observer positioned above the center of the spot is
\begin{equation}
F = \frac{I_0}{(1+z)^2} \left(\frac{R}{d}\right)^2\int^{2\pi}_0\dd\phi\int^1_{\cos\Delta\theta}\cos\alpha~\left(\frac{\dd\cos\alpha}{\dd\cos\psi}\right)\dd\cos\theta  \;,
\end{equation}
where $I_0$ is the specific intensity measured at the stellar surface, $1+z=\left(1-{2M}/{R}\right)^{-{1}/{2}}$ is the redshift, $R$ is the radius of the star, $d$ is the distance to the star, $\alpha$ is the angle that a ray initially makes with the surface normal, as measured in the static frame, and $\psi=\psi\left(\alpha, M/R\right)$ is the deflection angle. In the limit of vanishing $M/R$, $z \rightarrow 0$, $\cos\alpha \rightarrow \cos\theta$, $\dd\cos\alpha/\dd\cos\psi \rightarrow 1$, and we recover the familiar expression
\begin{equation}
F_0 = I_0 \left(\frac{R}{d}\right)^2\int^{2\pi}_0\dd\phi\int^1_{\cos\Delta\theta}\cos\theta~\dd\cos\theta = \pi I_0\left(\frac{R}{d}\right)^2\sin^2\Delta\theta \;.
\end{equation}

A relationship between the angle $\theta$ of the emission point, measured from the symmetry axis of the spot, and the angle $\alpha$ a ray makes with the local surface normal is established by the requirement that such a ray must be deflected by $\psi=\theta$ in order to be seen by the observer. In other words,
\begin{equation}
\theta = \int^1_0 \frac{\sin\alpha~\dd x}{\sqrt{\left(1-2M/R\right)-\left(1-2Mx/R\right)x^2\sin^2\alpha}}\;.
\end{equation}
The condition $\psi=\theta$ allows us to simplify the expression for the flux considerably, namely,
\begin{align}
F &= \frac{I_0}{(1+z)^2}\left(\frac{R}{d}\right)^2 \int^{2\pi}_0\dd\phi\int^1_{\cos\Delta\theta}\cos\alpha~\left(\frac{\dd\cos\alpha}{\dd\cos\theta}\right)\dd\cos\theta \nonumber \\
&= \frac{I_0}{(1+z)^2}\left(\frac{R}{d}\right)^2 \int^{2\pi}_0\dd\phi\int^1_{\cos\alpha(\Delta\theta)}\cos\alpha~\dd\cos\alpha \nonumber \\
&= \frac{\pi I_0}{(1+z)^2}\left(\frac{R}{d}\right)^2\sin^2\alpha(\Delta\theta)
\label{eqn:tests:waveform:absolute_flux}
\end{align}
where $\alpha(\Delta\theta)$ is the value of $\alpha$ for which $\psi(\alpha)=\Delta\theta$.

\begin{figure*}[!t]
\begin{center}
\includegraphics[height=.26\textheight]{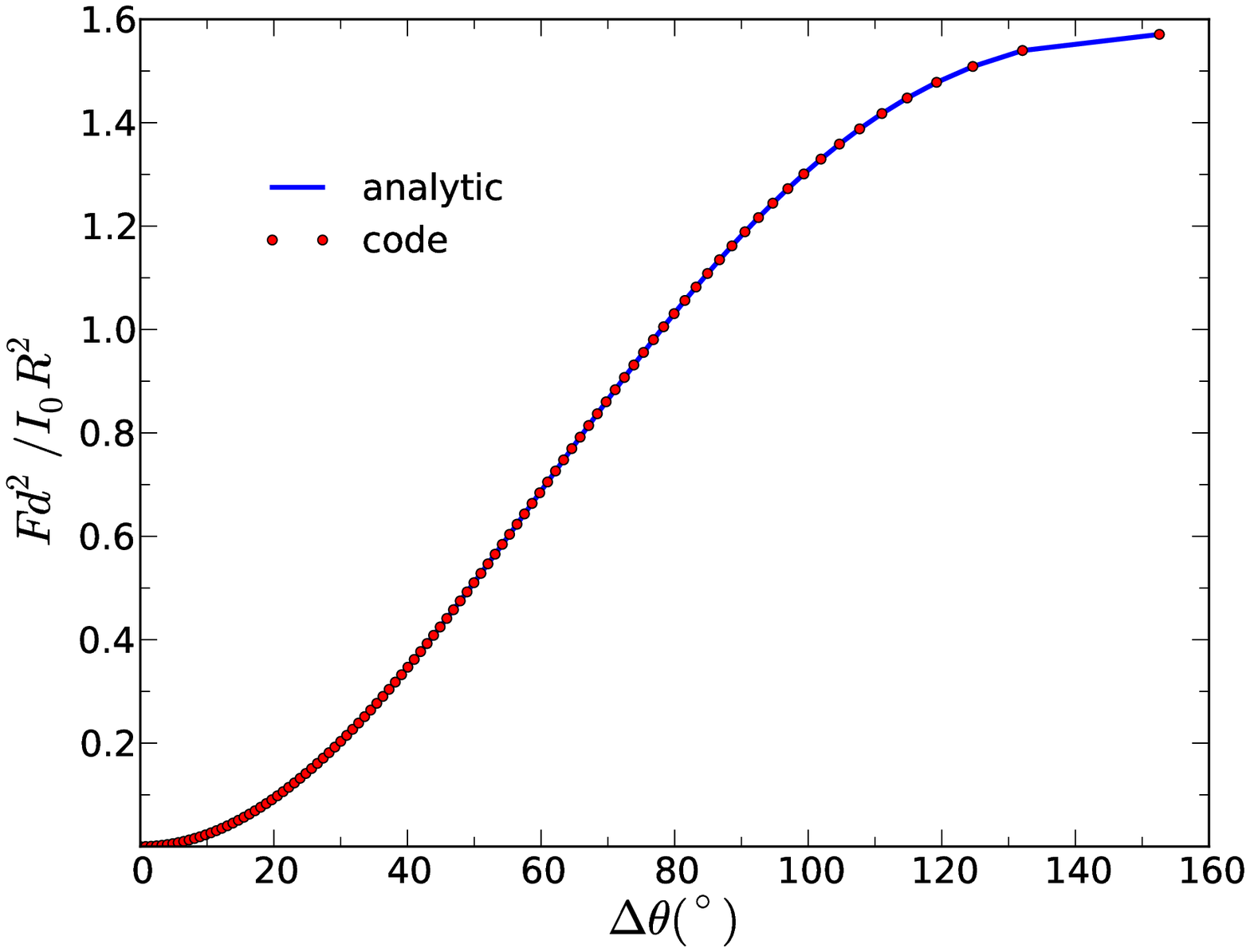}
\includegraphics[height=.26\textheight]{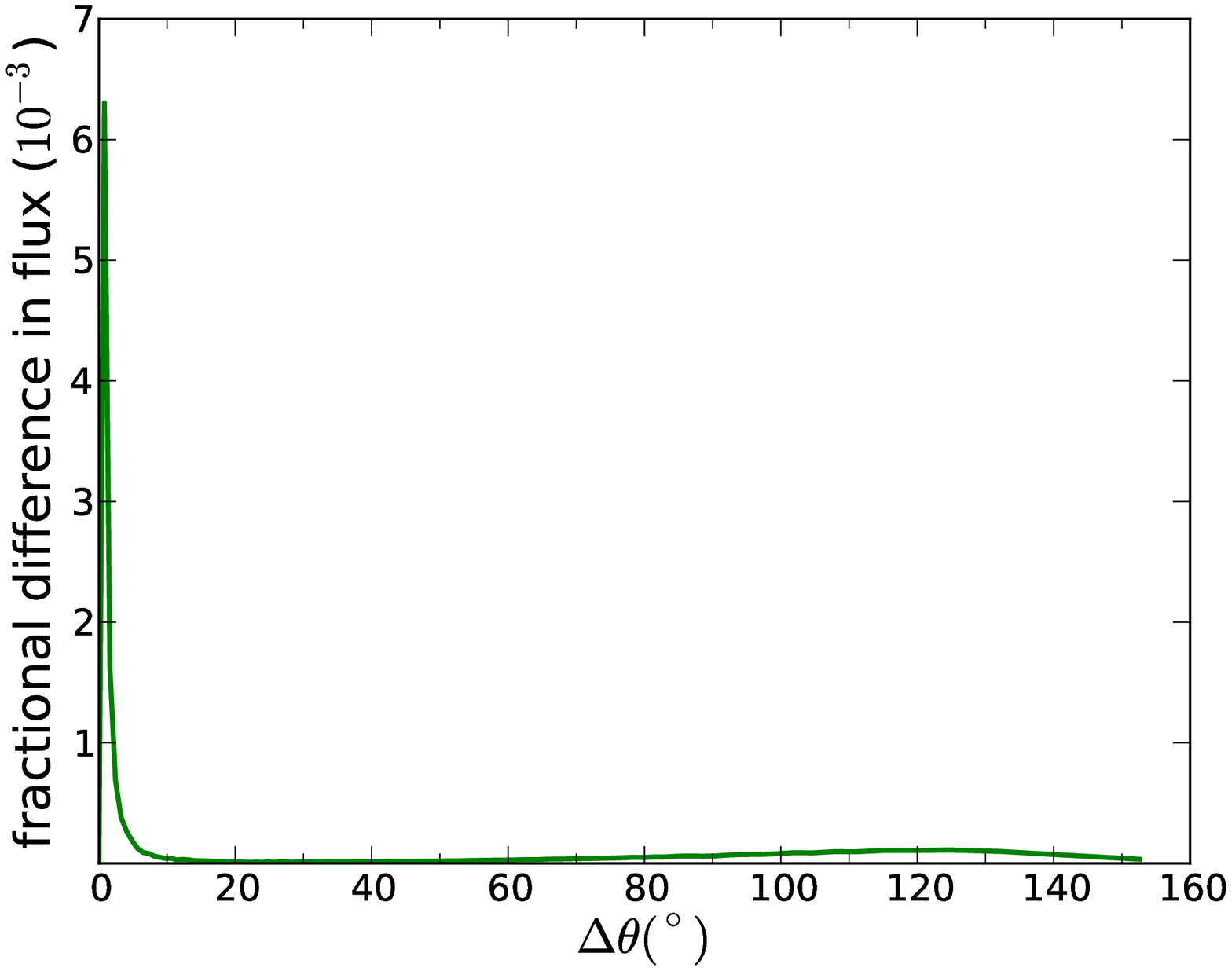}
\end{center}
\vspace{-0.5cm}
\caption{
Scaled bolometric flux (left) and fractional error in the bolometric flux (right) seen by a distant observer, as a function of $\Delta\theta$. The agreement between the results given by our waveform code (red dots) and the analytic results (blue curve) is excellent. These results are for $R/M=4.0$, $M=1.4M_{\odot}$, and $\theta_{\rm spot} = \theta_{\rm obs} = 0$. 
See Section~\ref{sec:tests:WF:absolute_flux} for further details.
\label{fig:tests:WF:absolute_flux}
}
\vspace{1cm}
\end{figure*}

We perform this test of our code as follows. For a given value of $M/R$ and a range of values of $\alpha$, we compute the value of $\psi$ that corresponds to each value of $\alpha$. Next, for each $\psi(\alpha)$ (as long as $\psi<\pi$), we use the relation 
$\Delta\theta=\psi(\alpha)$ to construct the relation $\alpha(\Delta\theta)$ and then use this value of $\alpha$ in expression~(\ref{eqn:tests:waveform:absolute_flux}) to determine the bolometric flux that is predicted by this analytical expression for each value of $\Delta\theta$. On the other hand, we use the angular radius $\Delta\theta(\alpha)$ to compute the bolometric flux for each value of $\alpha$, using our waveform code. We then compare these two values of the flux at each value of $\Delta\theta$. 
Figure~\ref{fig:tests:WF:absolute_flux} shows an example of such a comparison. The results given by the code differ from the results predicted by the analytical expression by less than $5\times10^{-4}$ for $\Delta\theta > 5^\circ$. Even for $\Delta\theta < 5^\circ$, the difference is less than $7\times10^{-3}$.

\subsubsection{Radiation pattern produced by a small spot on a non-rotating star in flat spacetime}
\label{sec:tests:WF:nonrot-flat}

We test the accuracy with which the waveform code computes the radiation pattern from the star by using it to compute the flux seen by a distant observer moving around a non-rotating star in flat spacetime that has a very small spot that emits isotropically. We then compare this result with the analytic result for this situation. In addition to testing the computation of the radiation pattern, this comparison also tests the accuracy of the fast Fourier transform (FFT) that we use to represent waveforms.

The radiation pattern for a star and hot spot with the assumed properties is
\begin{equation}
f(\phi) = \max\{\cos\phi, 0\}  \;,
\label{eqn:tests:WF:nonrot-flat:waveform}
\end{equation}
where $\phi$ is the phase of the waveform (equivalent to the azimuth of the distant observer, for a non-rotating star). The Fourier representation of this waveform is
\begin{equation}
f(\phi) = c_0 + \sum_{n=1}^{\infty} c_n \cos\left[n\left(\phi - \phi_n\right)\right]  \;,
\end{equation}
where
\begin{align}
c_0 &= \frac{1}{\pi} \nonumber\\
c_1 &= \frac{1}{2} \nonumber \\
c_n &= \left\{ \begin{array}{cc} 0, & n~{\rm is~odd}~(n>1) \\ \frac{2}{\pi}\frac{1}{n^2-1}, & n~{\rm is~even} \end{array} \right.
\end{align}
and
\begin{align}
\phi_1 &= 0 \nonumber\\
\phi_n &= \left\{\begin{array}{cc} {\rm indeterminate}~({\rm since}~c_n=0), & n~{\rm is~odd} \\ 0, & n=2, 6, 10, ... \\ \pi/n, & n=4, 8, 12, ...  \end{array}\right.
\end{align}

\begin{figure*}[!t]
\begin{center}
\includegraphics[height=.26\textheight]{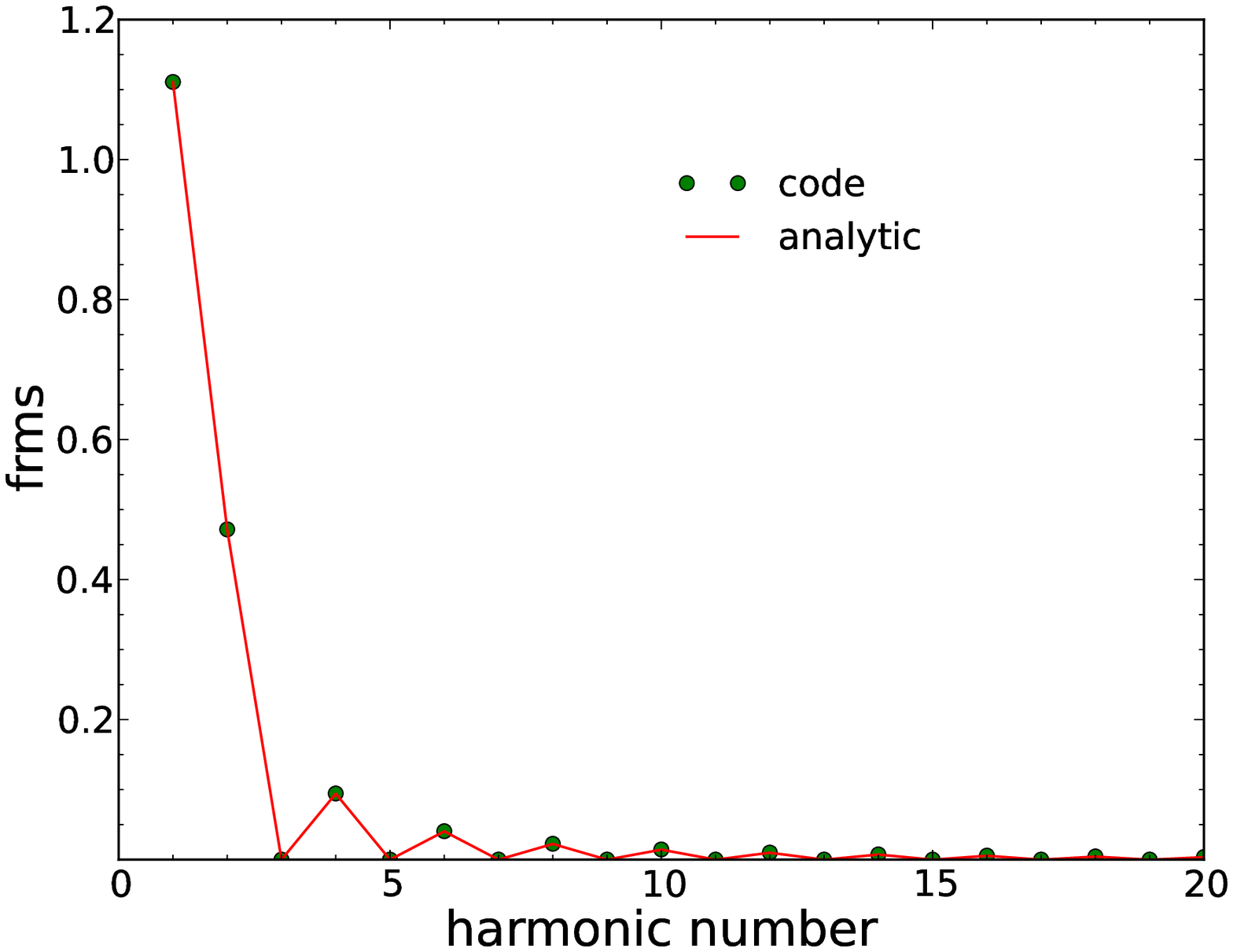}
\includegraphics[height=.26\textheight]{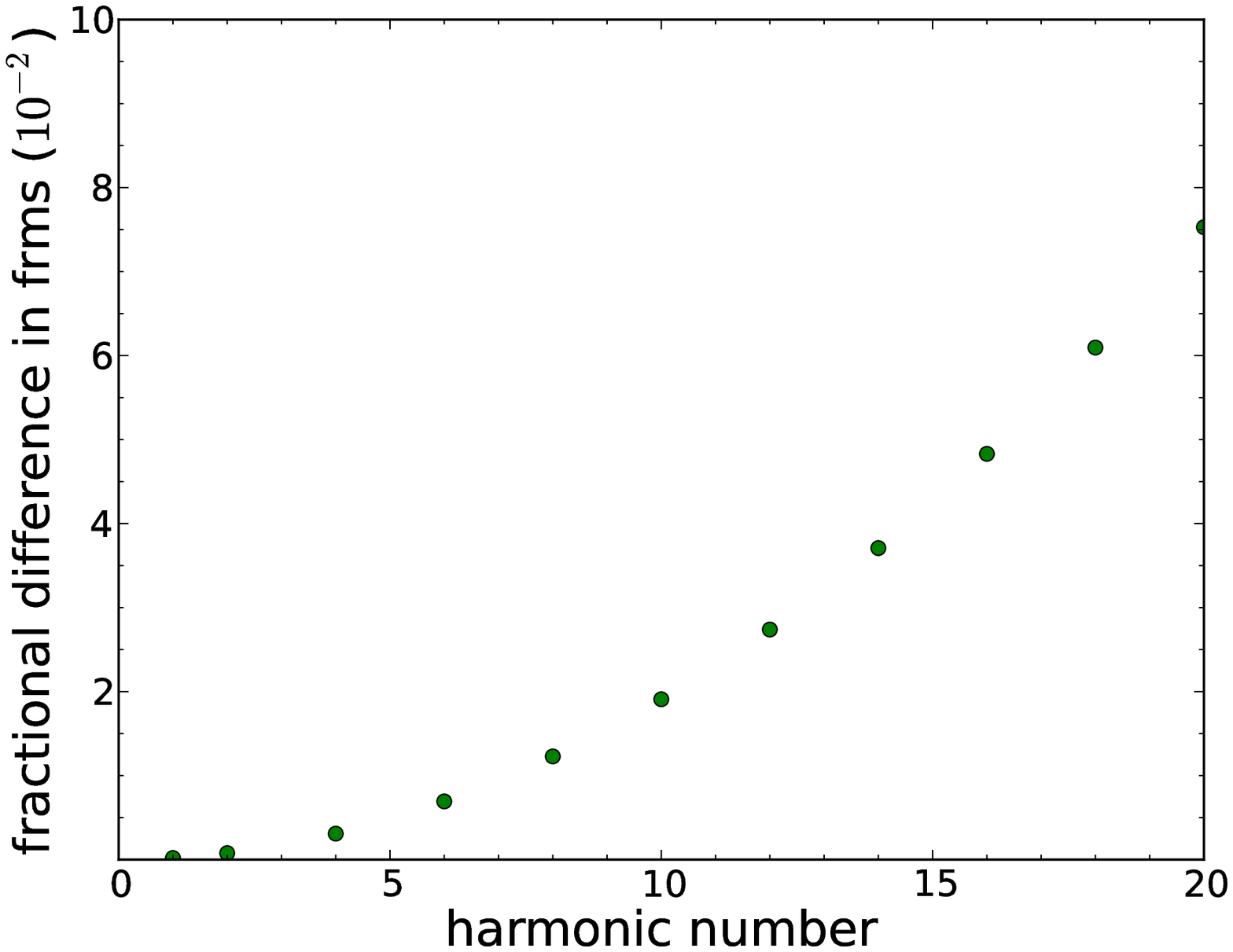}
\end{center}
\vspace{-0.5cm}
\caption{
Fractional rms amplitudes of the first 20 harmonics of the test waveform, showing good agreement between the results given by our waveform code and the analytic results. 
\textit{Left}: Amplitudes given by our waveform code (green dots) compared with the analytic amplitudes (red curve). 
\textit{Right}: Fractional differences between the amplitudes given by our waveform code and the analytic amplitudes. Only the differences for harmonic components with nonzero analytic amplitudes are shown. 
See Section~\ref{sec:tests:WF:nonrot-flat} for further details.
\label{fig:tests:WF:nonrot-flat:hfrms}
}
\end{figure*}

\begin{figure*}[!t]
\begin{center}
\includegraphics[height=.26\textheight]{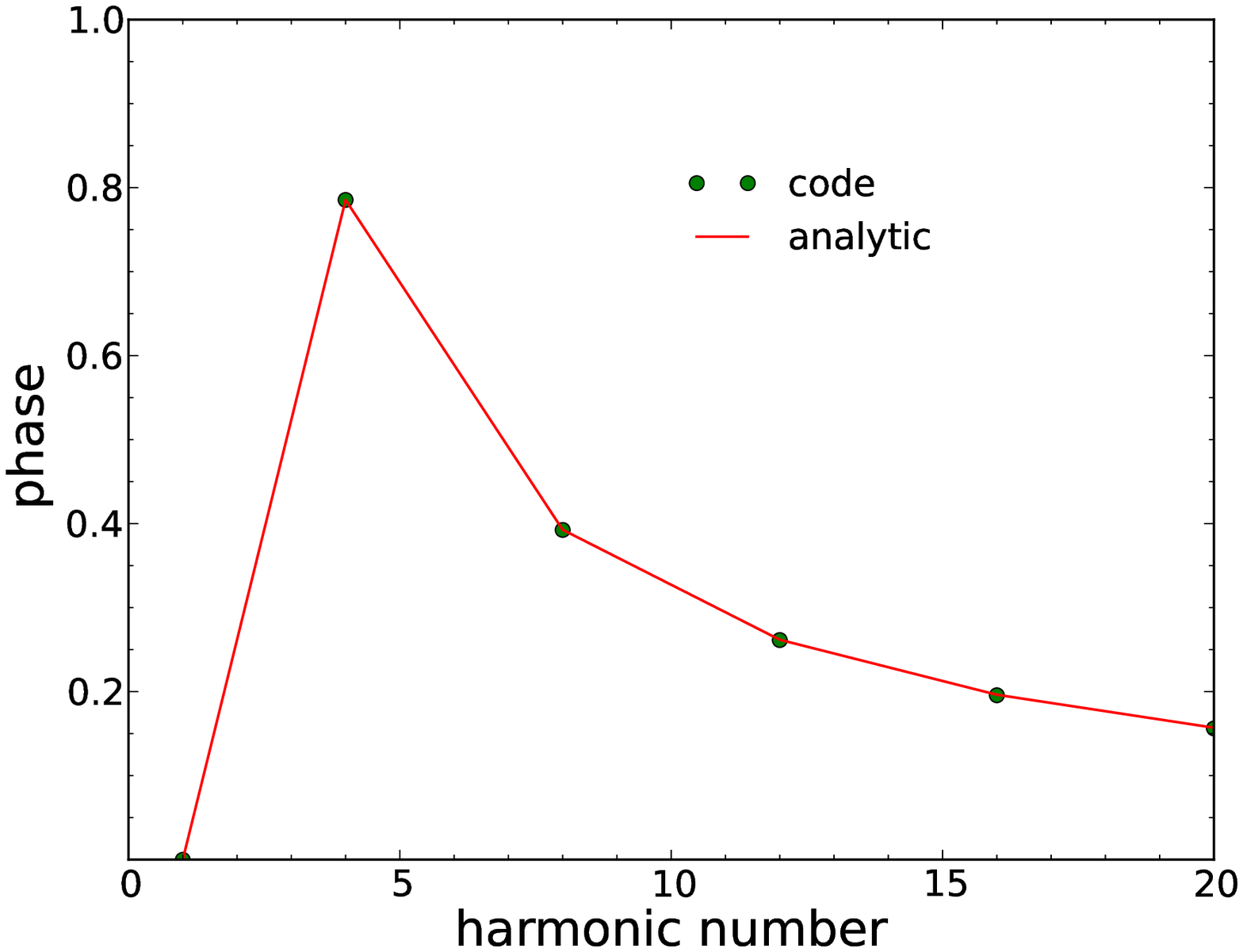}
\includegraphics[height=.26\textheight]{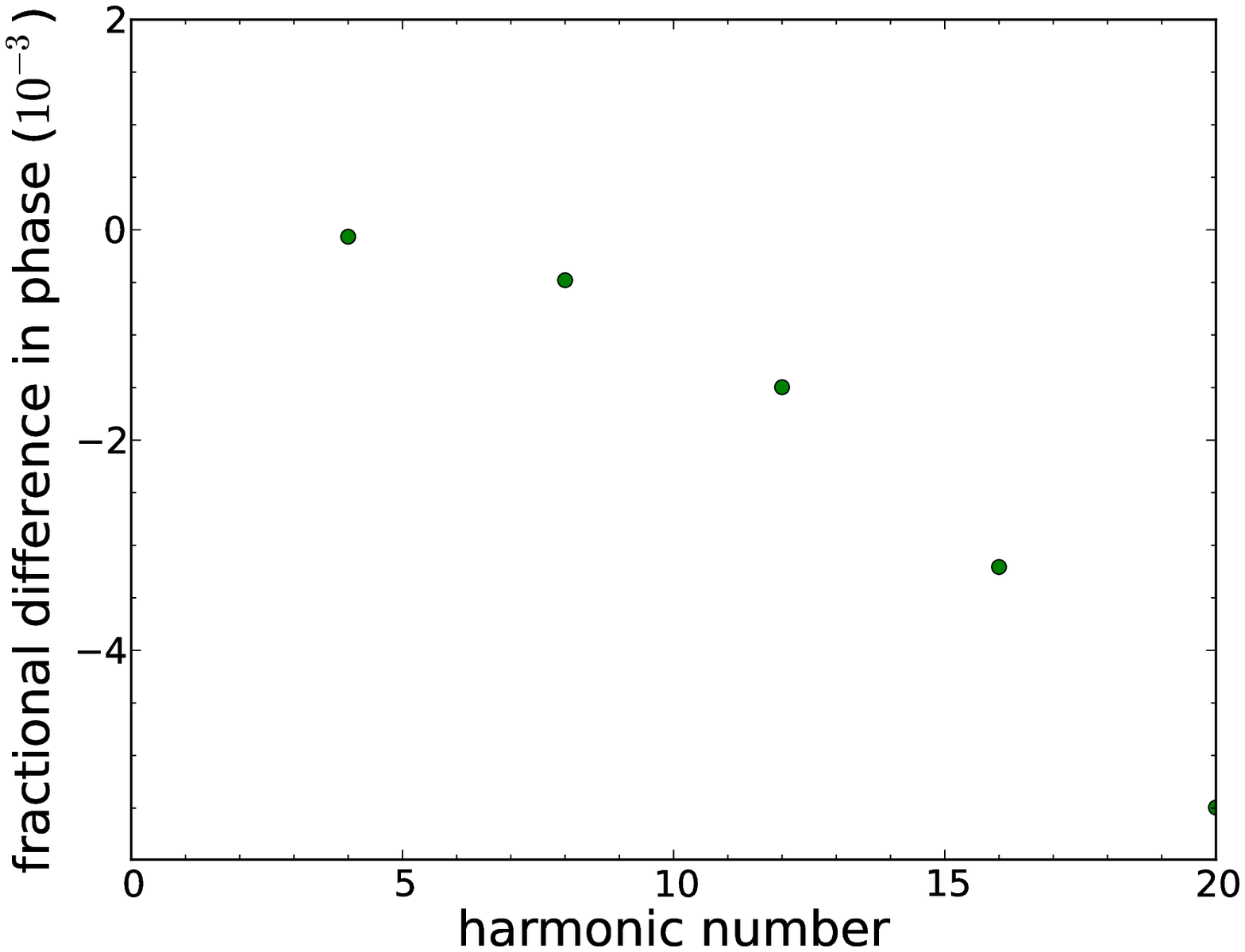}
\end{center}
\vspace{-0.5cm}
\caption{
Phases of the first 20 harmonics of the test waveform, showing good agreement between the results given by our waveform code and the analytic results. 
\textit{Left}: Phase of the $n^{\rm th}$ harmonic component of the waveform as a function of harmonic number $n$ (green dots) compared with the analytic phases (red curve). 
\textit{Right}: Fractional differences between the phases given by our waveform code and the analytic phases.
Only the results for harmonic components with nonzero analytic amplitudes are shown.
See Section~\ref{sec:tests:WF:nonrot-flat} for further details.
\label{fig:tests:WF:nonrot-flat:hphase}
}
\end{figure*}

Figures~\ref{fig:tests:WF:nonrot-flat:hfrms} and \ref{fig:tests:WF:nonrot-flat:hphase} compare these analytic predictions for the fractional rms amplitudes and the phases of the first 20 harmonics with the results for these quantities given by our waveform code for $R/M=10^6$, $M=1.4M_{\odot}$, $\theta_{\rm spot}=60^{\circ}$, $\Delta\theta=1^{\circ}$, and $\theta_{\rm obs}=20^{\circ}$ and the values of the resolution parameters listed in Table~\ref{table:tests:WF:resolution-parameters}. These figures show that the results given by our waveform code are in good agreement with the analytic predictions.

\subsubsection{Waveform produced by a small spot on a rotating star in flat spacetime}
\label{sec:tests:WF:rot-flat}

We test the accuracy with which our waveform code computes the Doppler boost and the photon travel time by using it to compute
the bolometric waveform produced by a small spot on a rotating star in flat spacetime and comparing it with the analytic waveform for this case.

The analytic expression for this waveform is
\begin{equation}
f(\phi) \propto \max \{\delta^4\cos\alpha, 0\} \;,
\end{equation}
where $\phi$ is the observed phase (i.e., after correction for the time delay) and 
\begin{equation}
\delta(\phi) = \frac{1}{\gamma [1+(v/c)\sin\theta_{\rm spot}\sin\theta_{\rm obs}\sin\Phi]}
\end{equation}
is the Doppler factor. Here $v$ is the linear velocity at the rotational equator and $\Phi \equiv \phi - \Omega\Delta t$ is the phase prior to correction for time delay in terms of the angular rotation frequency $\Omega$ of the hot spot and the time delay
\begin{equation}
\Delta t = \frac{R}{c}(1-\cos\alpha) \;,
\end{equation}
where $\alpha$ is the angle that the ray makes with the local normal to the stellar surface, as measured in the static frame, and is related to other angles via
\begin{equation}
\cos\alpha = \cos\theta_{\rm spot}\cos\theta_{\rm obs} + \sin\theta_{\rm spot}\sin\theta_{\rm obs}\cos\Phi \;.
\end{equation}
To compute the flux at each phase $\phi$ analytically, one must simultaneously solve for $\Phi$ and $\cos\alpha$ using this set of equations. 

\begin{figure*}[!t]
\begin{center}
\includegraphics[height=.26\textheight]{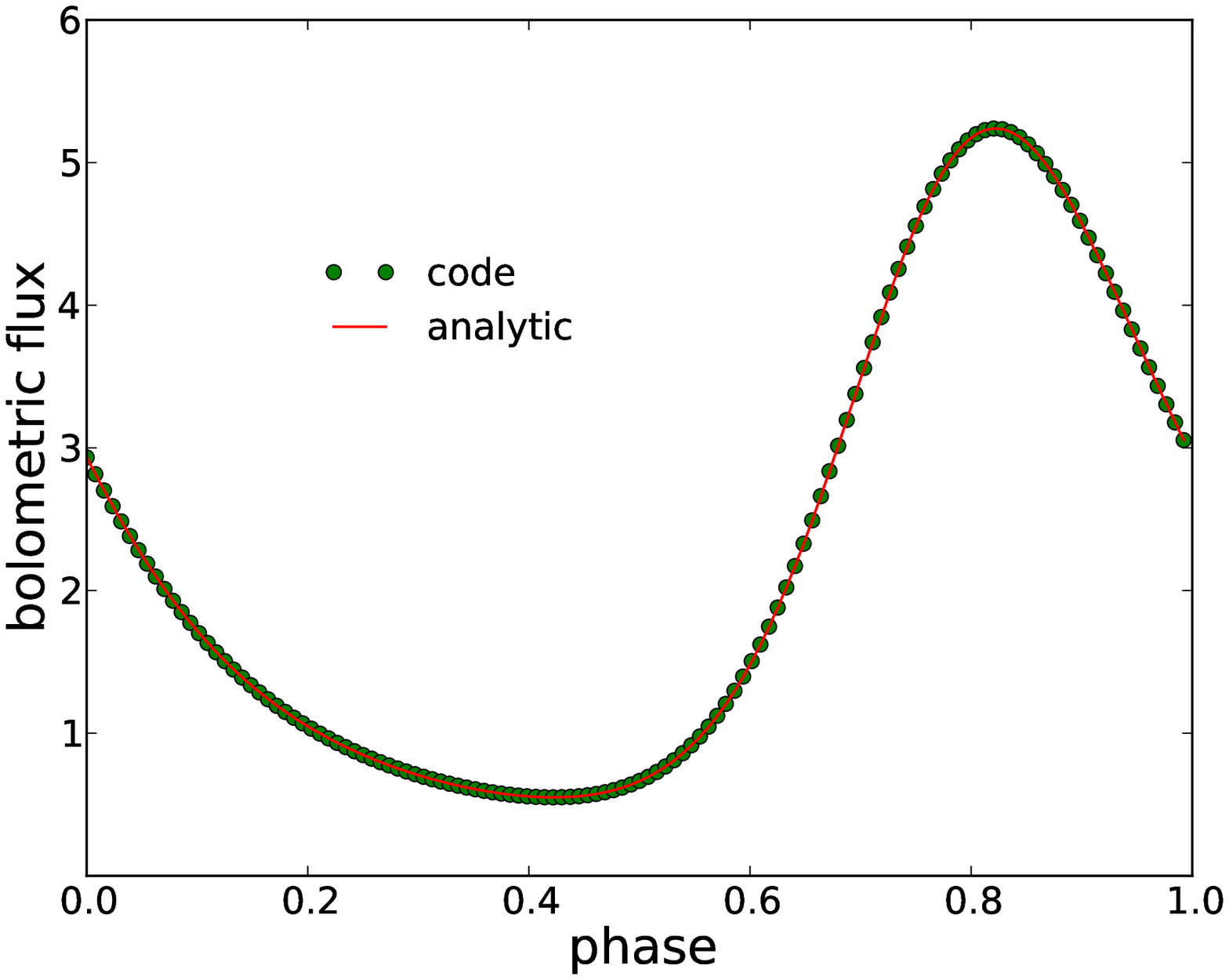}
\includegraphics[height=.26\textheight]{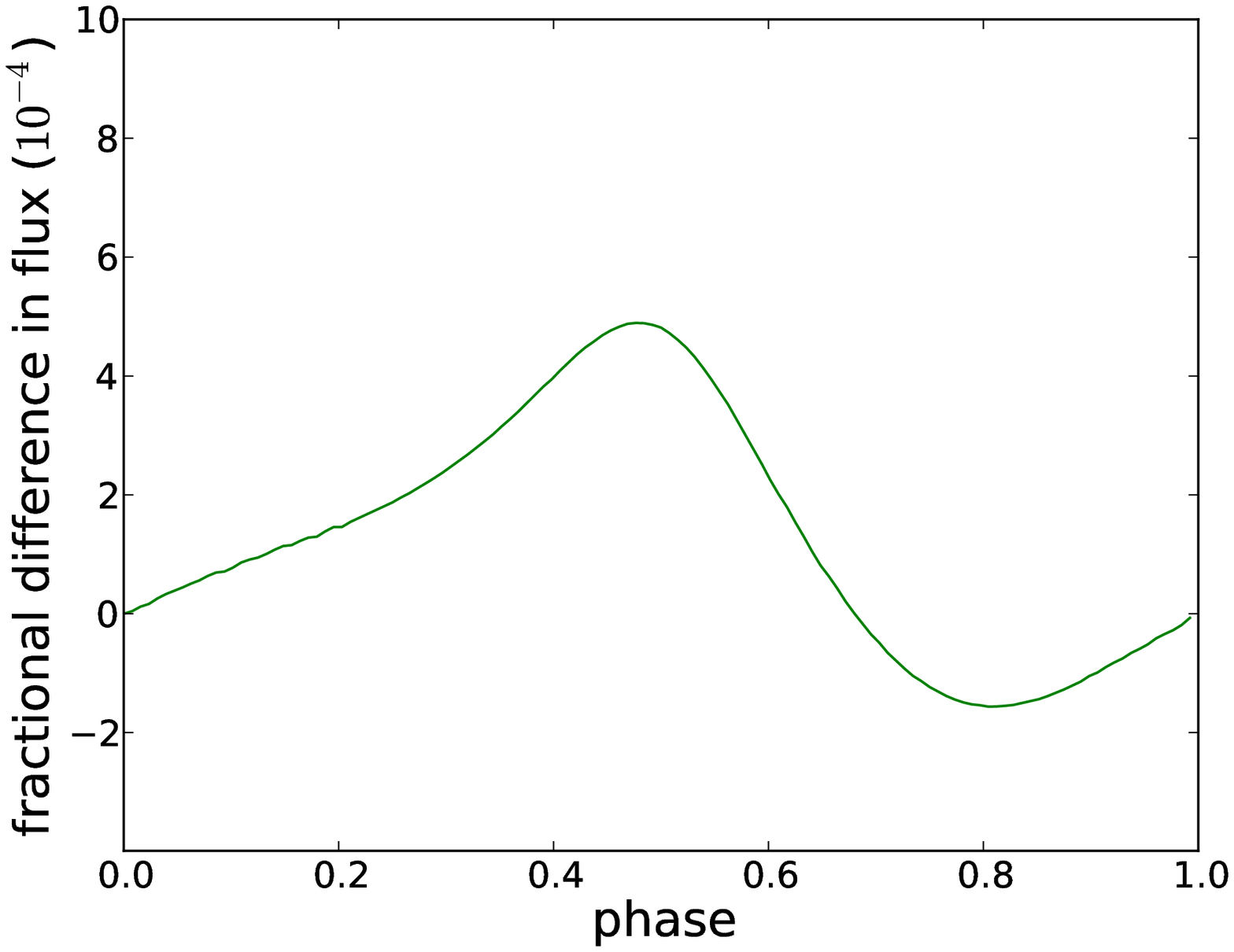}
\includegraphics[height=.26\textheight]{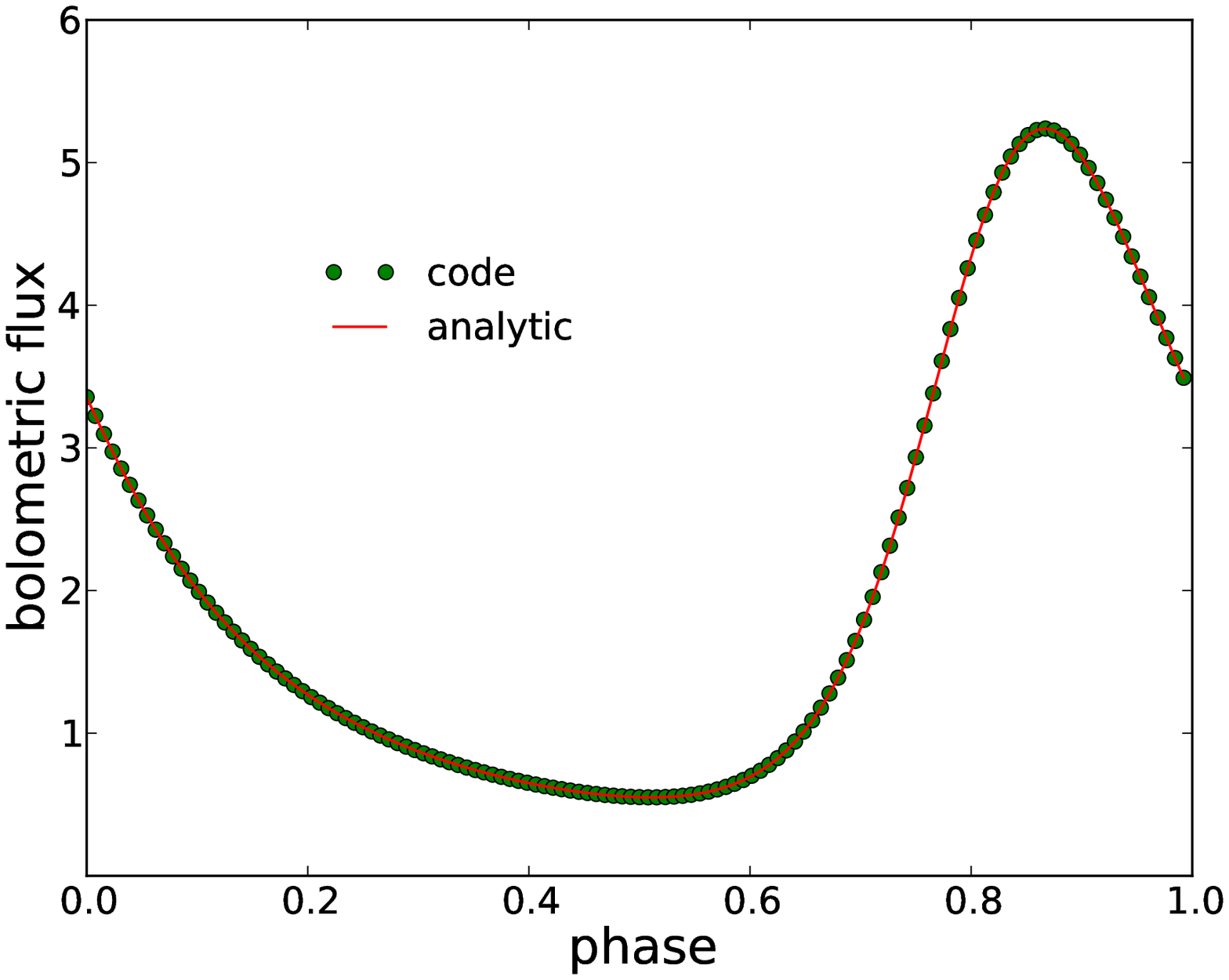}
\includegraphics[height=.26\textheight]{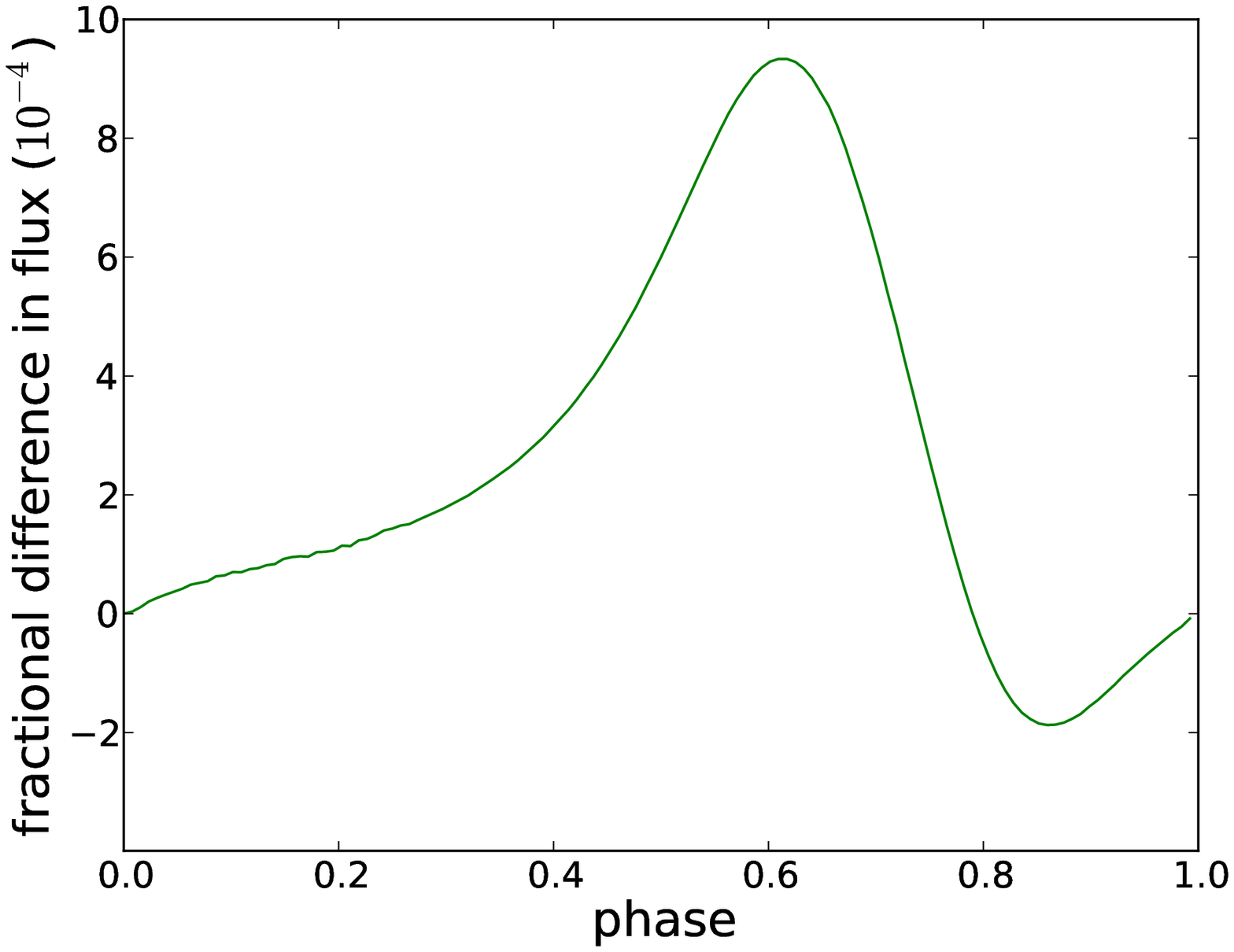} \\
\end{center}
\vspace{-0.5cm}
\caption{
\textit{Top left}: Bolometric flux waveform produced by a small spot on the surface of a rotating star in flat spacetime, computed using our waveform code (green dots) and the corresponding analytic result (red curve), with the time delay set to zero.
\textit{Top right}: Fractional differences (green curve) between the fluxes given by our code and the analytic fluxes.
\textit{Bottom panels}: Same as in the top panels, but with the time delay included. These results are for $R/M=10^6$, $M/M_{\odot}$=1.6, $\theta_{\rm spot}=60^{\circ}$, $\theta_{\rm obs}=20^{\circ}$, $\Delta\theta=1^{\circ}$, $\nu$=0.0141357027464 Hz. 
See Section~\ref{sec:tests:WF:rot-flat} for further details.
\label{fig:tests:WF:rot-flat}
}
\end{figure*}

Figure~\ref{fig:tests:WF:rot-flat} shows that the bolometric waveform given by our waveform code for this case is in good agreement with the analytic result, both when the time delay set to zero and when it is included.

\subsubsection{Angular deflection of a pencil beam from a small spot on a rotating star}
\label{sec:tests:WF:pencil}

We test the accuracy with which our waveform code computes the relativistic aberration and gravitational light-bending effects by computing the rotational phase at which a beam of radiation emitted from the stellar surface in a given direction would be seen by a distant observer and comparing this result with the phase given by Mathematica~8 (the ``predicted'' phase). If the star is not rotating, the rotational phase is undefined. In this case what is computed is the azimuth at which the beam of radiation would be seen by a distant observer.

In these tests, we assume that the radiation is emitted in a narrow ``pencil'' beam from a very small ($\Delta\theta = 1^{\circ}$) spot located on the star's rotational equator ($\theta_{\rm spot} = 90^{\circ}$) and that the observer is in the plane defined by the star's rotational equator ($\theta_{\rm obs} = 90^{\circ}$). Because of the spherical symmetry of the Schwarzschild spacetime, the radiation will remain in the equatorial plane as it propagates. The phase (azimuth) at which the radiation is seen by a distant observer depends on the angle $\alpha'_0$ of the pencil beam relative to the local normal to the stellar surface, as measured in the corotating frame, the linear velocity $v/c$ of the hot spot produced by its rotation, which determines the amount of relativistic aberration, and the compactness $M/R$ of the star, which determines the amount of light bending.

We first use a Lorentz transformation to determine the angle $\alpha_0$ of the ray in the static frame that corresponds to the angle $\alpha'_0$ of the ray in the comoving frame. We then compute the angular deflection of this ray using Mathematica, which allows us to determine the predicted phase (azimuth) at which the beam would be seen by a distant observer. Next we use our waveform code to compute the phase (azimuth) at which such a beam would be seen by a distant observer. To do this we set the beaming function $g(\alpha')$ equal to 1 for $|\alpha'-\alpha'_0| < \sigma_{\alpha'_0}$ and equal to 0 for $|\alpha'-\alpha'_0| > \sigma_{\alpha'_0}$, with $\sigma_{\alpha'_0} = 0.01$. In order to test the computation of the aberration and the gravitational bending separately from the computation of the time delay, we set the light travel time to be zero for all rays. The phase (azimuth) at which the beam would be seen by a distant observer is then equal to the total angular deflection of the ray produced by the relativistic aberration and light bending. For simplicity, we trace only rays that propagate in the prograde direction relative to the rotation of the hot spot. Finally, we compare the phase (azimuth) given by the waveform code to the phase predicted using Mathematica.

Tables~\ref{table:tests:waveform:pencil:flat-rot}, \ref{table:tests:waveform:pencil:curved-nonrot}, and \ref{table:tests:waveform:pencil:curved-rot} show the results for the computed and predicted phases (or azimuths) at which the pencil beam would be seen by a distant observer. These results are for emission from the surface of a rapidly rotating star in flat spacetime, from the surface of a non-rotating relativistic star, and from the surface of a rapidly rotating relativistic star, for a range of stellar compactnesses and hot spot rotation rates. All these results are for $M=1.4M_{\odot}$.

\begin{deluxetable}{ c c c c }
\tablewidth{0pt}
\tablecaption{
Comparison of pencil-beam phases\tablenotemark{a}
\label{table:tests:waveform:pencil:flat-rot}
}
\tablehead{
\colhead{$\alpha_0$} & \colhead{$v/c$} & \colhead{phase (code)} & \colhead{phase (predicted)}
}
\startdata
$15^{\circ}$ & 0.1 & 0.958 & 0.958 \\
$30^{\circ}$ & 0.1 & 0.917 & 0.917 \\
		   & 0.4 & 0.917 & 0.917 \\
$45^{\circ}$ & 0.1 & 0.875 & 0.875 \\
		   & 0.4 & 0.875 & 0.875 \\
		   & 0.7 & 0.877 & 0.875 \\
$60^{\circ}$ & 0.1 & 0.833 & 0.833 \\
		   & 0.4 & 0.833 & 0.833 \\
		   & 0.7 & 0.833 & 0.833 \\
$75^{\circ}$ & 0.1 & 0.792 & 0.792 \\
		   & 0.4 & 0.791 & 0.792 \\
		   & 0.7 & 0.791 & 0.792 \\
$90^{\circ}$ & 0.1 & 0.751 & 0.750 \\
		   & 0.4 & 0.751 & 0.750 \\
		   & 0.7 & 0.751 & 0.750 \\
\enddata
\vskip-10pt
\tablenotetext{a}{
Comparison of the computed and predicted phases at which a narrow beam of radiation emitted from a small spot on the equator of a rotating star at an angle $\alpha_0$ relative to the local surface normal (as measured in the static frame) is seen by a distant observer who is also located in the equatorial plane. In this test, $R/M=10^6$, so the spacetime is essentially flat. The stellar rotation frequency is chosen so the linear velocity of the stellar surface in the rotation equator is $0.1c$, $0.4c$, or $0.7c$. The values of $v/c$ that have been omitted from the table would produce beams that point in the retrograde direction in the comoving frame, a geometry not considered in this test. 
See Section~\ref{sec:tests:WF:pencil} for further details.
}
\end{deluxetable}

\begin{deluxetable}{ c c c c }
\tablewidth{0pt}
\tablecaption{
Comparison of pencil-beam phases\tablenotemark{a}
\label{table:tests:waveform:pencil:curved-nonrot}
}
\tablehead{
\colhead{$\alpha_0$} & \colhead{$R/M$} & \colhead{azimuth (code)} & \colhead{azimuth (predicted)}
}
\startdata
$0^{\circ}$   & 4.0 & 0.998 & 1.000 \\
		   & 5.0 & 0.998 & 1.000 \\
		   & 6.0 & 0.999 & 1.000 \\
$15^{\circ}$ & 4.0 & 0.941 & 0.941 \\
		   & 5.0 & 0.946 & 0.946 \\
		   & 6.0 & 0.948 & 0.949 \\
$30^{\circ}$ & 4.0 & 0.881 & 0.881 \\
		   & 5.0 & 0.891 & 0.892 \\
		   & 6.0 & 0.897 & 0.897 \\
$45^{\circ}$ & 4.0 & 0.818 & 0.818 \\
		   & 5.0 & 0.835 & 0.836 \\
		   & 6.0 & 0.844 & 0.845 \\
$60^{\circ}$ & 4.0 & 0.751 & 0.751 \\
		   & 5.0 & 0.777 & 0.777 \\
		   & 6.0 & 0.790 & 0.790 \\
$75^{\circ}$ & 4.0 & 0.674 & 0.674 \\
		   & 5.0 & 0.713 & 0.713 \\
		   & 6.0 & 0.733 & 0.733 \\
$90^{\circ}$ & 4.0 & 0.579 & 0.576 \\
		   & 5.0 & 0.642 & 0.640 \\
		   & 6.0 & 0.671 & 0.669 \\
\enddata
\vskip-10pt
\tablenotetext{a}{
Comparison of the computed and predicted azimuths at which a narrow beam of radiation emitted from a small spot on the equator of a non-rotating star at an angle $\alpha_0$ relative to the local surface normal (as measured in the static frame) would be seen by a distant observer who is also located in the equatorial plane, for three values of the stellar compactness. 
See Section~\ref{sec:tests:WF:pencil} for details.
}
\end{deluxetable}

\begin{deluxetable}{ c c c c c }
\tablewidth{0pt}
\tablecaption{
Comparison of pencil-beam phases\tablenotemark{a}
\label{table:tests:waveform:pencil:curved-rot}
}
\tablehead{
\colhead{$\alpha_0$} & \colhead{$R/M$} & \colhead{$v/c$} & \colhead{phase (code)} & \colhead{phase (predicted)}
}
\startdata
$30^{\circ}$   & 3.5 & 0.4 & 0.870 & 0.871 \\
 		     & 6.0 & 0.4 & 0.897 & 0.897 \\
$45^{\circ}$   & 3.5 & 0.4 & 0.801 & 0.802 \\
		     &      & 0.7 & 0.802 &  0.802 \\
 		     & 6.0 & 0.4 & 0.845 & 0.845 \\
		     &      & 0.7 & 0.790  & 0.790 \\
$60^{\circ}$   & 3.5 & 0.4 & 0.725 & 0.725 \\
		     &       & 0.7 & 0.725 & 0.725 \\
		     & 6.0 & 0.4 & 0.790 & 0.790 \\
		     &      & 0.7 & 0.790 & 0.790 \\
$75^{\circ}$   & 3.5 & 0.4 & 0.632 & 0.633 \\
		     &      & 0.7 & 0.632 & 0.633 \\
		     & 6.0 & 0.4 & 0.733 & 0.733 \\
		     &      & 0.7 & 0.732 & 0.733 \\
$90^{\circ}$   & 3.5 & 0.4 & 0.498 & 0.495 \\
		     &      & 0.7 & 0.498 & 0.495 \\
		     & 6.0 & 0.4 & 0.671 & 0.669 \\
		     &      & 0.7 & 0.670 & 0.669 \\
\enddata
\vskip-10pt
\tablenotetext{a}{
Comparison of the computed and predicted phases at which a narrow beam of radiation emitted from a small spot on the equator of a rotating star at an angle $\alpha_0$ relative to the local surface normal (as measured in the static frame) is seen by a distant observer who is also located in the equatorial plane, for two values of the stellar compactness. The stellar rotation frequency is chosen so the linear velocity of the stellar surface in the rotation equator is either $0.4c$ or $0.7c$. The values of $v/c$ that have been omitted from the table would produce beams that point in the retrograde direction in the comoving frame, a geometry not considered in this test. 
See Section~\ref{sec:tests:WF:pencil} for further details.
}
\end{deluxetable}

\subsubsection{Observed temperature of thermal emission from a rotating star in flat spacetime, as a function of rotational phase}
\label{sec:tests:WF:doppler-temperature}

We test the accuracy with which our waveform code computes the relativistic Doppler effect by computing the observed temperature of thermal emission from the surface of a rotating star in flat spacetime as a function of the rotational phase and then comparing the result with the analytic result for the observed temperature as a function of phase. 

\begin{figure*}[!t]
\begin{center}
\includegraphics[height=.26\textheight]{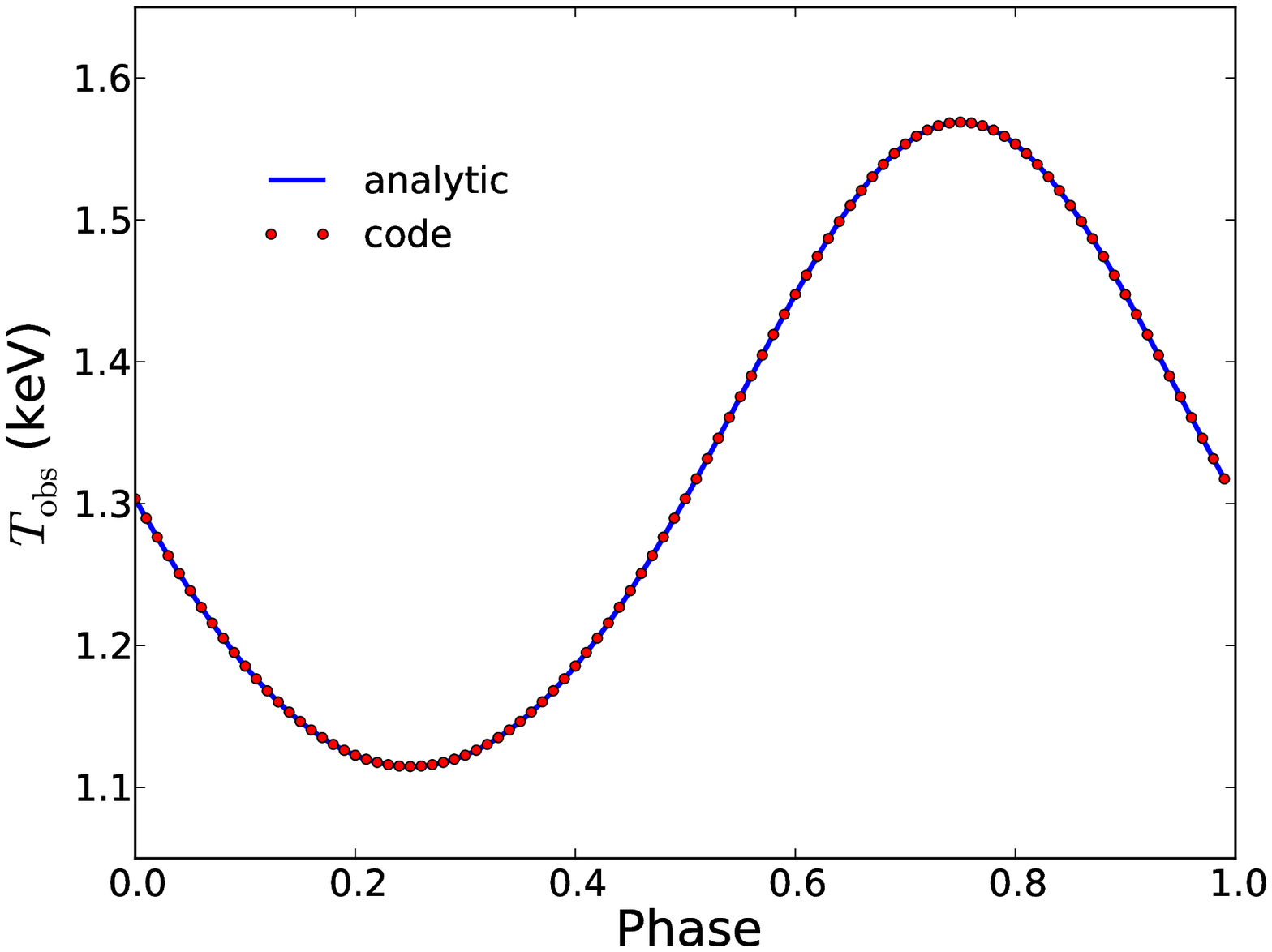}
\includegraphics[height=.26\textheight]{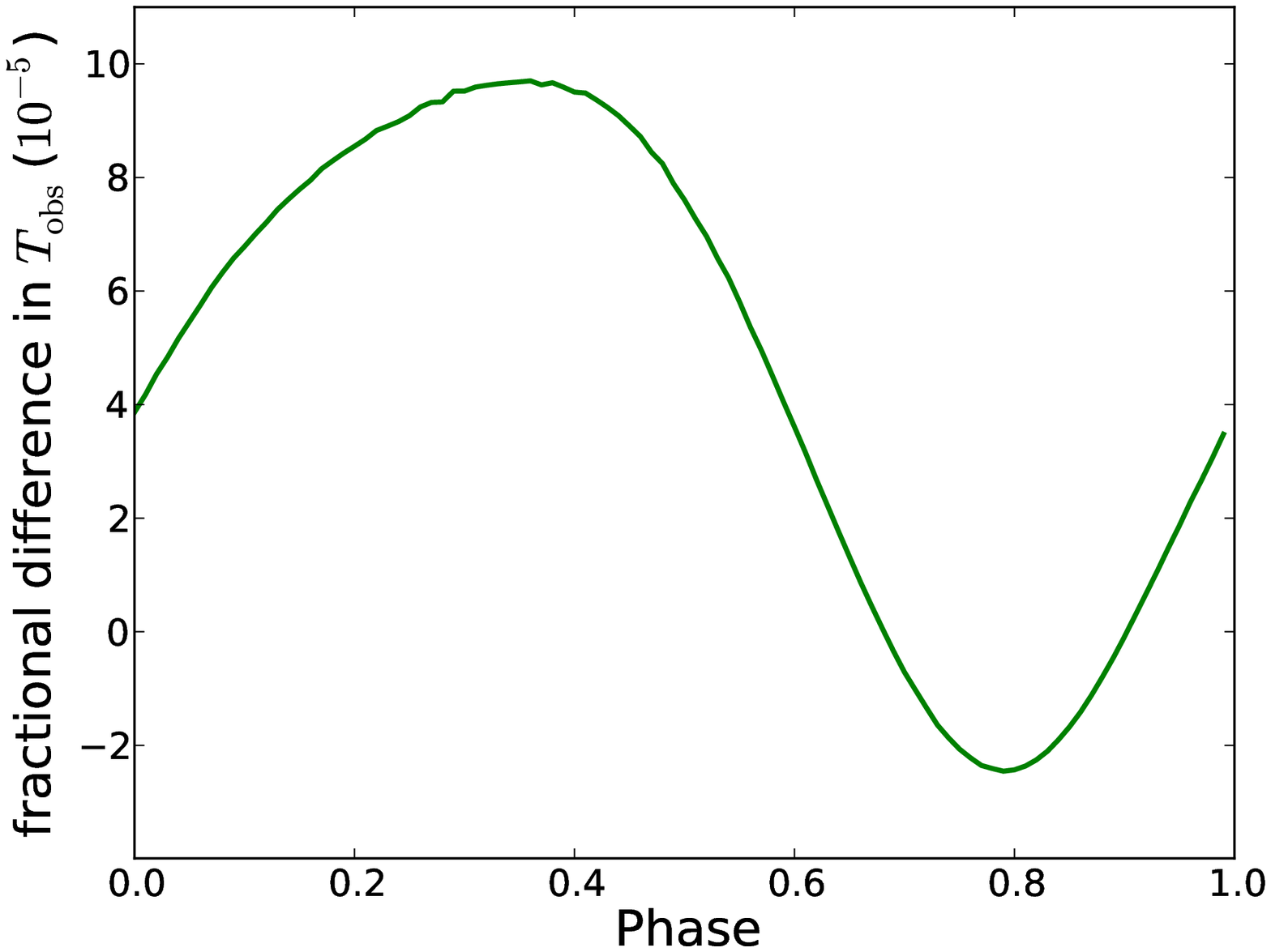}
\end{center}
\vspace{-0.5cm}
\caption{
\textit{Left}: Temperature of the Planck spectrum seen by a distant observer as a function of the star's rotational phase computed using our waveform code (red dots) and the analytic expression (blue curve).
\textit{Right}: Fractional difference (green curve) between the observed temperature computed using our waveform code and the analytic expression. The results given by our waveform code and the analytic results are in excellent agreement. These results are for a star with a rotation rate such that the linear velocity at the rotational equator is $v/c=0.7$, $R/M=10^6$, for $M=1.4M_{\odot}$, $\theta_{\rm spot}=60^{\circ}$, $\theta_{\rm obs}=20^{\circ}$, and $\Delta\theta=1^{\circ}$. 
See Section~\ref{sec:tests:WF:doppler-temperature} for further details.
\label{fig:tests:WF:doppler-temperature}
}
\end{figure*}

For simplicity, we consider isotropic emission from a small spot at an inclination $\theta_{\rm spot}$ on the stellar surface, with a Planck spectrum having a temperature $T_0$ as measured in the comoving frame. Then the spectrum seen by a distant observer with an inclination $\theta_{\rm obs}$ is a Planck spectrum with temperature
\begin{equation}
T(\phi) = \frac{T_0}{\gamma[1-(v_0/c)\sin\theta_{\rm spot}\sin\theta_{\rm obs}\sin\phi]} \;.
\end{equation}
Also for simplicity, we set the light travel time to zero. The observed phase is then the azimuthal position of the spot relative to the azimuth of the observer. To test a range of Doppler boosts, we choose values of the stellar rotational frequency that produce a linear velocity at the rotational equator of $0.1c$, $0.4c$, and $0.7c$. 
Figure~\ref{fig:tests:WF:doppler-temperature} shows the comparison of the temperature between the code and the analytic result, for the case of $v/c=0.7$. It is clear that the agreement is excellent.

\subsection{Tests of the Waveform Fitting Code}
\label{sec:tests:fitting}

In this subsection, we outline the tests we use to validate the set of codes that fit model waveforms to waveform data. An important part of this code set is the code that computes the likelihood of an observed waveform, given a set of model parameters. The accuracy of this step depends crucially on the accuracy with which the model waveforms are computed. We have shown in the previous subsection that our code computes model waveforms with high accuracy. Here we discuss three relatively simple end-to-end tests that we use to validate the waveform fitting process.

\subsubsection{Determining the compactness, temperature, and distance given uniform emission from a non-rotating star with no background}
\label{sec:tests:fitting:nonrot-nobkg-R_T}

In this exercise, we test our model-fitting code by fitting a model of emission from a non-rotating star to the synthetic spectrum produced by a non-rotating star that emits radiation uniformly over its entire surface, with a Planck spectrum. We assume there is no background. The observed spectrum is therefore a Planck spectrum with temperature
\begin{equation}
T_{\rm obs} = T_{\rm co} \left(1-\frac{2M}{R}\right)^{1/2}  \;,
\label{eqn:validation:fitting:nonrot-nobkg-R_T:tempscaling}
\end{equation}
where $T_{\rm co}$ is the temperature of the emission measured in the comoving frame at the stellar surface.
In this case, the goodness of the fit is determined entirely by the shape and normalization of the model spectrum. If the distance is not independently known, the normalization of the spectrum can be adjusted by changing the distance. Hence the likelihood of the data given the model has the same value (its maximum value) for all $T_{\rm co}$-$M/R$ pairs related by 
equation~(\ref{eqn:validation:fitting:nonrot-nobkg-R_T:tempscaling}).
Consequently, any change in the surface temperature of the emission in the model being fit can be perfectly compensated by appropriate changes in the compactness $M/R$ of the model star and its distance. This is one of the strongest degeneracies in the waveform fitting problem.

\begin{figure*}[!t]
\begin{center}
\includegraphics[height=.26\textheight]{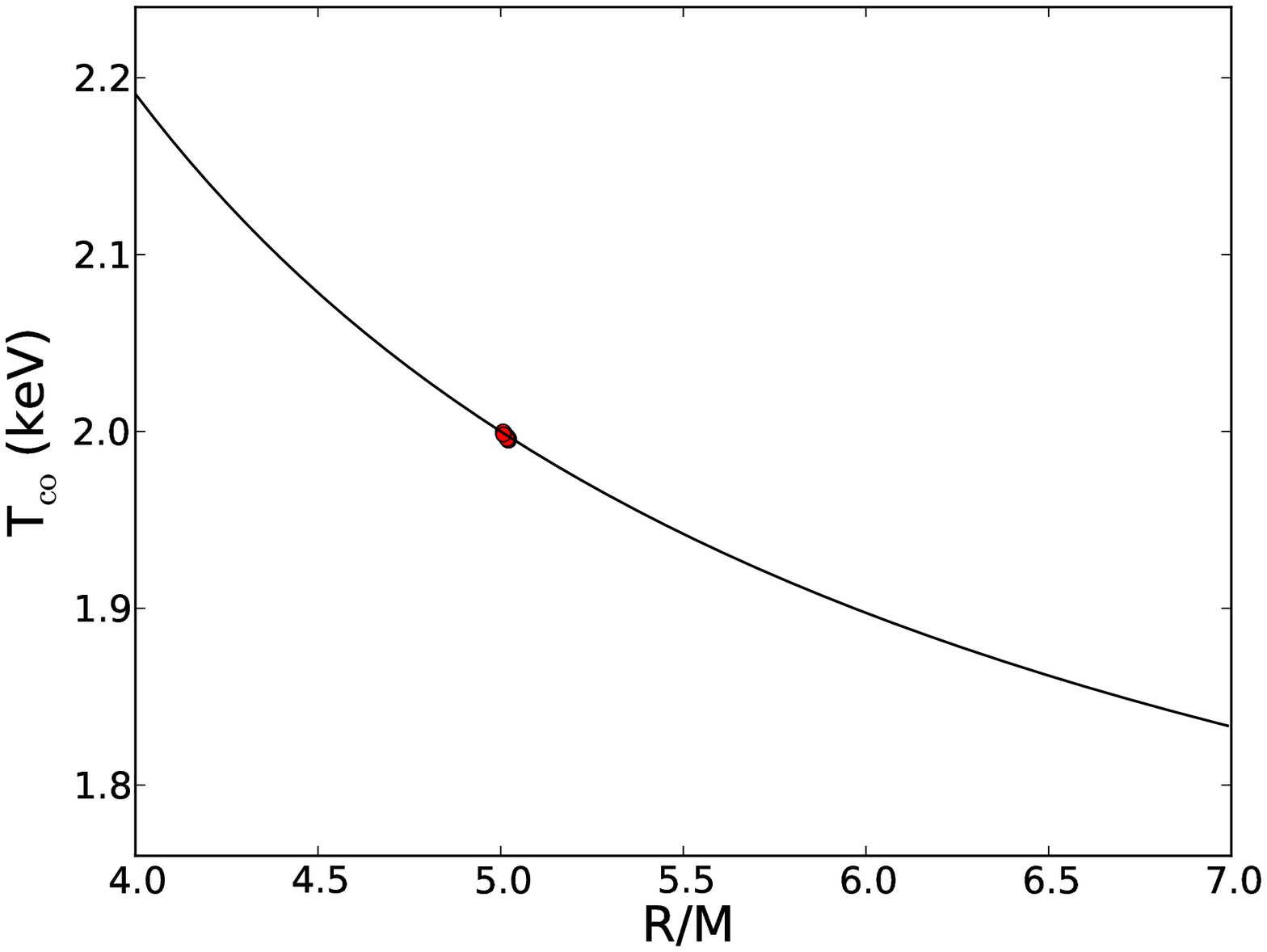}
\includegraphics[height=.26\textheight]{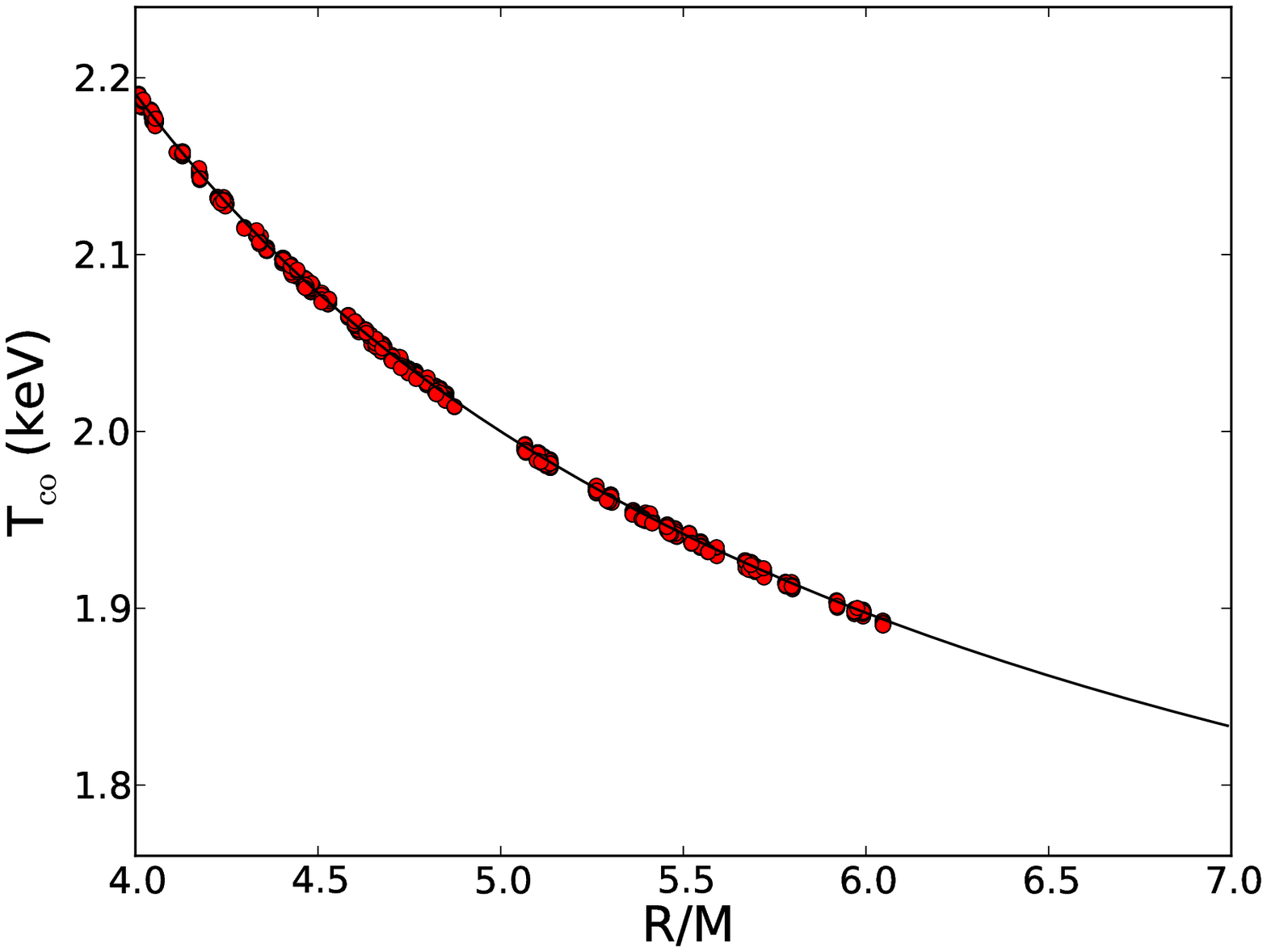}
\end{center}
\vspace{-0.5cm}
\caption{
Code test illustrating the strong degeneracy between the temperature $T_{\rm co}$ of the emission in the comoving frame and the stellar compactness $M/R$.
\textit{Left}: All parameters in the model except  $T_{\rm co}$ and $M/R$ were fixed in advance at the values used to generate the synthetic observed spectrum. 
The plotted points show the values of  $T_{\rm co}$ and $R/M$ that give log likelihoods within 1.15 (i.e., $1\sigma$ for 2 degrees of freedom) of the log likelihood given by the best-fit model. The models that give high likelihoods cluster tightly around $T_{\rm co}=2.0~{\rm keV}$ and $R/M=5.0$, the values used in generating the synthetic observed spectrum. 
\textit{Right}: All parameters except $T_{\rm co}$, $R/M$, and $d$ were fixed in advance at the values used to generate the synthetic spectrum.
The plotted points show the values of the parameters  $T_{\rm co}$ and $R/M$ that give log likelihoods within 1.73 (i.e., $1\sigma$ for 3 degrees of freedom) of the log likelihood given by the best-fit model. In this case, the models that give high likelihoods cluster tightly along the curve given by equation~(\ref{eqn:validation:fitting:nonrot-nobkg-R_T:tempscaling}) with $T_{\rm obs}$ equal to the observed value, since the normalization of each model can be freely adjusted by varying the distance to the star. The results shown here are for $M=1.6M_{\odot}$, $\Delta\theta=180^{\circ}$, $d=35~{\rm kpc}$, $\approx 10^6$ counts from hot spot, and no background. 
See Section~\ref{sec:tests:fitting:nonrot-nobkg-R_T} for further details.
\label{fig:tests:fitting:nonrot-nobkg-R_T}
}
\end{figure*}

This strong degeneracy can be broken for the synthetic data considered in this example if the mass $M$ of the star and the distance $d$ to the star are both known independently of the fitting process, because knowledge of these two quantities fixes the normalization of the observed spectrum. To see this, note that the normalization of the spectrum is proportional to $A/d^2$, where $A$ is the projected area of the emitting region. In the case considered here, $A=\pi R^2$ and is related to the stellar compactness and the mass through $R = (R/M) \times M$. Hence if $M$ and $d$ are both known independently, the temperature and flux of the observed radiation determines the flux at the stellar surface. This then fixes $R$ and hence $M/R$, since $M$ is assumed known. This result is illustrated in the left-hand panel of 
Figure~\ref{fig:tests:fitting:nonrot-nobkg-R_T}. If, however, $d$ is not known, the normalization of the spectrum from the stellar surface is unknown and can be freely adjusted, even if $M$ is known. In this case, we expect that all models having $T_0$ and $M/R$ values that fall on the curve given by equation~(\ref{eqn:validation:fitting:nonrot-nobkg-R_T:tempscaling}) will give good fits. The right-hand panel of 
Figure~\ref{fig:tests:fitting:nonrot-nobkg-R_T} illustrates this.

\subsubsection{Determining the spot inclination, observer inclination, and distance, given thermal emission from a small spot on a non-rotating star in flat spacetime with no background}
\label{sec:tests:fitting:nonrot-nobkg-ispot_iobs}

In this exercise, we test our model fitting code by fitting a model of emission from a non-rotating star in flat spacetime to the synthetic flux produced by a non-rotating star that emits radiation from a small spot, with a Planck spectrum. We assume that all the parameters in the model are independently known except the inclination of the spot and the observer, or the inclination of the spot and the observer and the distance to the star.

In flat spacetime, if $\theta_{\rm spot}+\theta_{\rm obs} < \frac{\pi}{2}$ (so that the spot is not occulted by the star), the flux seen by an observer at the stellar azimuth $\phi$ is
\begin{equation}
f(\phi) = \cos\theta_{\rm spot}\cos\theta_{\rm obs} + \sin\theta_{\rm spot}\sin\theta_{\rm obs}\cos\phi \;.
\end{equation}
This radiation pattern is a purely sinusoidal function of $\phi$ and can therefore be uniquely specified by the mean flux and the fractional rms amplitude of the flux variation with $\phi$. The
mean flux is
\begin{equation}
f_{\rm mean} = \cos\theta_{\rm spot}\cos\theta_{\rm obs} \;,\label{eqn:tests:fitting:nonrot-nobkg-ispot_iobs-fmean}
\end{equation}
while the fractional rms amplitude of the flux variation with $\phi$ is 
\begin{equation}
f_{\rm rms} = (1/\sqrt2)\tan\theta_{\rm spot}\tan\theta_{\rm obs} \;.
\label{eqn:tests:fitting:nonrot-nobkg-ispot_iobs-frms}
\end{equation}
For a given fractional rms amplitude,
equation~(\ref{eqn:tests:fitting:nonrot-nobkg-ispot_iobs-frms}) defines a curve in the $\theta_{\rm spot}$-$\theta_{\rm obs}$ plane along which the amplitude remains constant. 

\begin{figure*}[!t]
\begin{center}
\includegraphics[height=.26\textheight]{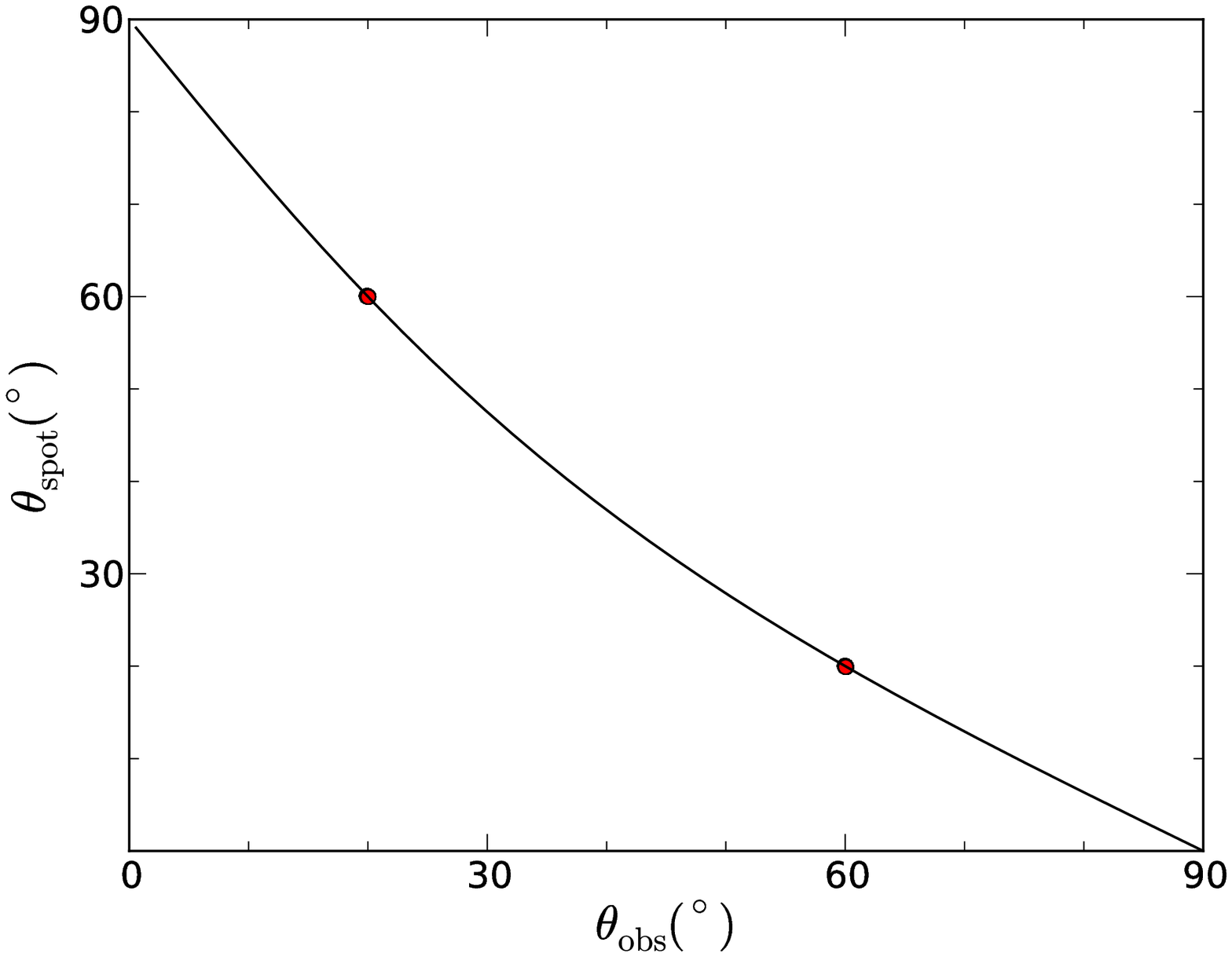}
\includegraphics[height=.26\textheight]{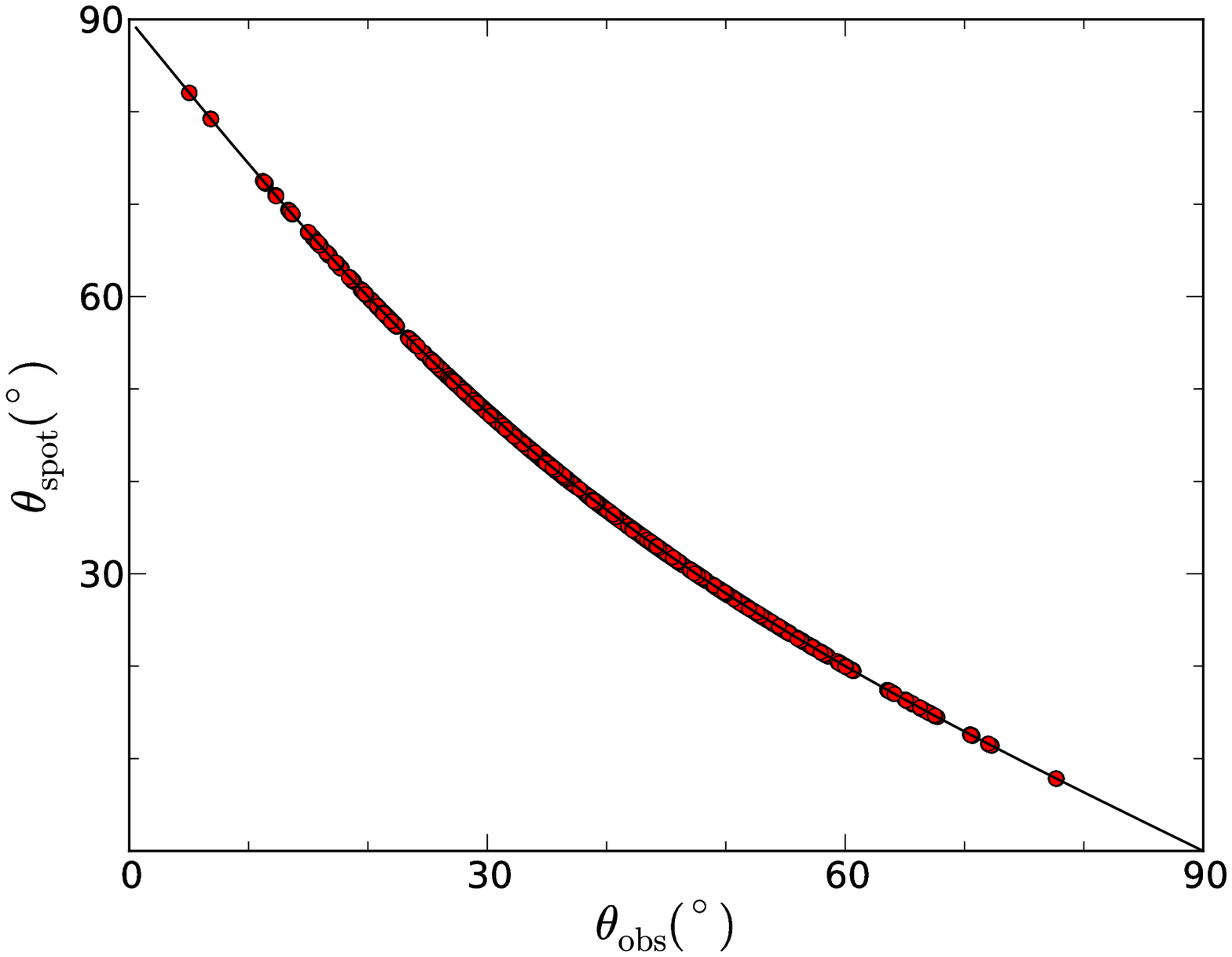}
\end{center}
\vspace{-0.5cm}
\caption{
Code test illustrating the strong degeneracy between the spot and observer inclinations $\theta_{\rm spot}$ and $\theta_{\rm obs}$.
\textit{Left}: All parameters in the model except $\theta_{\rm spot}$ and $\theta_{\rm obs}$ were fixed in advance at the values used to generate the synthetic radiation pattern. The plotted points show the values of $\theta_{\rm spot}$ and $\theta_{\rm obs}$ that give log likelihoods within 1.15 (i.e., $1\sigma$ for 2 degrees of freedom) of the log likelihood given by the best-fit model. 
The models that give high likelihoods cluster tightly around two points, one corresponding to the values of $\theta_{\rm spot}=20^\circ$ and $\theta_{\rm obs}=60^\circ$ used in generating the synthetic observed radiation pattern and the other given by interchanging the two values of $\theta_{\rm spot}$ and $\theta_{\rm obs}$, which leaves the mean flux, $f_{\rm mean}$, and the rms variation of the flux with $\phi$, $f_{\rm rms}$, unchanged 
[see equations~(\ref{eqn:tests:fitting:nonrot-nobkg-ispot_iobs-fmean}) and~(\ref{eqn:tests:fitting:nonrot-nobkg-ispot_iobs-frms})].
\textit{Right}: All parameters in the model except $\theta_{\rm spot}$ and $\theta_{\rm obs}$ and $d$ were fixed in advance at the values used to generate the synthetic radiation pattern.
The plotted points show the values of $\theta_{\rm spot}$ and $\theta_{\rm obs}$ that give log likelihoods within 1.73 (i.e., $1\sigma$ for 3 degrees of freedom) of the log likelihood given by the best-fit model. In this case, the models that give high likelihoods cluster tightly along the curve given by 
equation~(\ref{eqn:tests:fitting:nonrot-nobkg-ispot_iobs-frms}), where $f_{\rm mean}$ and $f_{\rm rms}$ remain unchanged, but are spread along the curve, because the normalization of each model can be freely adjusted by varying the distance to the star. 
The results shown here are for $R/M=10^6$, $M=1.6M_{\odot}$, $\theta_{\rm spot}=60^{\circ}$, $\theta_{\rm obs}=20^{\circ}$, $\Delta\theta=0.01$~radians, $T_{\rm co}=2.0~{\rm keV}$, $d=6~{\rm kpc}$, $\approx 10^6$ counts from the hot spot, and no background. 
See Section~\ref{sec:tests:fitting:nonrot-nobkg-ispot_iobs} for further details.
\label{fig:tests:fitting:nonrot-nobkg-ispot_iobs}
}
\end{figure*}

If the distance to the star is independently known, the model will provide a good fit to the data only close to two points on the curve given by 
equation~(\ref{eqn:tests:fitting:nonrot-nobkg-ispot_iobs-frms}). One of these is the point where $\theta_{\rm spot}$ and $\theta_{\rm obs}$ have the values that were used to generate the synthetic observed radiation pattern. The other is the point where the values of these two inclinations are interchanged, because the values of $f_{\rm mean}$ and $f_{\rm rms}$ are not changed by interchanging the two inclinations. Other points on the curve are strongly disfavored, because the value $f_{\rm mean}$ at these points will be very different from the observed value. If the distance is not independently known and must be included as a parameter in the fit, good fits can be found all along the curve, because the mean observed flux can be adjusted by varying the distance. These results are illustrated by the numerical results shown in 
Figure~\ref{fig:tests:fitting:nonrot-nobkg-ispot_iobs}.

\subsubsection{Determining the compactness, temperature, and distance, given thermal emission from a small spot on a non-rotating star in flat spacetime with a background}
\label{sec:tests:fitting:nonrot-withbkg-R_T}

This test is the same as the test discussed in 
Section~\ref{sec:tests:fitting:nonrot-nobkg-R_T}, except that here we add a background that is uniform in azimuth and has a thermal spectrum. In these fits, we treat the background in one of two ways:
\begin{enumerate}\itemsep0pt
\item we add to each model spectrum exactly the background that was used in generating the synthetic observed spectrum (i.e., we do not determine the background as part of the fitting procedure), or
\item we determine the background as part of the fitting procedure, using the code module that implements the procedure described in Section~\ref{sec:methods:estimatingMR:computational-procedure}.
\end{enumerate}
As subcases, we either set the distance to the value used in generating the synthetic observed spectrum during the fitting process (i.e., we assume that the distance is independently known), or we include the distance as one of the parameters to be determined in the fitting process. As explained before, assuming that the distance is independently known amounts to fixing the normalization of the spectrum, which more strongly constrains the fit.

\begin{figure*}[t!]
\begin{center}
\includegraphics[height=.26\textheight]{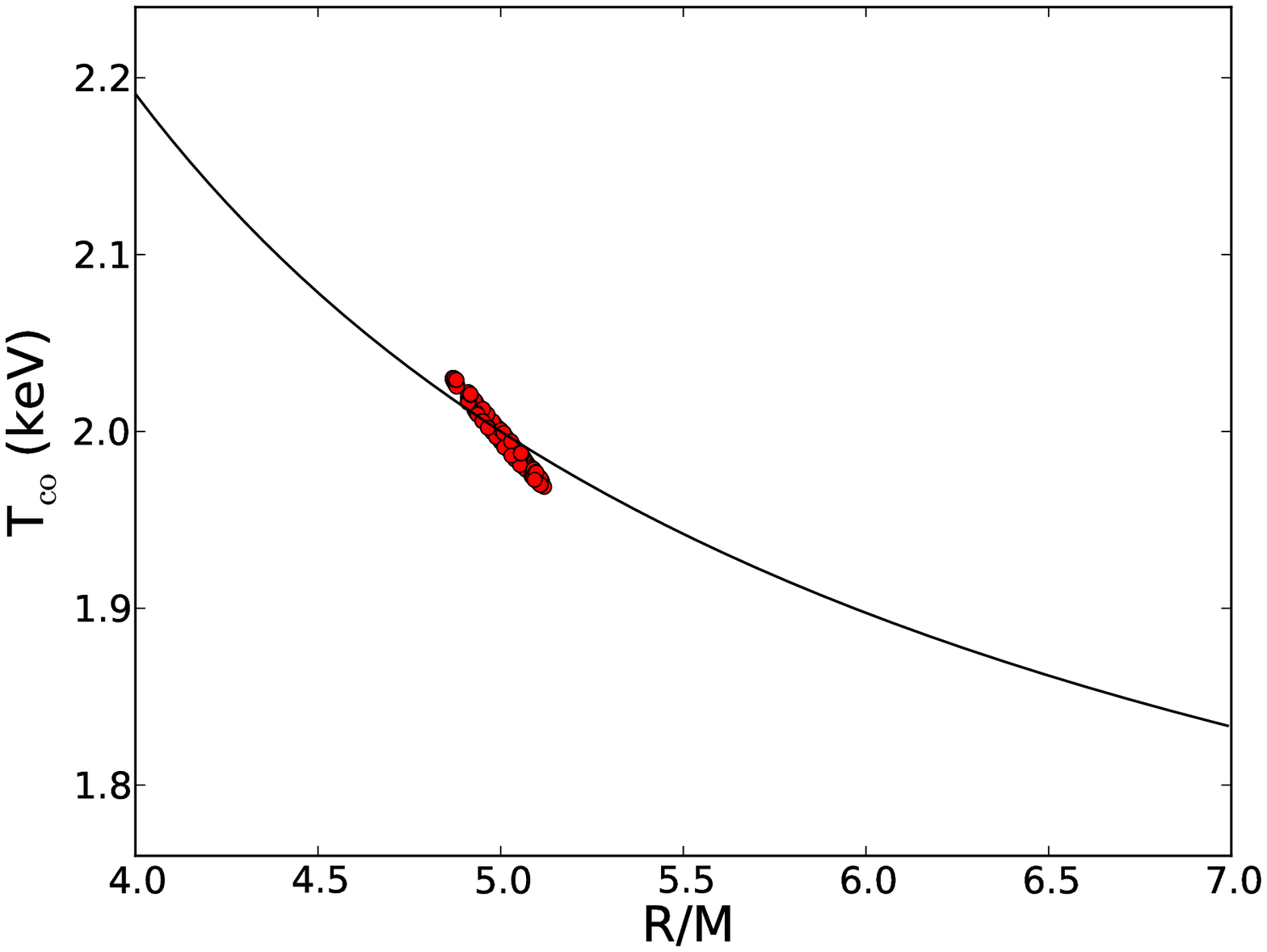}
\includegraphics[height=.26\textheight]{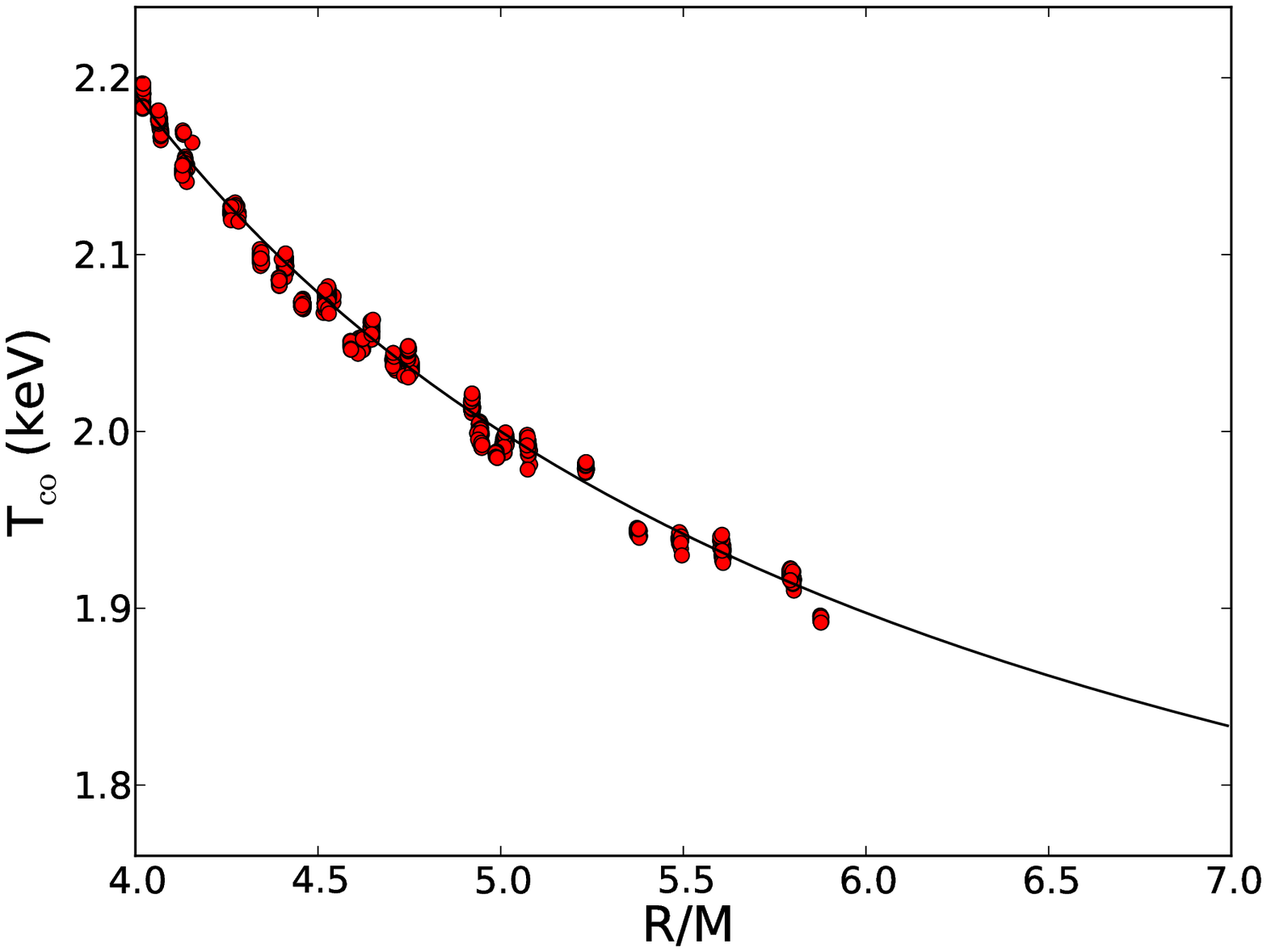} 
\end{center}
\vspace{-0.5cm}
\caption{
Code test illustrating the increased uncertainty in the fitted parameters induced by the additional fluctuations in the number of counts produced by a background component in the observed flux.
\textit{Left}: All parameters in the model except $T_{\rm co}$ and $M/R$ were fixed in advance at the values used to generate the synthetic observed spectrum. 
The plotted points show the values of $T_{\rm co}$ and $R/M$ that give log likelihoods within 1.15 (i.e., $1\sigma$ for 2 degrees of freedom) of the log likelihood given by the best-fit model. The models that give high likelihoods cluster around $T_{\rm co}=2.0~{\rm keV}$ and $R/M=5.0$, the values used in generating the synthetic observed spectrum, but not nearly as tightly as when there is no background. 
\textit{Right}: All parameters except $T_{\rm co}$, $R/M$, and $d$ were fixed in advance at the values used to generate the synthetic spectrum.
The plotted points show the values of the parameters $T_{\rm co}$ and $R/M$ that give log likelihoods within 1.73 (i.e., $1\sigma$ for 3 degrees of freedom) of the log likelihood given by the best-fit model. In this case, the models that give high likelihoods cluster along the curve given by equation~(\ref{eqn:validation:fitting:nonrot-nobkg-R_T:tempscaling}) with $T_{\rm obs}$ equal to the observed value, since the normalization of each model can be freely adjusted by varying the distance to the star. Again, they are not nearly as tightly clustered as when there is no background. The results shown here are for $M=1.6M_{\odot}$, $\Delta\theta=180^{\circ}$, $d=35~{\rm kpc}$, $\approx 10^6$ counts from the hot spot, $T_{\rm bkg}=1.5~{\rm keV}$, and $\approx 9 \times 10^6$ counts from the background. 
See Section~\ref{sec:tests:fitting:nonrot-withbkg-R_T} for further details.
\label{fig:tests:fitting:nonrot-withbkg-R_T}
}
\end{figure*}

Figure~\ref{fig:tests:fitting:nonrot-withbkg-R_T} shows that the increase in the total number of counts due to the additional counts contributed by the background, increases the uncertainties in $T_{\rm co}$ and $M/R$. In producing the fits shown in this figure, we added to each model spectrum exactly the background that was used in generating the synthetic observed spectrum. The left-hand panel in the figure shows the fit assuming that the distance to the star is independently known, whereas the right-hand panel shows the fit obtained if the distance is determined as part of the fitting procedure. Compared to the case without any background
(see Figure~\ref{fig:tests:fitting:nonrot-nobkg-R_T}), the scatter around the expected relationship between $T_{\rm co}$ and $M/R$ is larger when a background is present that when there is no background, even though the exact background has been added to each model. The reason for the increased scatter is the increase in the \textit{fluctuation} in the total number of counts, due to the counts contributed by the background (the fluctuation scales as the square root of the total number of counts).

\subsection{Convergence tests}
\label{sec:tests:convtests}

We wish to use values for the waveform code resolution parameters listed in Table~\ref{table:tests:WF:resolution-parameters} that will provide a resolution fine enough to meet our accuracy requirements, but no finer, so that our code runs as fast as possible. In order to determine these values of the resolution parameters, we perform a series of convergence tests using different values of these parameters. In these tests, we compare bolometric waveforms computed for different values of the resolution parameters, assuming the following values of the physical parameters: $R/M=5.0$, $M/M_{\odot}=1.6$, $\nu_{\rm rot}=600$~Hz, $\theta_{\rm spot}=90^{\circ}$, $\Delta\theta=25^{\circ}$, and $\theta_{\rm obs}=20^{\circ}$. We use the values of the resolution parameters listed in 
Table~\ref{table:tests:WF:resolution-parameters}, except that for clarity in the figures we plot the flux at only 16 values of the waveform phase. 

\begin{figure*}[!t]
\begin{center}
\includegraphics[height=.35\textheight]{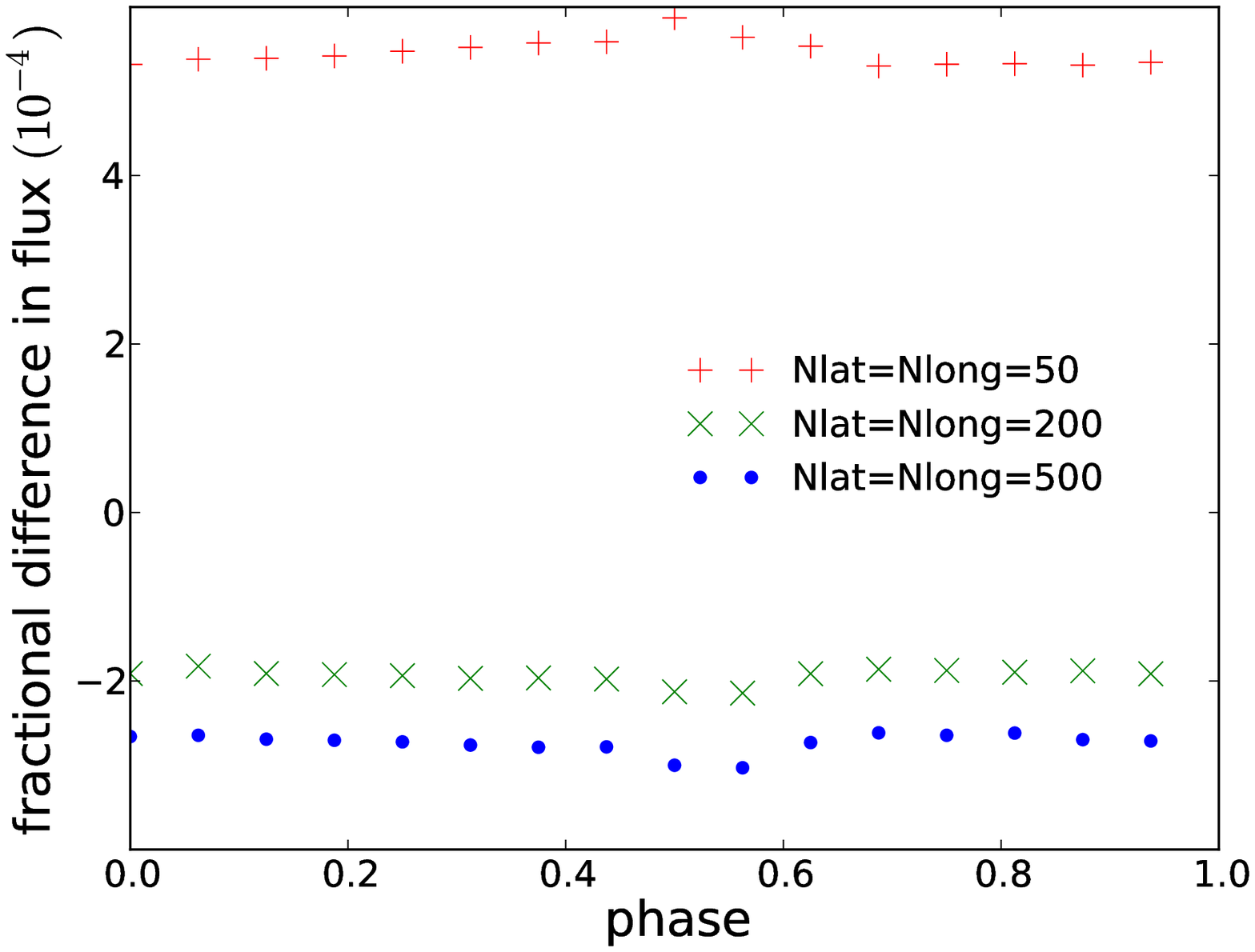}
\end{center}
\vspace{-0.8cm}
\caption{
Percentage differences between three bolometric waveforms computed using three different values of the spot resolution parameters \texttt{Nlat} and \texttt{Nlong} and a reference waveform computed using $\texttt{Nlat} = \texttt{Nlong} = 100$, the value we use in our waveform analyses and the code tests described in this appendix. This figure shows that for the values of these resolution parameters that we use, the fractional error in the bolometric waveform is $\lesssim3\times10^{-4}$.
See Section~\ref{sec:tests:convtests} for further details.
\label{fig:tests:convtest:nlat_nlong}
}
\end{figure*}

\begin{figure*}[!t]
\begin{center}
\includegraphics[height=.35\textheight]{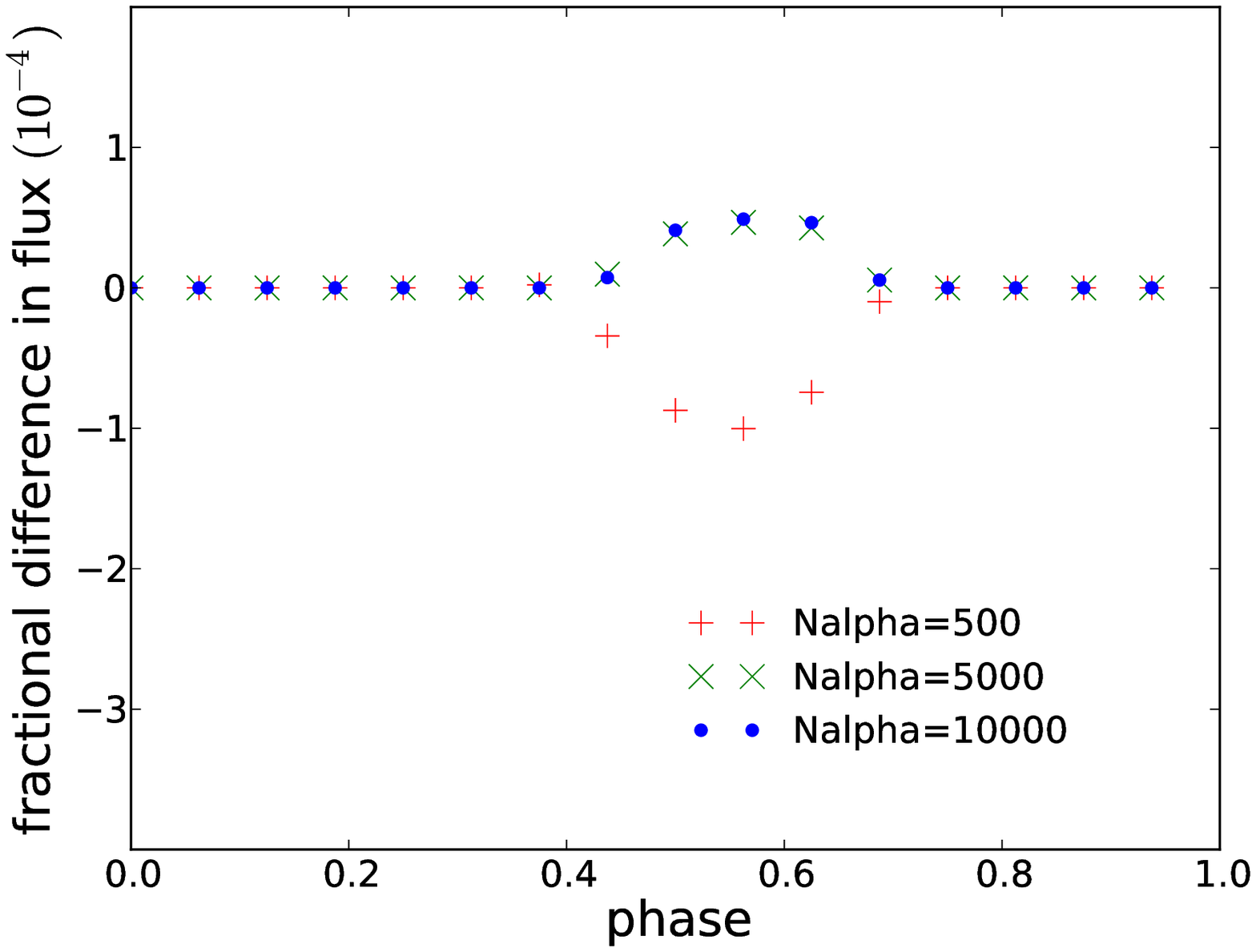}
\end{center}
\vspace{-0.8cm}
\caption{
Percentage differences between three bolometric waveforms computed using three different values of the ray direction resolution parameter \texttt{Nalpha} and a reference waveform computed using $\texttt{Nalpha} = 1000$, the value we use in our waveform analyses and the code tests described in this appendix. This figure shows that for the values of these resolution parameters that we use, the fractional error in the bolometric waveform is $\lesssim10^{-4}$.
See Section~\ref{sec:tests:convtests} for further details.
\label{fig:tests:convtest:nalpha}
}
\end{figure*}

\begin{figure*}[!t]
\begin{center}
\includegraphics[height=.35\textheight]{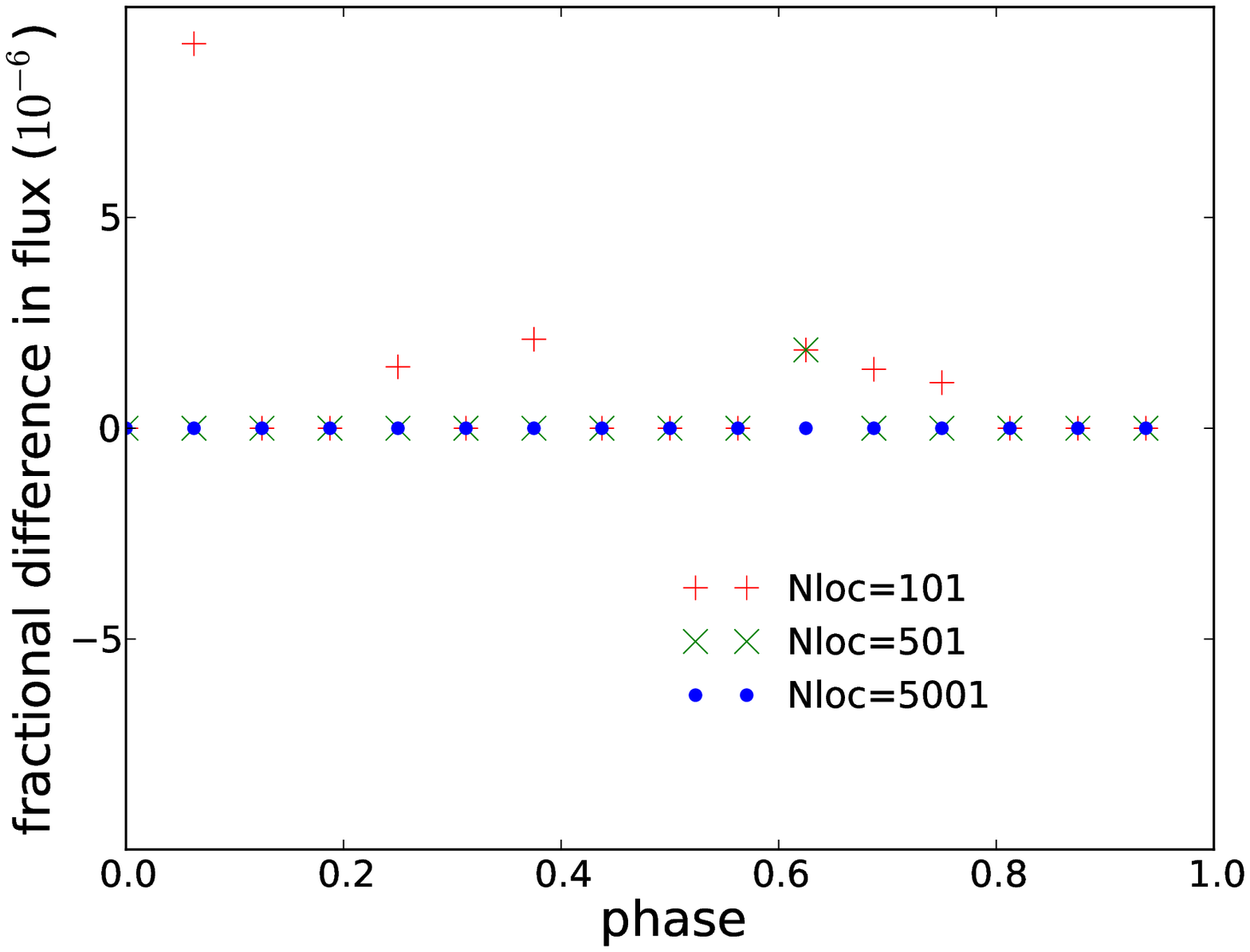}
\end{center}
\vspace{-0.8cm}
\caption{
Percentage differences between three bolometric waveforms computed using three different values of \texttt{Nloc}, the grid spacing in the waveform phase with  time delays included, and a reference waveform computed using $\texttt{Nloc} = 1001$, the value we use in the code tests described in this appendix. This figure shows that the fractional error in the bolometric waveform for this value of \texttt{Nloc} is $\lesssim10^{-5}$. 
See Section~\ref{sec:tests:convtests} for further details.
\label{fig:tests:convtest:nloc}
}
\end{figure*}

Figure~\ref{fig:tests:convtest:nlat_nlong} shows that our choice of 100 grid points in latitude and longitude ($\texttt{Nlat} = \texttt{Nlong} = 100$) resolves the hot spot well enough that the bolometric flux as a function of phase has a fractional error $\le 3\times10^{-4}$, which is adequate for our purposes.
Figure~\ref{fig:tests:convtest:nalpha} shows that 1000 grid points in the angle $\alpha'$ between the ray and the local normal ($\texttt{Nalpha} = 1000$) resolves this angle well enough that the bolometric flux has a fractional error $\le10^{-4}$, adequate for our purposes.
Figure~\ref{fig:tests:convtest:nloc} shows that 1001 grid points in the final value of the waveform phase ($\texttt{Nloc} = 1001$) resolves the phase well enough that the bolometric flux has a fractional error $\le10^{-5}$, which is more than adequate for our purposes. Indeed, $\texttt{Nloc} = 101$ provides essentially the same accuracy.

The tests of the values of the resolution parameters used in computing the deflection angle and the light travel time have been presented in 
Figures~\ref{fig:tests:WF:deflection:compare_n}
\ref{fig:tests:WF:deflection:compare_eps},
\ref{fig:tests:WF:dtime:compare_n},
and~\ref{fig:tests:WF:dtime:compare_eps_delt}.

\clearpage
\newpage


\section {JOINT FITS}
\label{app:jointfits}

\setcounter{table}{0}
\renewcommand\thetable{\Alph{section}\arabic{table}}

\setcounter{figure}{0}
\renewcommand\thefigure{\Alph{section}\arabic{figure}}

In this Appendix we discuss the constraints on the waveform model parameters that can be obtained by jointly fitting a collection of different, independent sets of waveform data. These sets could be different segments of a single burst or different bursts from the same star. 

We show here that the constraints on the interesting parameters that can be obtained by jointly fitting many data sets are often comparable to the constraints that would have been obtained by fitting a single data set that has an oscillation profile which is the same as the average oscillation profile of the multiple data sets and has the same total number of counts as the multiple data sets have when combined. This is true even though one might intuitively expect that the extra parameters required to fit many data sets in which the uninteresting parameters change from set to set would compromise the constraints on the (unchanging) interesting parameters.

We assume that the values of some of the parameters in the model waveform, such as the mass $M$ and radius $R$ of the star, the inclination $\theta_{\rm obs}$ of the observer, and the distance $d$ to the star remain unchanged during a burst and from burst-to-burst, whereas the values of other parameters, such as the angular radius $\Delta\theta$ of the hot spot, the inclination $\theta_c$ of the spot center, and the color temperature $T_{\rm co}$ of the emission from the spot, may change during a burst and from burst-to-burst. The parameters that remain unchanged must be treated differently from the parameters that may change.

Suppose that we have several independent data sets and wish to extract joint constraints on the subset of waveform parameters $\alpha_i$ that have the same values in all the data sets. We denote the other waveform parameters, which may change between data sets, by $\beta_j$. We can derive the correct approach to jointly fitting multiple data sets by starting from the Bayesian expression for the unnormalized posterior probability density 
\begin{equation}
q(\alpha_i)=\int d\beta_j P(\alpha_i,\beta_j){\cal L}(\alpha_i,\beta_j) \;.
\end{equation}
Here $P$ is the joint prior probability density over all the parameters and ${\cal L}$ is the likelihood of all the data, given the model and specific values for all the model parameters. If there are multiple data sets, then the joint likelihood is the product of the individual likelihoods, i.e.,
\begin{equation}
{\cal L}=\prod_k {\cal L}_k(\alpha_i,\beta_{jk}) \;,
\end{equation}
where here we write $\beta_{jk}$ to indicate the set of potentially variable parameters that is being evaluated for data set $k$.

We assume that the prior probability density distribution for the fixed parameters $\alpha_i$ is the same for each data set. Then we can write the full prior probability distribution as
\begin{equation}
P(\alpha_i,\beta_{j1},\beta_{j2},\ldots)=P(\alpha_i)P_1(\beta_{j1}|\alpha_i)P_2(\beta_{j2}|\alpha_i)\cdots
\end{equation}
where the product is over all data sets and the vertical bar indicates a conditional probability. For example, $P_1(\beta_{j1}|\alpha_i)$ is the prior probability of the variable parameters $\beta_{j1}$ in the first data set given that the fixed parameters have the values $\alpha_i$.

The posterior probability density of our fixed parameters is thus proportional to
\begin{equation}
q(\alpha_i)\propto P(\alpha_i)\prod_k\left[\int d\beta_{jk}P_k(\beta_{jk}|\alpha_i){\cal L}_k(\alpha_i,\beta_{jk})\right]\; .
\label{eqn:posterior}
\end{equation}
Thus the posterior probability density is equal to the prior probability density times the product of the integrals over the variable parameters for each data set, of the prior probability density of the variable parameters times the likelihood of the data given the values of all the parameters. The process of integrating over the uninteresting parameters (called nuisance parameters in this context) is called marginalization. Marginalization over the nuisance parameters means that in some sense they do not count in the tally of parameters that are adjusted during the fitting process. It is this part of the procedure that ultimately yields constraints from many data sets that are not muddled by the introduction of many fit parameters.

It is important to note that the posterior probability density is usually \textit{not} the product of the marginalized probability densities $Q_k(\alpha_i)$ for each of the data sets.  That is, 
\begin{equation}
q(\alpha_i)\neq \prod_k \left[\int d\beta_{jk}
P(\alpha_i) P(\beta_{jk}|\alpha_i) {\cal L}_k(\alpha_i,\beta_{jk})\right]\; .
\end{equation}
For $n$ data sets, this incorrect expression for the posterior probability density would yield 
\begin{equation}
q(\alpha_i)\propto [P(\alpha_i)]^n \prod_k \left[\int d\beta_{jk} P_k(\beta_{jk}) {\cal L}_k(\alpha_i,\beta_{jk})\right]\; .
\end{equation}
In this incorrect expression for $n$ data sets, there are $n$ factors of the prior probability density, compared with only one in the correct expression. To see that $n$ factors is incorrect, imagine a case in which the prior probability distribution is a multivariate Gaussian and that only one of a large number $n$ of data sets is informative. The incorrect expression would nevertheless produce very tight constraints, because it raises the Gaussian to a high power.

The posterior probability density (\ref{eqn:posterior}) is not in general proportional to the posterior probability density one would obtain from an equivalent number of counts in a profile that has the average shape of the profile in the multiple data sets. To see this, let's perform a thought experiment in which we have obtained a single, continuous segment of data during which all the waveform parameters are known to be fixed, but we analyze it as some number of contiguous data sets where the values of the parameters $\alpha_i$ are fixed but the values of the $\beta_{jk}$ need not be. Then the joint analysis gives equation~(\ref{eqn:posterior}), whereas the analysis that assumes all the parameters are fixed gives
\begin{equation}
q(\alpha_i)\propto P(\alpha_i)\int d\beta_j P(\beta_j|\alpha_i){\cal L}(\alpha_i,\beta_j) \;,
\label{eqn:all-parameters-fixed}
\end{equation}
where as before ${\cal L}=\prod_k {\cal L}_k(\alpha_i,\beta_{j})$. Note that instead of allowing the $\beta_j$ to take on different values $\beta_{jk}$ in each data set, here they are assumed to have the same values. In general, 
equation~(\ref{eqn:all-parameters-fixed}) does not give the same posterior as equation~(\ref{eqn:posterior}).

There are, however, circumstances in which the posteriors are identical. Consider a situation in which (1)~the prior for the variable parameters is independent of the fixed parameters, and (2)~the likelihood is the product of two independent functions, one for $\alpha_i$ and one for $\beta_{jk}$:
\begin{equation}
{\cal L}_k(\alpha_i,\beta_{jk})={\cal L}_{k\alpha}{\cal L}_{k\beta}\; .
\label{eqn:independent-likelihoods}
\end{equation}
Then
\begin{equation}
\begin{array}{rl}
q(\alpha_i)&\propto P(\alpha_i)\prod_k\left[\int d\beta_{jk}P_k(\beta_{jk}){\cal L}_k(\alpha_i,\beta_{jk})\right]\\
&=\left[P(\alpha_i)\prod_k{\cal L}_{k\alpha}\right]\prod_k\left[\int d\beta_{jk}P_k(\beta_{jk}){\cal L}_{k\beta}\right]\; .
\end{array}
\label{eqn:posterior-independent-likelihoods}
\end{equation}
The second factor in 
equation~(\ref{eqn:posterior-independent-likelihoods}) is independent of $\alpha$ and therefore enters the posterior as a constant, which will be normalized away. The product $\prod_k{\cal L}_{k\alpha}$ is just the total likelihood, and the posterior will therefore be unchanged by dividing the original single data set into multiple data sets.

In order to investigate whether this general result applies to the problem studied in this paper, namely, analysis of multiple segments of burst oscillation data, we have carried out a number tests. In these tests we compared the constraints obtained by analyzing a set of waveform data as single segment and as a sequence of segments. We find that in practice the constraints obtained by jointly analyzing multiple data sets are usually comparable to the constraints obtained by analyzing a single data set with an oscillation profile similar to the average profile of the oscillations in the multiple data sets and with the same total number of counts as there are in the multiple data sets.

\begin{figure*}[!t]
\begin{center}
\includegraphics[height=.26\textheight]{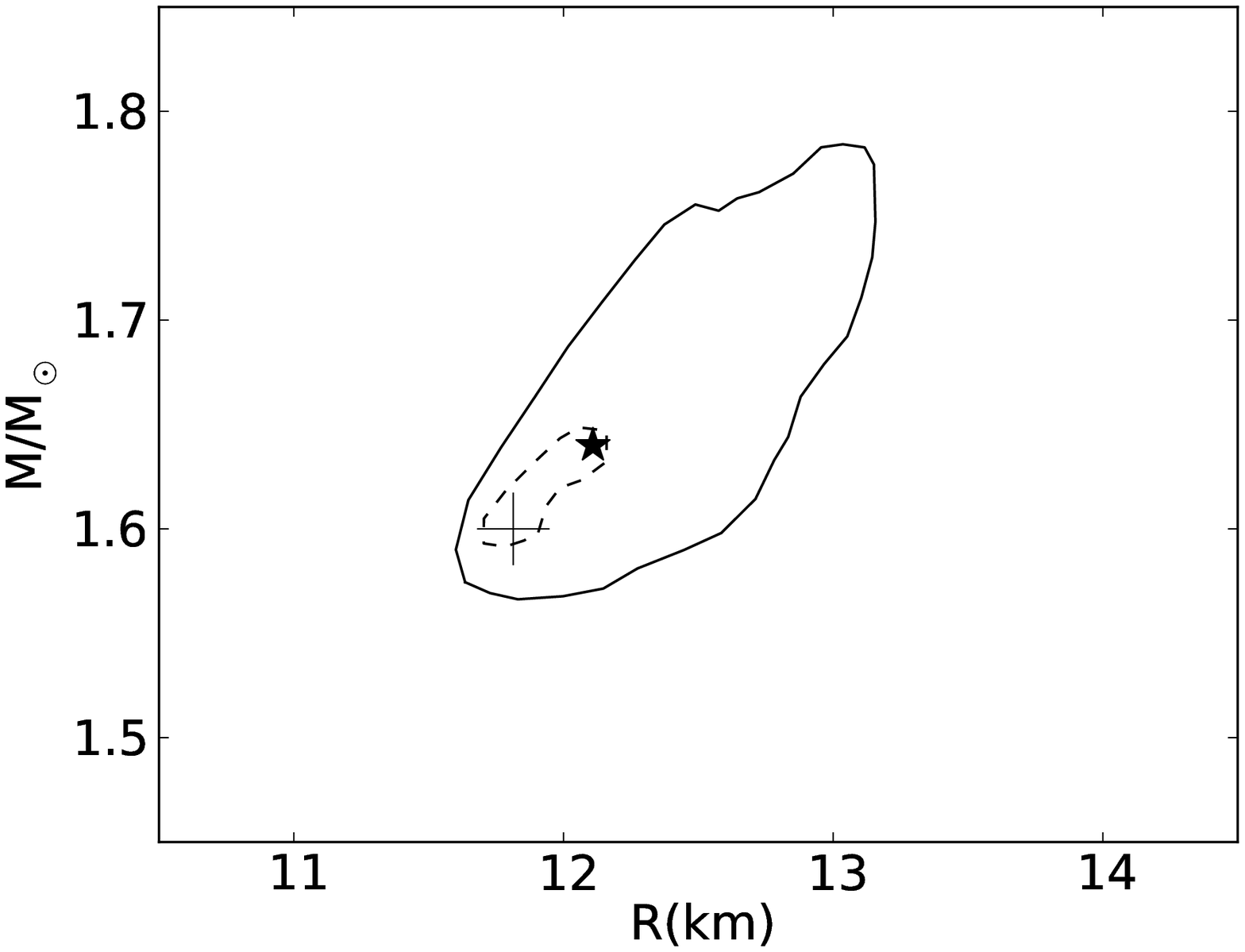}
\includegraphics[height=.26\textheight]{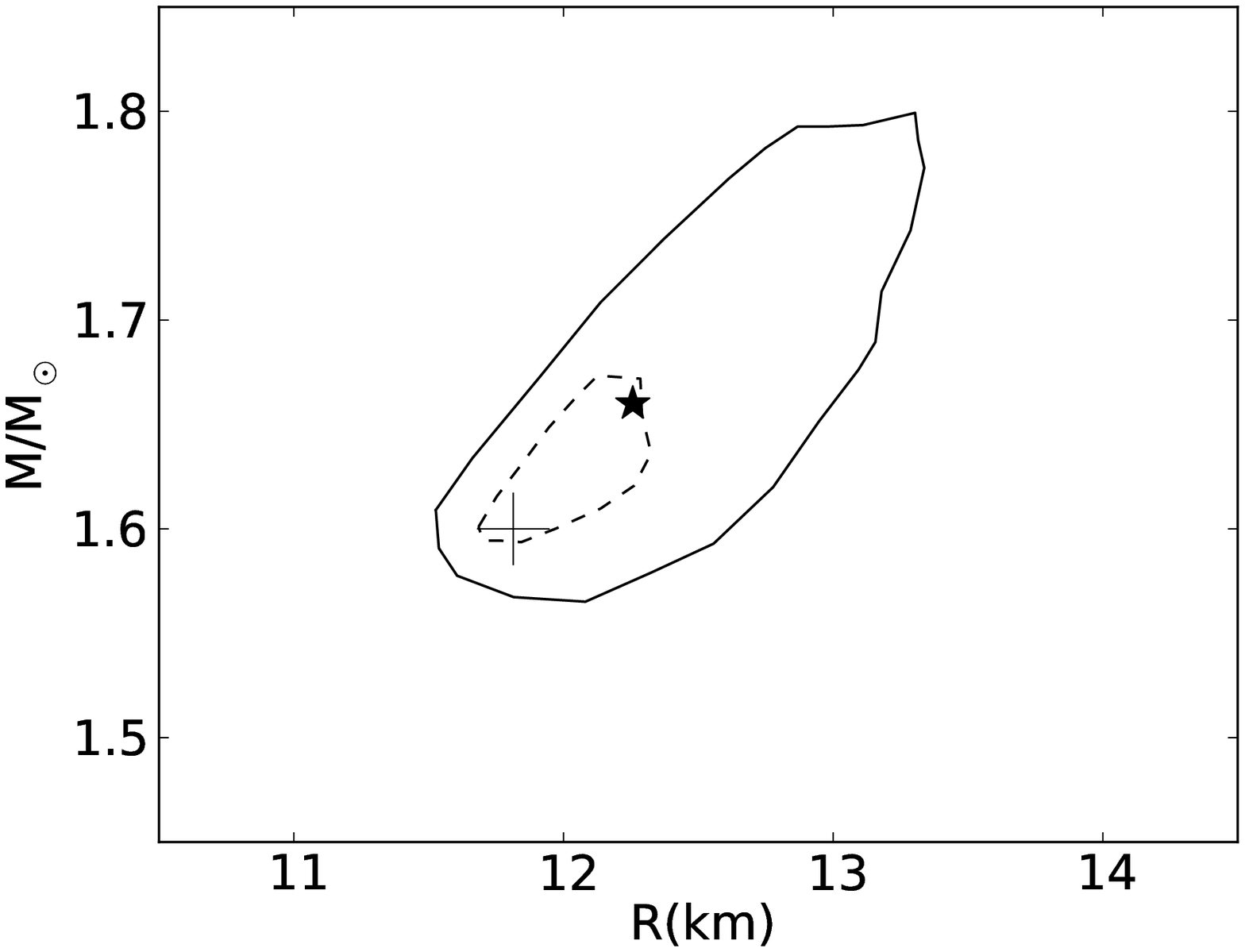}
\end{center}
\vspace{-0.5cm}
\caption{
An example showing that the constraints on $M$ and $R$ obtained by jointly analyzing multiple data sets as described in the text are usually comparable to the constraints obtained by analyzing a single data set with an oscillation profile similar to the average profile of the oscillations in the multiple data sets and with the same total number of counts as there are in the multiple data sets. The results shown here are for the high-inclination reference case with our medium background and no independent knowledge of the parameters in the model.
\textit{Left}: Constraints obtained by analyzing a single data set with about $10^6$ counts from the hot spot. 
\textit{Right}: Constraints obtained by breaking up the single data set into five segments of equal length and then analyzing the segments jointly.
\label{fig:app:joint-fits:unknown-distance}
}
\end{figure*}

\begin{figure*}[!t]
\begin{center}
\includegraphics[height=.26\textheight]{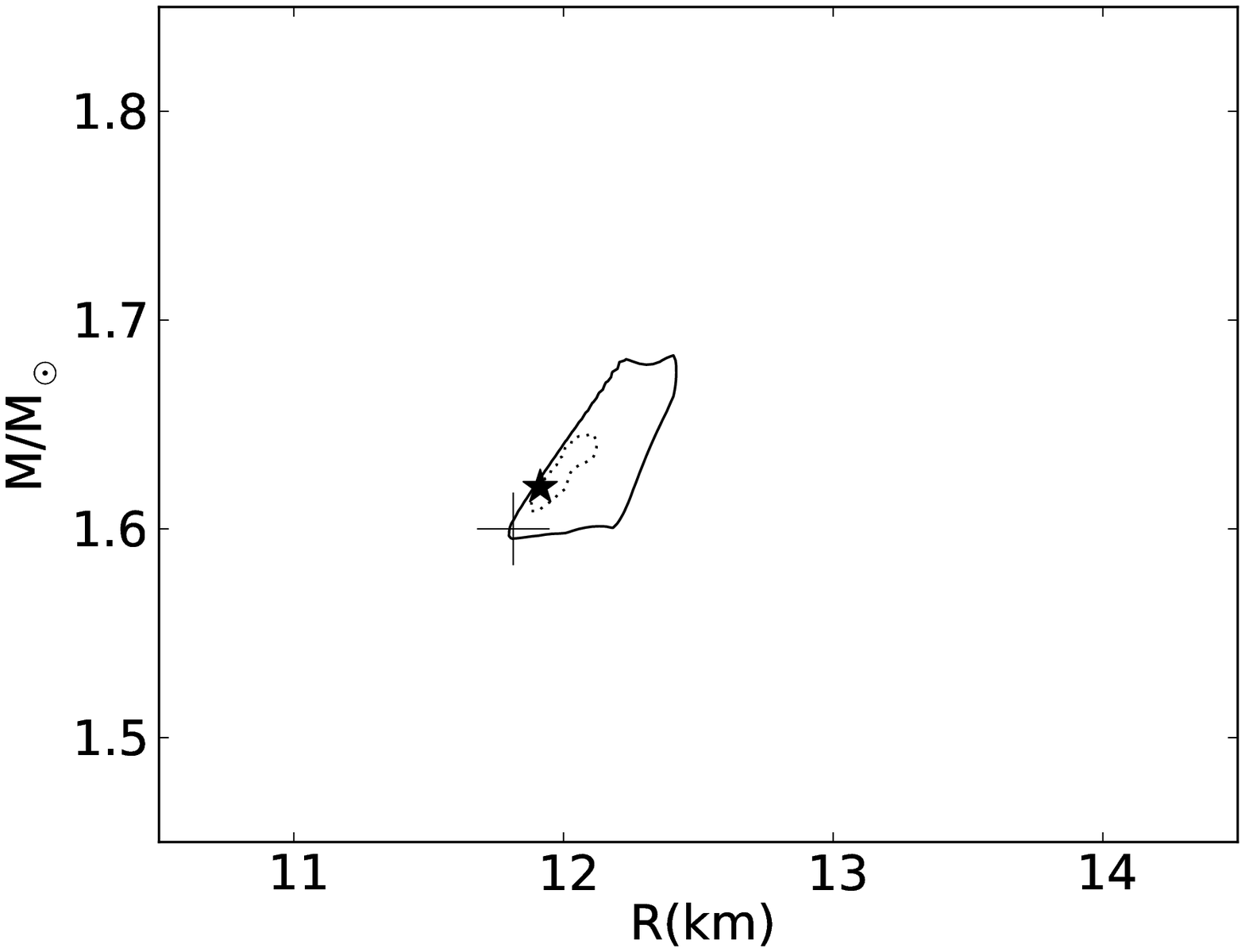}
\includegraphics[height=.26\textheight]{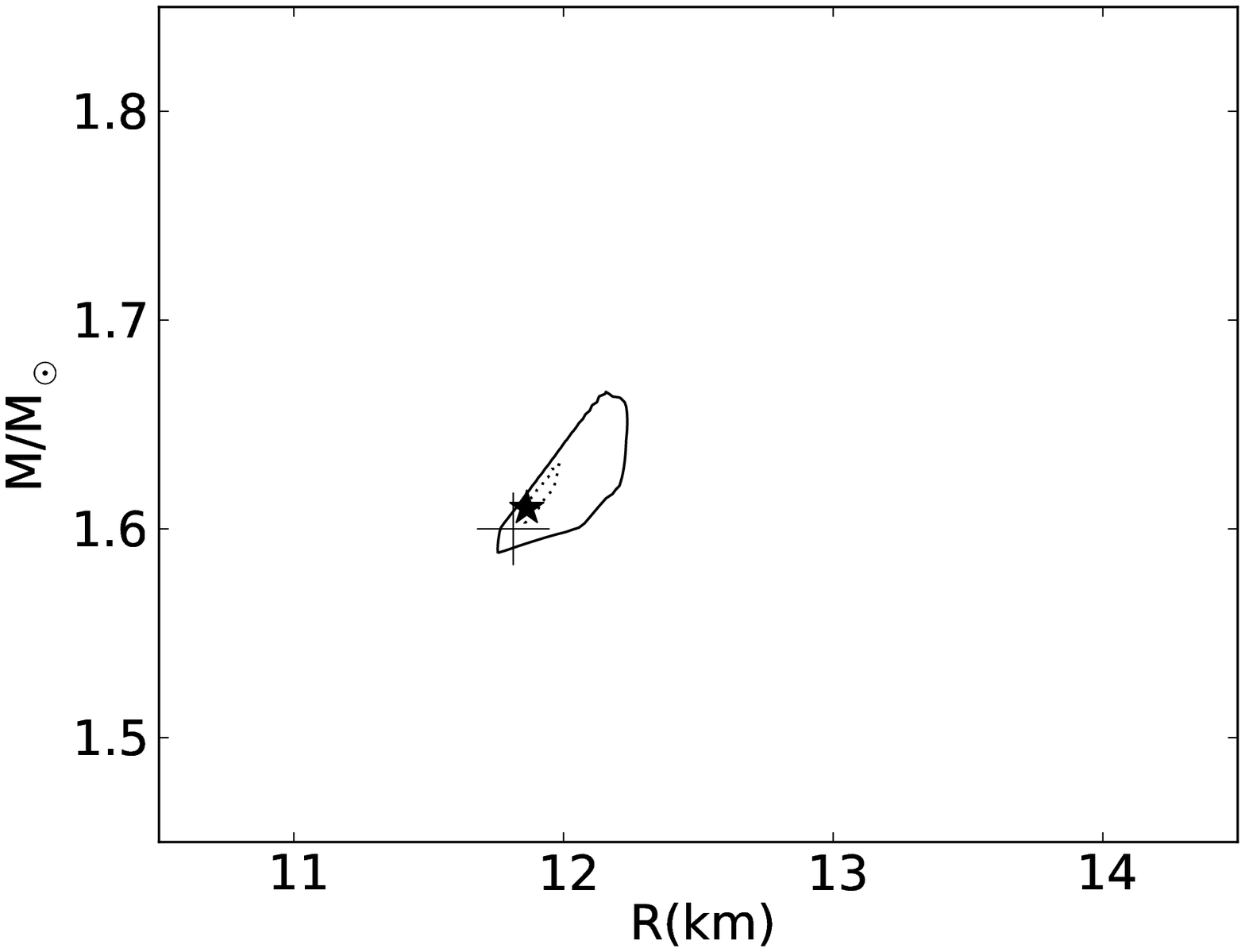}
\end{center}
\vspace{-0.5cm}
\caption{
Another example showing that the constraints on $M$ and $R$ obtained by jointly analyzing multiple data sets as described in the text are usually comparable to the constraints obtained by analyzing a single data set with an oscillation profile similar to the average profile of the oscillations in the multiple data sets and with the same total number of counts as there are in the multiple data sets. The results shown here are for the high-inclination reference case with our medium background and independent knowledge of the distance to the star.
\textit{Left}: Constraints obtained by analyzing a single data set with about $10^6$ counts from the hot spot. 
\textit{Right}: Constraints obtained by breaking up the single data set into five segments of equal length and then analyzing the segments jointly.
\label{fig:app:joint-fits:known-distance}
}
\end{figure*}

We illustrate these results by the following two examples. Both compare the constraints on $M$ and $R$ obtained by analyzing a single data set that has about $10^6$ counts from the hot spot with the constraints obtained by breaking up the single data set into five segments of equal length and then analyzing the segments jointly, using the method described above. All the confidence regions shown here are for the high-inclination reference case and our medium background.
Figure~\ref{fig:app:joint-fits:unknown-distance} compares the constraints obtained when no independent information about the parameters in the model is available.
Figure~\ref{fig:app:joint-fits:known-distance} compares the constraints obtained when the distance to the star is known independently of the waveform analysis.
In both cases, the constraints on $M$ and $R$ obtained by jointly analyzing multiple data sets as described in the text are comparable to the constraints obtained by analyzing a single data set with an oscillation profile similar to the average profile of the oscillations in the multiple data sets and with the same total number of counts as there are in the multiple data sets.

\clearpage
\newpage

\bibliography{paper}

\end{document}